\documentclass[article, shortnames]{jss}

\usepackage{textcomp,mathrsfs}
\usepackage{graphicx,amssymb,amsmath,amsthm,mathrsfs}
\usepackage{multirow,makeidx,algorithmic,algorithm}
\usepackage{latexsym,graphicx,amssymb,amsmath,amsthm,mathrsfs}
\usepackage{mathrsfs}
\usepackage{amssymb}
\usepackage{booktabs,fancyhdr} 
\usepackage{rotating}
\usepackage{pdflscape}
\usepackage{graphicx}
\usepackage{array}
\usepackage{bigints}
\usepackage{listings}
\usepackage{mathtools}
\usepackage{amsthm}
\usepackage{amsmath}
\usepackage{caption}
\usepackage{subcaption}
\usepackage{outlines}

\usepackage{hyperref}

\newcounter{xxx}
\usepackage{todonotes}

\newcommand{\longline}{\makebox[\linewidth]{\rule{15.5cm}{0.4pt}}}

\newcommand{\blang}{\proglang{Blang} }

\newcommand{\api}[3]{\url{https://www.stat.ubc.ca/~bouchard/blang/javadoc-#1/#2.html}}

\renewcommand\P{{\mathbb P}}        
\newcommand\I{{\mathbf 1}}        

\newcommand{\ud}{\,\mathrm{d}}    

\definecolor{purple}{rgb}{0.50, 0.0, 0.33}
\lstdefinelanguage{blang}{
  keywords={package, model, laws, generate, random, param, for, import, it, logf, class, is, var, val, int, if, else, static, return, throw, new, try, catch, finally, def, implements, boolean, true, false, override, void},
  keywordstyle=\color{purple}\bfseries,
  keywords=[2]{number, objectid},
  keywordstyle=[2]\color{green}\bfseries,
  identifierstyle=\color{black},
  sensitive=true,
  comment=[l]{//},
  morecomment=[s]{/*}{*/},
  commentstyle=\color{gray}\ttfamily,
  stringstyle=\color{blue}\ttfamily,
  morestring=[b]',
  morestring=[b]"
}

\usepackage{textcomp}

\usepackage{thumbpdf,lmodern}

\usepackage{tikz}
\usetikzlibrary{bayesnet}
\usepackage{amsmath,amssymb}
\usepackage{graphicx, subcaption}
\usepackage{float}

\usepackage{framed}

\author{Alexandre Bouchard-C\^{o}t\'{e}\\University of British Columbia\And Kevin Chern\\University of British Columbia\AND Davor Cubranic\\University of British Columbia\And Sahand Hosseini\\University of British Columbia\AND Justin Hume\\Third Foundation Labs Inc.\And Matteo Lepur\\University of British Columbia\AND Zihui Ouyang\\University of British Columbia\And Giorgio Sgarbi\\University of British Columbia}
\Plainauthor{Alexandre Bouchard-C\^{o}t\'{e}, Kevin Chern, Davor Cubranic, Sahand Hosseini, Justin Hume, Matteo Lepur, Zihui Ouyang, Giorgio Sgarbi}

\title{Blang: Bayesian Declarative Modelling of General Data Structures and  Inference via Algorithms Based on Distribution Continua}
\Plaintitle{Blang: Bayesian Declarative Modelling of General Data Structures and  Inference via Algorithms Based on Distribution Continua}
\Shorttitle{Bayesian Modelling of General Data Structures}

\Abstract {
Consider a Bayesian inference problem where a variable of interest does not take values in a Euclidean space. These ``non-standard'' data structures are in reality fairly common. They are frequently used in problems involving latent discrete factor models, networks, and domain specific problems such as sequence alignments and reconstructions, pedigrees, and phylogenies. In principle, Bayesian inference should be particularly well-suited in such scenarios, as the Bayesian paradigm provides a principled way to obtain confidence assessment for random variables of any type. However, much of the recent work on making Bayesian analysis more accessible and computationally efficient has focused on inference in Euclidean spaces.

In this paper, we introduce \proglang{Blang}, a domain specific language and library aimed at bridging this gap.
\proglang{Blang} allows users to perform Bayesian analysis on arbitrary data types while using a declarative syntax similar to \proglang{BUGS}.
\proglang{Blang} is augmented with intuitive language additions to create data types of the user's choosing.
To perform inference at scale on such arbitrary state spaces, \proglang{Blang} leverages recent advances in sequential Monte Carlo and non-reversible Markov chain Monte Carlo methods.
}

\Plainkeywords{Bayesian modelling language, Bayesian inference, non-standard data structures, Blang}

\Address{
  Alexandre Bouchard-C\^{o}t\'{e} \\
  Department of Statistics\\
  Faculty of Science\\
  University of British Columbia\\
  3182 Earth Sciences Building, 2207 Main Mall
  Vancouver, BC Canada V6T 1Z4\\
  E-mail: \email{bouchard@stat.ubc.ca}\\
  URL: \url{https://www.stat.ubc.ca/~bouchard/}
}

\newcommand{\updated}[1]{{\leavevmode #1}}

\begin{document}
\section[Introduction]{Introduction} \label{sec:intro}


\proglang{Blang} is a probabilistic programming language (PPL) and software development kit (SDK) for performing Bayesian data analysis.  
Its design supports scalable inference over arbitrary data types, in particular, combinatorial spaces which are of central importance in areas such as computational biology. 
The design philosophy is centred around the day-to-day requirements of real-world data science. 
In the following, we put \proglang{Blang} in the context of the rich PPL and Bayesian modelling ecosystem.

Probabilistic programming has revolutionized applied Bayesian statistics in the past two decades; now being a part of the core toolbox of applied statistics. 
For example, packages such as \proglang{BUGS} \citep{Lunn2000, Lunn2009, Lunn2012}, \proglang{JAGS} \citep{Plummer2003}, \proglang{Stan} \citep{Carpenter2017}, and \proglang{PyMC3} \citep{Salvatier2015} have been widely used in various applications ranging from ecology \citep{Semmens2009}, to astronomy \citep{Greiner2016}, and psychology \citep{Paul2019}. See \cite{van_de_meent_introduction_2018} for a recent survey.  

In recent years, research in the area of Bayesian modelling software has focused on two main directions. 
On one hand, considerable progress has been made in designing general-purpose PPLs \citep{anglican, probabilisticC, blog, church} which are able to represent any computable probability distributions \citep{Roy2017}. 
However, inference in these powerful languages often has to resort to  algorithms such as non-Markovian Sequential Monte Carlo that can have poor scalability. 
Rapid progress is being made to lift this limitation (e.g., \cite{paige_inference_2016,Yuan19,ronquist2020probabilistic}) but in typical applications that involve challenging combinatorial spaces, general purpose PPL inference engine are not yet able to match the performance of specialized samplers.  
A second area of active development (\cite{Carpenter2017,Salvatier2015,bingham2018pyro}, \emph{inter alia}) has been to use automatic differentiation combined with Hamiltonian Monte Carlo (HMC) sampling \citep{Duane1987,Neal2012}, which is highly efficient in problems defined on continuous state spaces.
Naturally, algorithms based on HMC are not necessarily well-suited for inference problems defined on discrete and combinatorial state spaces. 

In the past, efficient sampling in combinatorial spaces has been achieved by designing portfolios of specialized samplers in a case-by-case basis (see e.g., \cite{Lakner2008}).
This process is typically time consuming and error prone.
There is an opportunity to simplify this process, minimize manual intervention of tuning algorithms, and to speed-up and parallelize inference. 
This is possible with new developments in computational statistics such as non-reversible Markov chain Monte Carlo (MCMC) methods \citep{Syed2019} based on parallel tempering (PT) \citep{Geyer1991}, and a non-standard flavour of sequential Monte Carlo (SMC) method that we call Sequential Change of Measure (SCM; to avoid confusion with state-space SMC \citep{DelMoral2006,Neal2001}) augmented with adaptive schemes \citep{Zhou2016}. 
All these schemes are based on a continuum of probability distributions, all defined on the same space and interpolating between the prior and posterior. 
The benefit of these methods is that a simplistic set of sampling algorithms can still achieve high sampling efficiency while exploiting parallel architectures. 
\proglang{Blang} fully automates the construction of interpolating probability distributions and therefore democratizes the use of high-performance Monte Carlo schemes such as non-reversible PT and SCM.

\proglang{Blang} is designed to be efficient not only in computational terms but also for the user's development time. 
To achieve this goal, considerable effort has been put to facilitate model construction, testing, reuse and integration into existing data analysis pipelines, and to support reproducible data analysis.
Instead of creating a language from scratch, \proglang{Blang} is built using \proglang{Xtext} \citep{Efftinge2006}, a powerful framework for designing programming languages.
Owing to this infrastructure, \proglang{Blang} incorporates a feature set comparable to many modern, fully-fledged, multi-paradigm languages: functional, generic and object programming, static typing, just-in-time compilation, garbage collection, IDE support for static types, profiling, code coverage, and debugging.

\proglang{Blang} comes with a growing library of built-in models, which are themselves written in \proglang{Blang} (as done in \cite{murray_automated_2018}), moreover, users can share and maintain models via an established transitive dependency management and versioning system. \proglang{Blang} also implements a suite of existing and novel testing strategies for models and MCMC methods, blending them with unit testing and multiple testing tools.

One of the existing PPLs most closely related to \proglang{Blang} is \proglang{RevBayes} \citep{hohna_probabilistic_2014,RevBayes}. \proglang{RevBayes} is a declarative PPL which provides extensive support for Bayesian inference over phylogenetic trees, an archetypical example of a challenging combinatorial space. Moreover, \proglang{RevBayes} supports interactive usage, a functionality currently not supported in \proglang{Blang}. However, as phylogenetic inference is the primary domain targeted by \proglang{RevBayes}, users interested in combinatorial spaces other than phylogenetic trees will benefit from \proglang{Blang}'s abstractions which target arbitrary combinatorial spaces.

\updated{
The goal of this paper is to provide readers with an introduction to \proglang{Blang}.
We begin with an outline of the language's goals in Section \ref{sec:goals}, open source licence, Section \ref{sec:license}, and a first tutorial, Section \ref{sec:overview}.
Followed by a \textit{conceptual} overview in Section~\ref{sec:langoverview}, which itself is sandwiched by two examples of increasing complexity (Sections~\ref{sec:overview} and \ref{sec:gmm}).
With the big picture laid out, \proglang{Blang}'s declarative syntax and structure are formalized and detailed in Section~\ref{sec:syntax}.
A \textit{cheatsheet} highlighting the key ideas discussed in the preceeding sections is summarized in Section~\ref{sec:cheatsheet}.
The sections have been arranged in what the authors believe to be a pedagogical format.
However, readers may find it helpful to first skim the sections consisting of examples (Sections~\ref{sec:overview}, \ref{sec:gmm}, and \ref{sec:cheatsheet}).
The remaining sections are more advanced, but are nonetheless helpful for drawing context and understanding the motivation behind \proglang{Blang}'s design.
Section~\ref{sec:customSamplers} illustrates and discusses a key feature of \proglang{Blang}: the creation of custom data structures and custom samplers.
Section~\ref{sec:sdk} introduces \proglang{Blang}'s software development kit (SDK), which can be used to implement complex models and assist in testing the correctness of implementations.
Section~\ref{sec:design} consists of design patterns.
Finally, Section~\ref{sec:inference} describes \proglang{Blang}'s architecture and inference algorithms as a whole.
}

\section{Goals} \label{sec:goals}

\proglang{Blang}'s purpose is to provide Monte Carlo approximations of posterior distributions arising in Bayesian inference problems.
The design of the language and its software development kit is guided by the following high-level goals:

\begin{description}
  \item[Correctness:] Bayesian inference software is notoriously difficult to implement.
  An example from the tip of the iceberg is shown in \cite{Geweke2004}, which identifies software bugs and erroneous results in earlier published studies.
  We address this issue using a marriage of statistical theory and software engineering methodology, such as compositionality and unit testing.
  \item[Ease of use:] \proglang{Blang} uses a familiar \proglang{BUGS}-like syntax and it is designed to be integrated well in modern data science workflows (input in Tidy format \citep{Wickham2014}, samples output in Tidy format).
  \item[Generality:] 
  As a programming language, \proglang{Blang} is Turing-complete and equipped with an open type system, as well as facilities to quickly develop and test sampling algorithms for new types. By open type system, we mean that the set of types is not limited to integers and real numbers, and can be arbitrary classes. 
  \proglang{Blang} does not fully automate the process of posterior sampling from  user-defined types but 
  instead greatly facilitates the development, composition and sharing of custom sampling algorithms. 
  \item[Computational scalability.] The language is designed to ensure that state-of-the-art Monte Carlo methods can be utilized. In particular, we made certain trade-offs to ensure that a well-behaved continuum of distributions can be automatically created. This is complemented with methods that extend existing PPL strategies to combinatorial space, for example code scoping analysis to discover sparsity patterns with arbitrary types, as well as built-in support for parallelization to arbitrary numbers of cores. 
\end{description}

\section{License, source, version and documentation availability}\label{sec:license}

\proglang{Blang} is free and open source.
The language and SDK are available under a permissive BSD 2-Clause license.
The relevant GitHub repositories  are linked at \url{https://github.com/UBC-Stat-ML/blangDoc}. Online documentation is available at \url{https://www.stat.ubc.ca/~bouchard/blang/}, including Javadoc pages at \url{https://www.stat.ubc.ca/~bouchard/blang/Javadoc.html}.

\section[Tutorial]{Tutorial} \label{sec:overview}

This section aims to introduce readers to \proglang{Blang} by presenting a minimal working example.
We begin with instructions for performing inference on a simple model using the command-line interface (CLI).
Realistic applications are demonstrated in Sections~\ref{sec:gmm} and \ref{sec:customSamplers}.
Advanced tutorials can be found in Appendix~\ref{sec:tutorialAdvanced}.

\subsection[Installing Blang's command-line interface]{Installing \proglang{Blang}'s command-line interface} \label{subsec:install-cli}

We provide instructions here for installing and using \proglang{Blang} via CLI.
Alternative \proglang{Blang} interfaces include an integrated development environment (IDE), detailed in Section~\ref{sec:desktop-ide}, as well as a Web interface (Section~\ref{sec:web-ide}).
Instructions are also available from the documentation website under the link \texttt{Tools}.\footnote{Documentation for \proglang{Blang} is available at \url{https://www.stat.ubc.ca/~bouchard/blang/}} Additionally, an \proglang{R} and \proglang{Python} interface to \proglang{Blang} are currently under development.\footnote{The interfaces and associated instructions will be hosted on \url{https://github.com/UBC-Stat-ML}}

The prerequisites for the CLI installation process are:

\begin{enumerate}
\item A UNIX-compatible environment running \proglang{bash} \updated{ or \proglang{zsh}}. This includes, in particular, Mac OS X, Linux, and Windows Subsystem for Linux.
\item The \code{git} command.
\item \updated{The \proglang{Java} Software Development Kit (SDK), version 8, 11, 13, or 15. Other versions of \proglang{Java} may be incompatible with the version of \proglang{Xtext} our software builds upon.\footnote{Specifically, OpenJDK 8, 11, 13, and 15 have been tested at the time of writing. \proglang{Java} is typically backward compatible, but since the library \proglang{Xtext} performs bytecode manipulations it is more sensitive to  versioning than typical \proglang{Java} libraries. Managing and installing several versions of \proglang{Java} is greatly facilitated by the easy to install package \pkg{sdkman} available at \url{https://sdkman.io/}.}} The \proglang{Java} \emph{runtime environment} is necessary, and the \emph{runtime environment} is not sufficient, as compilation of models requires compilation into the Java Virtual Machine. Type \code{javac -version} to test if the \proglang{Java} SDK is installed. If not, the \proglang{Java} SDK is freely available at \url{https://openjdk.java.net/}. 
\item \updated{Optionally, if automatic plotting of posterior distributions, trace plots, diagnostics, etc is required, \proglang{R} as well as the packages \pkg{dplyr} and \pkg{ggplot2} should be installed. In particular, the command \code{Rscript} should be in the \code{PATH} variable for the optional plotting functionalities to work correctly.}
\end{enumerate}

The following installation process is most thoroughly tested on Mac OS X \updated{and Linux}, however users have reported installing it successfully on certain Windows configurations \updated{(using either Windows Subsystem for Linux or Cygwin).\footnote{Note also that the Eclipse IDE plug-in does not require a UNIX-compatible environment, see Section~\ref{sec:desktop-ide}.}}

To install the CLI tools, input the following commands in a \proglang{bash} \updated{ or \proglang{zsh}} terminal interpreter:

\longline
\begin{CodeChunk}
 %PARSESTART
\begin{CodeInput}
> git clone https://github.com/UBC-Stat-ML/blangSDK.git
> cd blangSDK
> source setup-cli.sh
\end{CodeInput}
%PARSEEND
\end{CodeChunk}
\longline

The \code{git clone} command downloads the \pkg{blangSDK} repository, \code{cd} changes the current working directory, and \code{source setup-cli.sh} compiles and installs  \proglang{Blang} (i.e., updates the \code{PATH} variable). If the user moves the \pkg{blangSDK} folder, the command \code{source setup-cli.sh} needs to be rerun. 

You may now use \proglang{Blang} from any directory by typing \code{blang} (use lower case for the CLI command as UNIX is case-sensitive).

\subsection{Posterior inference} \label{subsec:runModel}

Consider the simplified Doomsday Argument \citep{carter_brandon_anthropic_1983} \updated{for modelling the total number of humans that were ever or will ever be born.
Denote the estimated number of humans that have been born up to the present time as $y$, and the total number of humans that were ever or will ever be born (an unknown variable) as $z$.
The Doomsday Argument posits $y \mid z \sim \text{Uniform}(0, z)$.
With a prior belief of what values $z$ can take on encoded as an exponential distribution, we update our belief using a PPL to obtain an approximation of the posterior distribution of $z \mid y$.
}
Using a PPL for such a simple model is excessive but is useful for demonstrating the basic mechanics of Bayesian inference in \proglang{Blang}.

\longline

\code{Doomsday.bl}\vspace*{-10pt}\footnote{\url{https://github.com/UBC-Stat-ML/JSSBlangCode/blob/master/reproduction_material/example/jss/Doomsday.bl}. Note the package statements are different.}

\longline
\begin{lstlisting}
package toy

model Doomsday {
  param RealVar rate
  random RealVar y
  random RealVar z
  laws {
    z | rate ~ Exponential(rate)
    y | z ~ ContinuousUniform(0.0, z)
  }
}
\end{lstlisting}
\longline

The first line is a package declaration, which identifies the package in which the Doomsday model belongs to.
The remaining code illustrates four \proglang{Blang} keywords.
\updated{
\begin{itemize}
  \item \code{model}: there should be exactly one \code{model} per file.
        The keyword should be followed by an identifier, in this case \code{Doomsday}.
        \proglang{Blang} is a case-sensitive language and we use the convention that model names are capitalized.
  \item \code{random} and \code{param} are used to declare model variables.
        By default, \proglang{Blang} approximates the posterior distribution over the latent \code{random} variables conditioning on the observed \code{random} variables.
  \item Variables need to specify their types.
        For example, \code{random RealVar z} is of type \code{RealVar} and we give it the name \code{z}.
        As a convention, types are capitalized and variable names are not. 
  \item Briefly, \code{random} variables encompass all observed and unobserved random variables.
        \code{param} variables encompass all known constants.
        The distinction is further discussed in Section~\ref{sec:langoverview}. 
  \item Each model is required to have exactly one \code{laws} keyword followed by a code chunk surrounded by curly braces,
  called the \emph{laws block}. 
  The purpose of the laws block is to define joint distributions over the random variables.
  Here, we show one method to do so, which is inspired by the \proglang{BUGS} notation and its derivatives. 
  For example,
   \begin{center}
   \code{y | z } $\sim$ \code{ ContinuousUniform(0.0, z)}
   \end{center}
   denotes the conditional distribution of \code{y} given \code{z} is equal to a uniform distribution between \code{0} and \code{z}.
  In contrast to \proglang{BUGS}, we require specification of the random variables that we are conditioning on, here \code{| z}.\footnote{There are several motivations behind this design choice deviating from \proglang{BUGS}. Technically, static analysis could identify the list of variables we are conditioning on. However the notation used here is closer to a mathematical notation used for example in the Bayesian non-parametric literature (e.g., \cite{teh2006,griffiths_indian_2011}). More importantly however, the explicit conditioning allows us to generalize the notation to handle complex dependencies. This is demonstrated in Sections~\ref{sec:undirected} and \ref{sec:markovChainExample}.}
\end{itemize}
}

From the \code{project} directory, type the following command, in which we specify that \code{rate} and \code{y} are fixed to given values while \code{z} is unobserved:

\begin{CodeChunk}
\begin{CodeInput}
> blang --model toy.Doomsday --model.rate 1.0 --model.y 1.2 --model.z NA 
\end{CodeInput}
\end{CodeChunk}

The same model can be run via the Eclipse IDE (with the \code{-{}-model} argument omitted), following instructions from Section~\ref{sec:IDE},\footnote{Instructions hosted on \proglang{Blang}'s website will be continually updated \url{https://www.stat.ubc.ca/~bouchard/blang/}} or via a prepackaged repository of examples:

\longline
\begin{CodeChunk}
%PARSESTART  % oops sorry broke that, can help fix it tomorrow % fixed!
\begin{CodeInput}
> git clone https://github.com/UBC-Stat-ML/JSSBlangCode.git
> cd JSSBlangCode/reproduction_material/example/
> blang --model jss.Doomsday --model.rate 1.0 --model.y 1.2 --model.z NA 
\end{CodeInput}
%PARSEEND
\begin{CodeOutput}
Compilation {
  ...
} [ ... ]
Preprocess {
    ...
 } [ ... ]
Inference { 
   ... 
 } [ ... ]
executionMilliseconds : 1037
outputFolder: ./JSSBlangCode/.../results/all/2019-06-27-14-13-21-RL.exec
\end{CodeOutput}
\end{CodeChunk}

Samples approximating the posterior distribution of \code{z} given the observation \code{y} are outputted in Tidy format \citep{Wickham2014} to \code{samples/z.csv} located in the directory specified by \code{outputFolder}. 

By default, posterior inference is done in two stages.
The first stage, corresponding to the \code{Initialization} block in the standard output, uses SCM which attempts to automatically identify configurations of positive density.
In the second stage, an adaptive non-reversible PT algorithm is initialized from the output of the first stage and performs a series of adaptation rounds, corresponding to \code{Round(1/9)} through \code{Round(9/9)} blocks in the standard output.
PT algorithms are known to perform well even in the face of difficult sampling problems such as those arising in multimodal distributions or weakly identifiable models. 
We describe the inference algorithms and their configuration in detail in Section~\ref{sec:inferenceEngines}.

\section{Conceptual overview} \label{sec:langoverview}
We now describe more formally the semantics of our language's core construct: the \code{model}. The basic notation introduced here will be useful to describe the syntax in full detail in the next section.

\subsection{Models}\label{sec:model-overview}

A \proglang{Blang} \code{model} encodes a set of \textit{densities} $\{f_\theta(x) : \theta \in \Theta, x \in T\}$, and hence the distribution of a random object $X:\Omega \to T$.
We use the term density in a generalized sense, encompassing discrete, continuous, and mixed models, by allowing it to be defined with respect to customizable reference measures. 

We assume $x = (x_1, x_2, \dots, x_n)$ where $n < \infty$ is fixed. 
Despite $n$ being finite in this formalism, each $x_i$ is permitted to be of random or infinite dimensionality. 
The \textit{type} or space in which the $x_i$'s lie in, is denoted by $T_i$.
Hence $x_i \in T_i$ and $x \in T = T_1 \times T_2 \times \dots \times T_n$. 
We also assume each type $T_i$ is implicitly associated with a default reference measure $\mu_i$. 
These default choices can be changed using the \code{is} keyword defined in Section~\ref{subsubsec:constraints}. 
Once each reference measure $\mu_i$ is given, by definition the densities are turned into distributions as follows:

\begin{align}\label{eq:spec}
\P_\theta(X \in A) = \int_A f_\theta(x) \prod_{i=1}^n \mu_i(\ud x_i).
\end{align}

Where $A$ is some event, or more formally, an element of the $\sigma-$algebra of $T$.
We also assume a decomposition for the parameters $\theta = (\theta_1, \theta_2, \dots, \theta_m)$ where $m$ is fixed and each coordinate $\theta_j$ has its type denoted by $\Theta_j$.
Hence, $\theta_j \in \Theta_j$ and $\theta \in \Theta = \Theta_1 \times \Theta_2 \times \dots \times \Theta_m$.
We use the terminology \textit{model variables} to refer to $x$ and $\theta$ collectively.

To understand how these mathematical concepts translate into \proglang{Blang} syntax, let us relate them via the Doomsday example from Section~\ref{sec:overview}. The correspondence is shown in Figures~\ref{fig:doomsDayBlangDefn} and \ref{fig:doomsDayMathDefn}. The variables marked with the \code{random} keyword are concatenated to form $x$, while those marked with \code{param} keyword are concatenated to form $\theta$.

\begin{figure}[h]
\begin{minipage}[b]{0.45\textwidth} 
\begin{lstlisting}
model Doomsday {

  param RealVar rate
  random RealVar y
  random RealVar z

  laws { ... }

}
\end{lstlisting}
\caption{\proglang{Blang} Syntax.}
\label{fig:doomsDayBlangDefn}
\end{minipage}
\begin{minipage}[b]{0.5\textwidth}
\begin{align*}
\theta &= (\texttt{rate}) \\ 
x &= (\texttt{y}, \texttt{z}) \\
\{f_\theta\} &= \{\texttt{Doomsday(rate)}\} \\
\Theta_1 &= T_1 = T_2 = \texttt{RealVar}
\\
\\
\end{align*}
\caption{Mathematical notation.}
\label{fig:doomsDayMathDefn}
\end{minipage}
\label{code:doomsDay}
\end{figure}

\subsection[Interpretation of laws blocks]{Interpretation of \code{laws} blocks}

The \code{laws} block is responsible for computing the point-wise evaluation of $\text{log}(f_\theta(x))$ for any input $x$ and $\theta$. 
To do so, two methods are supported:
\begin{description}
  \item[Composite laws] use existing \proglang{Blang} models as building blocks to create a new one.
  \item[Atomic laws] provide an arbitrary algorithm to compute the log density.
\end{description}

Both composite and atomic laws allow the user to express a known factorization of the density
\begin{align}\label{eq:factorization}
f_\theta(x) = \prod_{k=1}^K f^{(k)}(x, \theta).
\end{align}
Such a factorization can then be used as the basis of automating key aspects of state-of-the-art Monte Carlo methods, such as the construction of a well-behaved continuum of auxiliary distributions and the detection of sparsity patterns.
Additionally this factorization enables efficient sampling of latent variables, as only a fraction of factors will require evaluation per variable.

\subsection{Interpretation of atomic laws}\label{sec:interpretation-atomic}

In the case of an atomic law, for each $k \in \{1, 2, \dots, K\}$, an expression or algorithm is provided to compute the value of factor $k$ in log scale, i.e., $\log\left(f^{(k)}(x, \theta)\right)$. 

For example, consider the continuous uniform distribution, which can be factorized as 
\[
f^\text{unif}_\theta(x) = \underbrace{\frac{1}{\theta_2 - \theta_1}}_{f^{(1)}(x)} \ \ \underbrace{\I[\theta_1 \le x \le \theta_2]}_{f^{(2)}(x)},
\]
where $\theta = (\theta_1, \theta_2) = (\texttt{min}, \texttt{max})$. The \code{model} defining a \code{ContinuousUniform} distribution in the \proglang{Blang} SDK, encodes this factorization as follows:

\longline

\code{ContinuousUniform.bl}\vspace*{-10pt}\footnote{\url{https://github.com/UBC-Stat-ML/JSSBlangCode/blob/master/reproduction_material/example/jss/others/ContinuousUniformExample.bl}}

\longline
\begin{lstlisting}
model ContinuousUniform {
  random RealVar realization
  param  RealVar min
  param  RealVar max
  
  laws {
    logf(min, max) {               
      if (max - min <= 0.0) return NEGATIVE_INFINITY
      return - log(max - min)
    }
    logf(realization, min, max) {  
      if (min <= realization && realization <= max) return 0.0
      else return NEGATIVE_INFINITY
    }
  }
  ...
}
\end{lstlisting}
\longline

\subsection{Interpretation of composite laws}\label{sec:interpretation-composite}

In the case of a composite law, the decomposition in Equation~\ref{eq:factorization} typically comes from an application of the chain rule. In the Doomsday example, this is just:
\begin{align}\label{eq:chain}
f^\text{Dooms}_\theta(x) \ \  = \ \ \underbrace{\theta_1 \exp(-\theta_1 x_2)}_{\tilde f^{(1)}(x, \theta)} \ \ \underbrace{\frac{\I[0 \le x_1 \le x_2]}{x_2}}_{\tilde f^{(2)}(x, \theta)}.
\end{align}
To understand composite laws, notice the factors in this decomposition can often be retrieved from another existing \code{model}.
In such a case, we say that a \code{model}, $\{f_\theta^\text{caller}(x) : x \in T, \theta \in \Theta\}$, \emph{calls} another model, $\{f_{\theta'}^\text{callee}(x') : x' \in T', \theta' \in \Theta'\}$.
This is illustrated in our running example as the \code{Doomsday} model, the caller, calls the \code{ContinuousUniform} model, the callee.
Consequently allowing us to write the second factor in Equation~(\ref{eq:chain}) using the previously defined \code{ContinuousUniform} model via 
\begin{align}
\tilde f^{(2)}(x, \theta) \ \  = \ \ f^\text{unif}_{t(x,\theta)}(s(x)),
\end{align}
for $t(x,\theta)$ and $s(x)$ defined as follows.

First, $t : T \times \Theta \to \Theta'$  is a transformation from the \emph{caller} model's variables into the \emph{callee} model's parameters, in this case $t(x, \theta) = (0, x_2)$.
The two entries in the list $(0, x_2)$ correspond to the two \code{param} variables, \code{min} and \code{max}, in the definition of \code{ContinuousUniform} shown in Section~\ref{sec:interpretation-atomic}. 
We see that the order in which the \code{param} are declared is important when a \code{model} is to be used in a composite fashion.

Second, $s:T \to T'$ is a selection of a subset $i_1, \dots, i_{|x'|}$ of coordinates in $x$, so that $s(x) = (x_{i_1}, \dots, x_{i_{|x'|}})$.
Hence, $s$ selects which of the calling model's random variables are used as the callee model's random variables.
Here $s(x) = (x_1)$, where the single entry, $(x_1)$, corresponds to the \code{random} variable, \code{realization}, in the definition of \code{ContinuousUniform}.
Again, if more than one random variable is selected, the order in which they are declared in the callee model determines how they are matched.

Considering now the \proglang{Blang} statement:

\longline
\begin{Code}
y | z ~ ContinuousUniform(0.0, z)
\end{Code}
\longline

we see that the left of the pipe symbol, \code{|}, encodes the selection $s$, and the expression in parentheses encodes the transformation $t$. 

In summary, the two lines in the laws block of the Doomsday model:

\longline
\begin{Code}
z | rate ~ Exponential(rate)
y | z ~ ContinuousUniform(0.0, z)
\end{Code}
\longline

have the same interpretation as they would in probability theory. 
However, our notation can also be extended to useful novel patterns (see Sections~\ref{sec:undirected} and \ref{sec:markovChainExample}).

\subsection{Model tree} \label{subsec:modeltree}

Composite laws induce a directed tree over models, where a directed edge denotes a \code{model} calling another \code{model}. We call this tree the \emph{model tree}. The root of this tree is called the \emph{root model}.

\subsection[Interpretation of generate blocks]{Interpretation of \code{generate} blocks} \label{subsec:interpretation-generate}

In addition to the atomic and composite constructs available to specify a mandatory \code{laws} block, \proglang{Blang} provides an optional orthogonal way to specify $\P_\theta(X \in A)$, called a \code{generate} block.
The \code{generate} block performs \textit{forward simulation}: it takes as input a random seed, $\omega \in \Omega$, and returns $X(\omega)$ such that Equation~(\ref{eq:spec}) holds.

The \code{generate} block is technically redundant, but is crucial to check software correctness by setting up statistical unit tests as described in Section~\ref{subsec:testing}.
It is also used for various purposes during posterior inference, for example, by providing a form of regeneration in PT, and to initialize SCM samplers. 

\subsection{Normal form}\label{sec:normalform}

\updated{
A laws block containing either only composite laws or only atomic laws is said to be in \emph{normal form}.
For example, the laws block in \code{Doomsday.bl} is in normal form, as it consists of composite and only composite laws.
Similarly, the laws block in \code{ContinuousUniform.bl} is also in normal form, as it consists of atomic and only atomic laws.
As a counterexample, the following laws block in a model is not in normal form:
}

\longline
\begin{Code}
z | rate ~ Exponential(rate)
logf(z) {
  return -log(z)
}
\end{Code}
\longline

\updated{
as it contains both composite and atomic laws.
Laws blocks in normal form are useful to automatically construct sequences of annealed distribution, used in certain samplers used by \proglang{Blang}'s runtime architecture (see \emph{Constructing a sequence of measures} in Section~\ref{sec:inferenceEngines}).

A model is said to be in \emph{generative normal form} if it satisfies the following conditions:
\begin{enumerate}
  \item All models in the model tree are in normal form.
  \item All models in the model tree based on atomic laws attached to unobserved variables are equipped with a generate block.
\end{enumerate}

Generative normal form is only required if the inference engine is PT or SCM, as samples from the prior are exploited for initialization and/or regeneration.
}

We show in Section~\ref{sec:undirected} how to rewrite a wide range of models into a generative normal form.
If a model cannot be written in generative normal form, the user may still apply standard MCMC methods but not the more advanced PT and SCM schemes.

\subsection[From Blang models to posterior inference]{From \proglang{Blang} models to posterior inference}

Any \proglang{Blang} model can be transformed into a posterior inference computer program.
The inputs of this computer program consists of variables in the root model.
All \code{param} variables in the root model become required inputs.
In contrast, \code{random} variables in the root model can either be specified or left missing as latent.
The target posterior distribution is then defined as the distribution of latent random variables given the variables that have been given an input value.

\section{Tutorial---a complete example}\label{sec:gmm}

We illustrate an example of posterior inference for a Gaussian mixture model (GMM).
We highlight and briefly discuss key components in implementing a model, and showcase a series of post-processed statistics and plots.
After a formal introduction of the syntax (Section~\ref{sec:syntax}), we will return to this example in the form of a summary in Section~\ref{sec:cheatsheet}.

Consider the following model:

\begin{align*}
  \text{concentration} && \alpha &= [1 , 1 ] \\
  \text{proportions} && \pi \mid \alpha & \sim \text{Dirichlet}(\alpha) \\
  \text{labels} && z_i \mid \pi & \sim \text{Categorical}(\pi) \\
  \text{means} && \mu_k & \sim \text{Normal}(0,10^2) \\
  \text{standard deviations} && \sigma_k \ & \sim \text{Uniform}(0, 10) \\
  \text{observations} && y_i \mid \mu, \sigma, z_i & \sim \text{Normal}(\mu_{z_i}, \sigma_{z_i}^2)
\end{align*}
for $i \in \{1,2,\dots,n\}$ and $k\in\{1,2\}$.

We encode this GMM in \proglang{Blang} as follows:\footnote{Complete and commented implementations in this section are available in the reproduction materials located in the directory \code{reproduction\_materials/example}.}

\longline

\code{MixtureModel.bl}\vspace*{-10pt}\footnote{\url{https://github.com/UBC-Stat-ML/JSSBlangCode/blob/master/reproduction_material/example/jss/gmm/MixtureModel.bl}}

\longline 

\begin{lstlisting}
package jss.gmm

model MixtureModel { 
  
  random List<RealVar>  y 
  param  Integer        n  ?: y.size
  param  Matrix         a  ?: fixedVector(1.0, 1.0)
  random List<IntVar>   z  ?: latentIntList(n) 

  param  Integer        K  ?: 2
  random Simplex        pi ?: latentSimplex(K)
  random List<RealVar>  mu ?: latentRealList(K)
  random List<RealVar>  sd ?: latentRealList(K)
  
  laws {

    pi | a ~ Dirichlet(a)
    
    for (int k : 0 ..< K) { 
      mu.get(k) ~ Normal(0.0, 100.0)
      sd.get(k) ~ ContinuousUniform(0.0, 10.0)
    }
    
    for (int i : 0 ..< n) {
      z.get(i) | pi ~ Categorical(pi)
      y.get(i) | mu, sd, IntVar k = z.get(i) 
        ~ Normal(mu.get(k), pow(sd.get(k), 2.0))
    }
  }
}

\end{lstlisting}
\longline

We began by declaring variables as we did in the Doomsday model.
In addition to declarations, we initialized them to their respective latent types. 
Default initializations are expressed using \code{?:} followed by an expression in a syntax called XExpression described in detail in Section~\ref{subsec:xexp}.\footnote{Those familiar with \proglang{Java} can think of XExpressions as ``shorthand \proglang{Java}'' for now.}
Default initializations can be overridden from the CLI (command line interface). 
We discuss this mechanism in detail in Sections~\ref{subsec:modvars}.
In this example, interpret initializations as creating instances of latent objects.\footnote{In Section~\ref{sec:customSamplers}, we create a constructor for objects of type permutation. Its application is helpful in painting a bigger picture on how these latent objects are used behind the scenes.}
A list of data types available for latent variables can be found in Figures~\ref{app:rvarfuncs} and \ref{app:linalgfuncs}.

In the next code block, the \code{laws} block, we declared the distribution of each latent variable.
We used \code{for} loops to encode a set of declarations.
For example, the following two implementations are equivalent:

\longline 

\begin{lstlisting}
for (int k : 0 ..< 2) { 
  mu.get(k) ~ Normal(0.0, 100.0)
  sd.get(k) ~ ContinuousUniform(0.0, 10.0)
}
\end{lstlisting}
\longline

and

\longline 

\begin{lstlisting}
mu.get(0) ~ Normal(0.0, 100.0)
mu.get(1) ~ Normal(0.0, 100.0)
sd.get(0) ~ ContinuousUniform(0.0, 10.0)
sd.get(1) ~ ContinuousUniform(0.0, 10.0)
\end{lstlisting}
\longline

To perform posterior inference on \code{MixtureModel} based on observed $y_i$'s, we invoke the following commands in the CLI:

\longline

\begin{CodeChunk}
\begin{CodeInput}
> git clone https://github.com/UBC-Stat-ML/JSSBlangCode.git
> cd JSSBlangCode/reproduction_material/example
> blang --model jss.gmm.MixtureModel \
	--model.y file data/obs1.txt \
	--engine PT \
	--engine.nChains 36 \
	--engine.nScans 30000 \
	--postProcessor DefaultPostProcessor
\end{CodeInput}
\begin{CodeOutput}
Preprocess {
   ...
 } [ ... ]
Inference { 
   ...
 } [ ... ]
Postprocess {
  Post-processing allLogDensities
  Post-processing energy
  Post-processing z
  Post-processing logDensity
  Post-processing mu
  Post-processing nOutOfSupport
  Post-processing pi
  Post-processing sd
  MC diagnostics
} [ ... ]
executionMilliseconds : ...
outputFolder :./JSSBlangCode/.../all/2020-12-31-23-59-03-N0PvDjdc.exec
\end{CodeOutput}
\end{CodeChunk}
\longline

In this example, \code{obs1.txt} is a new-line separated file formatted as follows:

\longline

\code{obs1.txt}\vspace*{-10pt}

\longline 

\begin{tabular}{r} 
$3.2$ \\
$-0.3$ \\
$1.7$ \\
\vdots
\end{tabular}

\longline

More generally, information on the format used to input data can be obtained by \emph{appending} \code{-{}-help} to the command line arguments (the command line help is contextual, so the information given by appending \code{-{}-help} to the model and inference engine specific arguments will be more detailed than only using \code{blang -{}-help}). 
A more sophisticated method to input data, based on the \emph{plate notation}, is discussed in Section~\ref{subsec:plates}.
We briefly summarize the key CLI arguments for the example below:

\begin{center}
\begin{tabular}{l l l} 
 \hline
Argument & Description \\ 
 \hline\hline
\code{-{}-model.y file} & Specifies the file path to a newline-separated file with \code{y}'s values.\\
\code{-{}-engine PT} & Specifies Parallel Tempering as the inference algorithm.\\
\code{-{}-engine.nChains} & Controls the number of (annealed) parallel Markov chains.\\
\code{-{}-engine.nScans} & Controls the number of posterior samples to draw.\\
\code{-{}-postProcessor} & Specifies the post-processor. \\
 \hline
\end{tabular}
\end{center}

The details of how \code{-{}-engine} arguments influence the performance of inference are discussed in Section~\ref{sec:inference}.

All experiment outputs are stored in a \code{results} directory, within the working directory in which the \proglang{Blang} CLI command is called.
Generally, there are three categories of outputs: samples (raw output), post-processed statistics/plots (summaries of the samples), and monitoring statistics/plots (to assess the quality of the posterior approximation).
Options for post-processing is handled via the \code{-{}-postProccesor} runtime argument, accepting \code{DefaultPostProcessor} or \code{NoPostProcessor} as arguments. Again use \code{-{}-postProccesor DefaultPostProcessor -{}-help} for more information.

Currently, the \code{DefaultPostProcessor} option produces trace and density plots,\footnote{The \code{DefaultPostProcessor} requires \proglang{R} as well as the packages \pkg{dplyr} and \pkg{ggplot2}.} and provides summary statistics including Highest Density credible Intervals (HDI, constructed using the method described in \cite{chen_monte_1999}) and effective sample size (ESS) estimates (based on a numerically robust version of the $\sqrt{n}$-size batch estimator described in \cite{Flegal2010}).
Type information is used to select appropriate plotting strategies (e.g., probability mass functions for \code{IntVar} types, density estimates for \code{RealVar}).
Examples of summary statistics for \code{MixtureModel}'s parameters are shown below, and can be found under the directory summaries in \code{results/latest}.\footnote{Numerical values are truncated to fit in the page width.}

\begin{center}
\begin{tabular}{c l r r r r r r r} 
 \hline
 index & parameter & mean & sd & min & median & max & HDI.lower & HDI.upper\\
 \hline\hline
 $0$ & mean & $0.66$ & $1.77$ & $-29.40$ & $1.21$ & $30.96$ & $-1.33$ & $2.52$ \\
 $1$ & mean & $0.63$ & $1.85$ & $-39.99$ & $1.20$ & $29.20$ & $-1.46$ & $2.38$ \\
 $0$ & variance & $1.17$ & $1.14$ & $0.01$ & $0.83$ & $9.95$ & $0.07$ & $2.57$ \\
 $1$ & variance & $1.16$ & $1.15$ & $0.00$ & $0.81$ & $9.98$ & $0.04$ & $2.53$  \\
 $0$ & pi &0$.50$ & $0.23$ & $0.00$ & $0.51$ & $0.99$ & $0.16$ & $0.83$\\ 
 $1$ & pi &0$.50$ & $0.23$ & $0.00$ & $0.49$ & $0.99$ & $0.16$ & $0.83$ \\
 \hline
\end{tabular}
\end{center}

\begin{figure}[H]
  \centering
  \begin{subfigure}[b]{0.45\linewidth}
    \includegraphics[width=\linewidth]{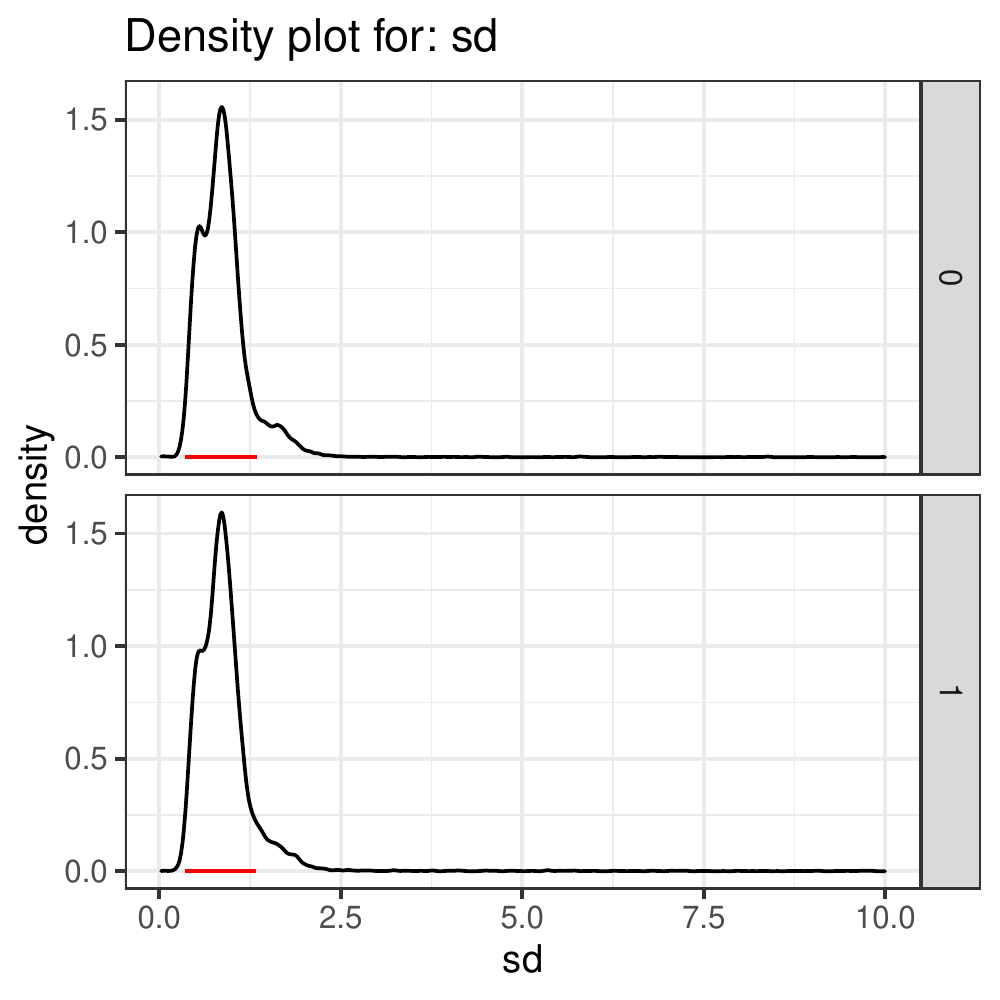}
  \end{subfigure}
  \begin{subfigure}[b]{0.45\linewidth}
    \includegraphics[width=\linewidth]{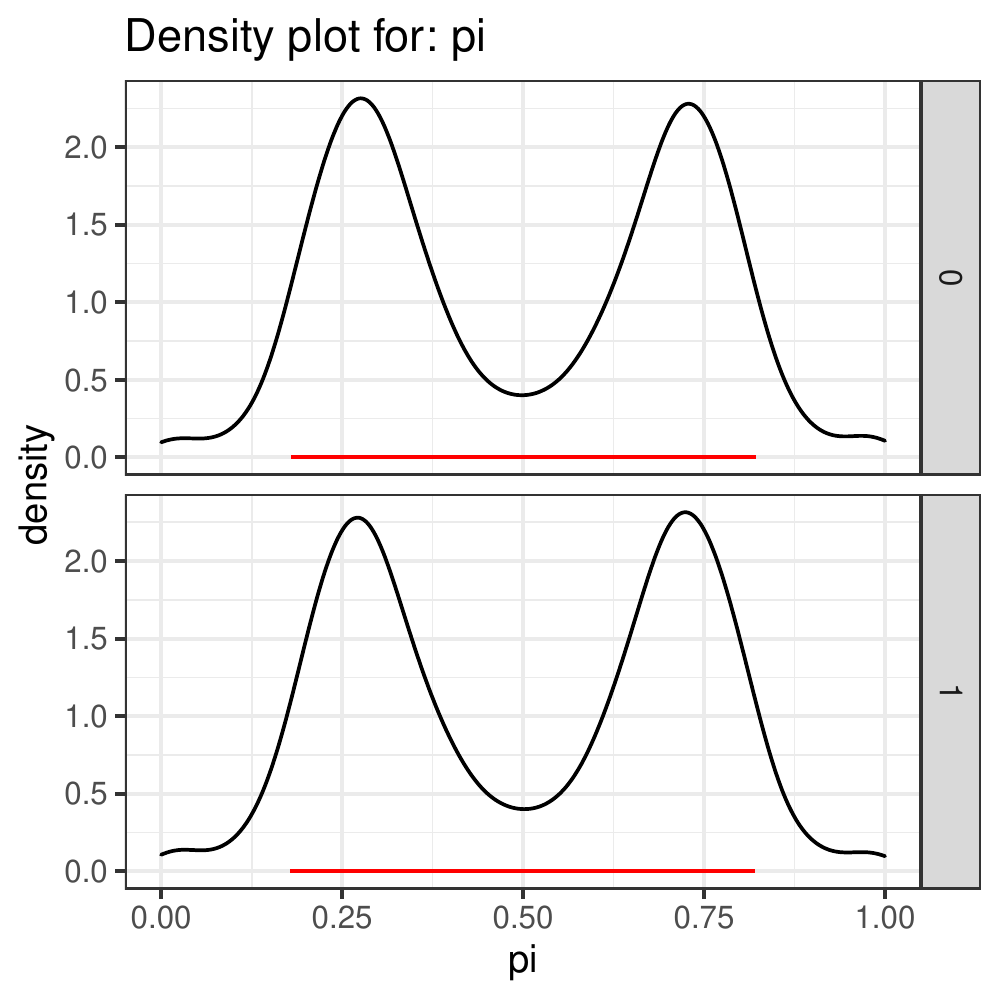}
  \end{subfigure}
  \caption{Posterior density plots for a subset of random variables in the GMM. The facets (rows) are indexed by the mixture components. Left: standard deviation parameters. Right: mixture proportion parameters.
  The two pairs of nearly-identical plots are indicative of successful label switching, showing that the multimodal posterior distribution is well approximated.
  By default, the 90\% highest density interval is underlined in red.  }
  \label{fig:posteriorPlots}
\end{figure}

Notice the posterior summaries are nearly identical for the two mixture components.  
Similarly, the marginal posterior plots in Figure~\ref{fig:posteriorPlots} also exhibit this symmetry. 
This symmetry is to be expected in this example: it arises from the unidentifiability of the GMM parameters known as label switching  \citep{jasra2005markov}.
Here the inference engine used, an adaptive non-reversible parallel tempering algorithm (abbreviated PT), is capable of capturing this symmetry despite the high-dimensional multimodality involved (the $z_i$'s of all variables have to be flipped to switch modes). 

\begin{figure}[H]
  \centering
 \begin{subfigure}[b]{0.45\linewidth}
    \includegraphics[width=\linewidth]{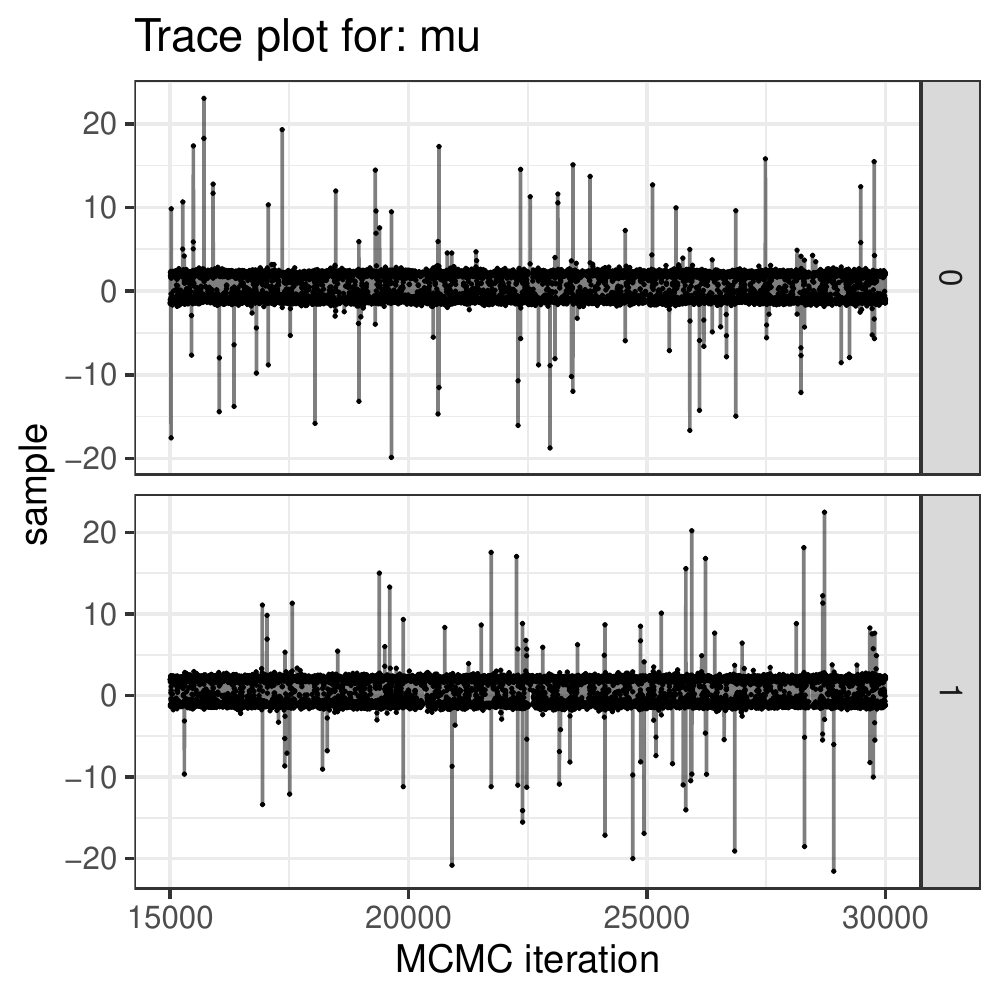}
  \end{subfigure}
  \begin{subfigure}[b]{0.45\linewidth}
    \includegraphics[width=\linewidth, trim={0 4877 0 0}, clip=true]{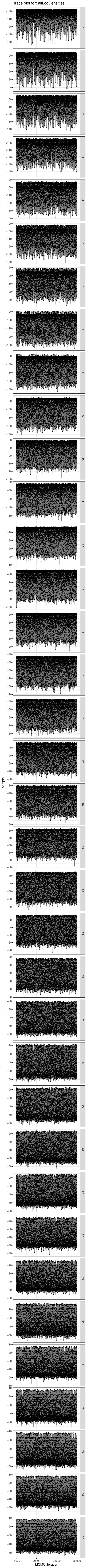}
  \end{subfigure}
  \caption{Left: Trace plot for cluster-specific location parameters. The two clusters are shown as facets. 
Right: log densities for two of the 36 tempered chains used in PT.
Notice that the ``jumps'' between modes are densely distributed along the traces, i.e., they occur very frequently in this example.
Other diagnostics produced will be discussed in Section~\ref{sec:inferenceEngines}.
}
  \label{fig:tracePlots}
\end{figure}

\begin{figure}[H]
  \centering
  \includegraphics[width=0.5\linewidth]{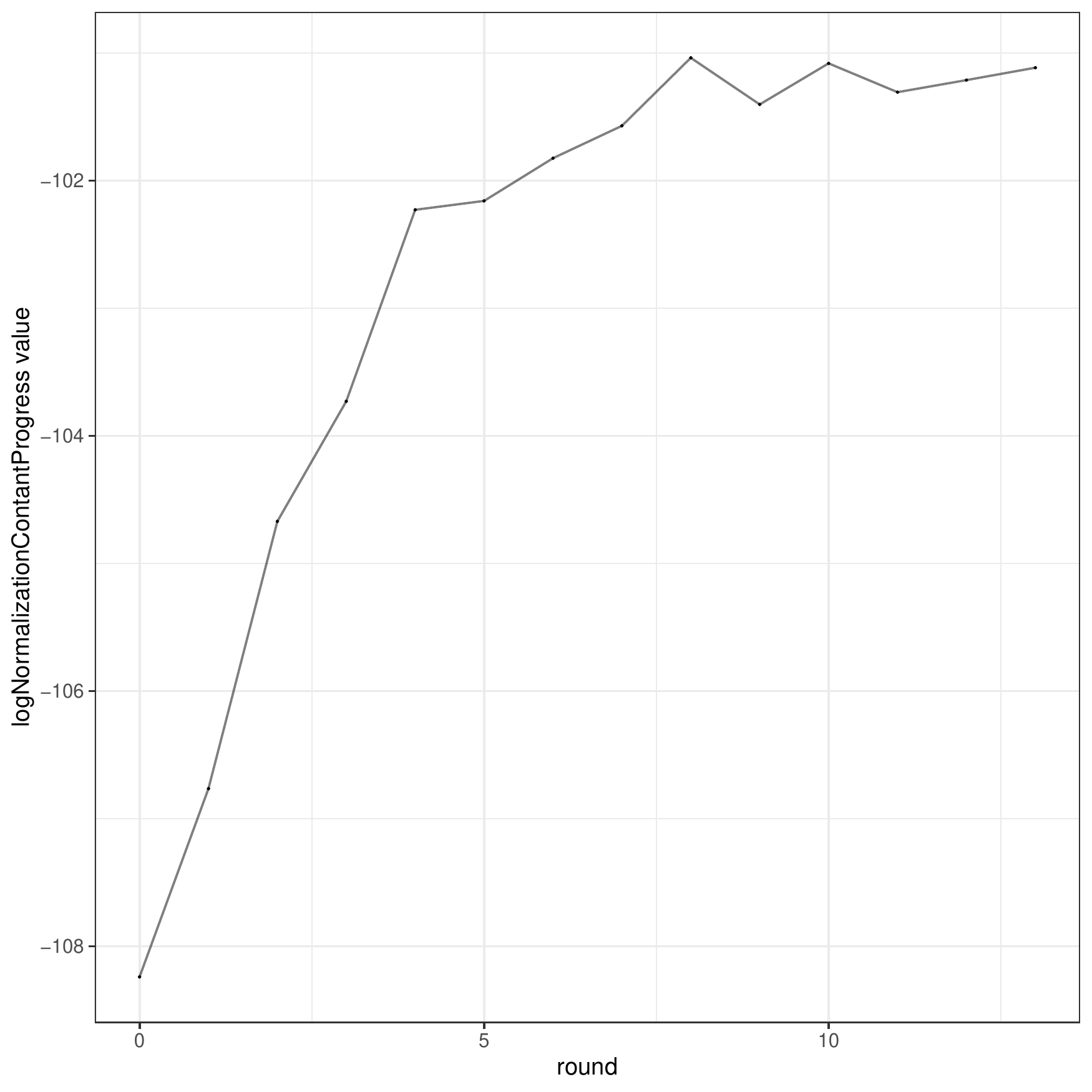}
  \caption{Log normalization constant estimates across adaptation rounds when the PT algorithm is used.
The fact that these estimates plateaued supports that the allocated computational budget is sufficient for this inference task.
}
  \label{fig:logNormConstProgress}
\end{figure}

Another statistic that is often of interest is the normalization constant (also known as model evidence, or marginal likelihood).
The logarithm of this value is automatically output in \code{logNormalizationEstimate.csv}.
The various methodologies available to estimate the log normalization constant are discussed in Section~\ref{sec:evidence}.
Figure~\ref{fig:logNormConstProgress} illustrates the progression of estimates across PT adaptation rounds.

Output files for diagnosing and monitoring the performance of inference algorithms are also produced. We will describe them in Section~\ref{sec:inferenceEngines}.

\section[A complete tour of Blang's syntax]{A complete tour of \proglang{Blang}'s syntax} \label{sec:syntax}

In this section we provide a more systematic survey of the \proglang{Blang} language. The formal definition of the language can be accessed in the \pkg{blangDSL} repository at \href{https://github.com/UBC-Stat-ML/blangDSL/blob/master/ca.ubc.stat.blang.parent/ca.ubc.stat.blang/src/ca/ubc/stat/blang/BlangDsl.xtext}{https://github.com/UBC-Stat-ML/blangDSL}.

\subsection{Project organization} \label{subsec:projorg}

\proglang{Blang} projects are composed of three types of files: \proglang{Blang} files (\code{.bl}), \proglang{Xtend} files (\code{.xtend}), and \proglang{Java} files (\code{.java}). 
This section is devoted to the syntax of \proglang{Blang} files. \proglang{Xtend} and \proglang{Java} files are used to create supporting code for non-standard data types, samplers, and user-defined functions. 
The user can choose either  \proglang{Xtend} or \proglang{Java} for creating supporting code. 
For users not familiar with \proglang{Java}, we recommend using \proglang{Xtend} because its syntax is consistent with \proglang{Blang}'s syntax. 
This is a consequence of both languages being constructed with the \proglang{Xtext} language development framework.

\subsection[Interoperability with Java]{Interoperability with \proglang{Java}}

\proglang{Blang}, \proglang{Xtend} and \proglang{Java} are seamlessly interoperable as the first two are transpiled into \proglang{Java}. More precisely, any \proglang{Java} type can be imported and used in \proglang{Blang}, and any model defined in \proglang{Blang} can be imported and used in \proglang{Java} with no extra work needed.

As such, types in \proglang{Blang} are equivalent to \emph{Java types}, a terminology that encompasses \proglang{Java} classes, interfaces, primitives, enumerations and annotation interfaces.
At a high level, a type can be thought of as a group of \emph{objects} (chunks of computer memory) that satisfy a certain set of properties (for example, they all support being passed in a certain function).
We do not assume prior knowledge of the \proglang{Java} language, in fact, \proglang{Blang} and \proglang{Xtend} syntax is often simpler compared to \proglang{Java}'s.

\subsection{Comments} 

Single line comments use the syntax 

\begin{Code}
// some comment spanning the rest of the line.
\end{Code}

Multi-line comments use 

\begin{Code}
/* 
many commented lines 
can go here 
*/
\end{Code}

\subsection[Blang models: high-level syntax]{\proglang{Blang} models: High-level syntax} \label{subsec:highlevelsyntax}

A \proglang{Blang} file is organized as follows:

\longline

\code{NameOfMyModel.bl} \vspace*{-10pt}

\longline
\begin{lstlisting}
// package and import statements

model NameOfMyModel {
  
  // variables declarations
  
  laws {
    // laws declaration
  }
  
  generate(nameOfMyRandomObject) {
    // generate block
  }
}
\end{lstlisting}
\longline

We briefly describe each code block as follows:
package statements are responsible for defining the package in which a \proglang{Blang} model belongs to.
Import statements are responsible for importing classes, functions, and models from other packages.
The variables declarations block is responsible for declaring model variables, i.e., observed (constant) variables, latent variables, unknown parameters, known (constant) parameters.
The laws block is used to declare the probability distribution associated with each of the (random) model variables (see Section~\ref{subsec:laws}).
The optional generate block is used for forward sampling from the model (see Section~\ref{subsubsec:generate}).
It will also be helpful to keep in mind that XExpressions (to be introduced) are imperative, while laws blocks are declarative.
\updated{
Declarative code blocks do not have a notion of order, in other words, permuting the order of two statements will have no observable effect on the program.
}

In the remainder, if a string such as \code{NameOfMyModel} contains the substring ``\code{My}'', or has an integer as suffix, it refers to an identifier that should be tailored to the context of the model being written.

\proglang{Blang} is case-sensitive. 
Identifiers (model names, variable names, etc) should start with a letter and only use letters, numbers, and underscores. Furthermore, as a convention we encourage users to capitalize model names.

\subsection{Packages and imports} \label{subsec:pkgimports}

The \emph{packages} construct deals with the rare, but unavoidable, situation of wanting to use code from two developers that used the same name for a \proglang{Blang} \code{model}.
Package declarations will disambiguate the two.

Packages in \proglang{Blang} work the same as in \proglang{Java}, and precede \code{import} statements.
To declare a \proglang{Blang} \code{model} as part of a hierarchical group of related code, place the following declaration at the very beginning of the \proglang{Blang} file:

\longline
\begin{Code}
package myOrganization.myPackageName
\end{Code}
\longline

This \emph{package declaration} line is optional but recommended if you plan to share your code.
The dot in \code{myOrganization.myPackageName} denotes a hierarchical organization going from broader to more specific from left to right.
As a convention, package names are generally not capitalized.

To use another \proglang{Blang} \code{model} called \code{AnotherModel} from a package named \code{some.other.pack}, we can use \code{import} statements of the form:

\longline
\begin{Code}
import some.other.pack.AnotherModel
\end{Code}
\longline

after the package declaration line.
The same syntax can be used to import \proglang{Java} or \proglang{Xtend} classes, where \code{import static} is used to import a function, while a standalone \code{import} statement is used for types.

Package declarations effectively enable users to refer to specific objects of a package explicitly through import statements.
In the example below we see why this would be useful.
Suppose our model requires two data types from \code{package1} and \code{package2}, each of which contain an identically named but different implementation of \code{DupedType}.
In the unlikely event of having to use two types with duplicated names within the same file, importing should be avoided (i.e., do not \code{import package1} nor \code{import package2}).
Instead each instance of the type should be prefixed with the package name within the code, as such:

\newpage
\longline

\code{MyModel.bl}\vspace*{-10pt}

\longline
\begin{lstlisting}
model MyModel{
  random package1.DupedType var1
  random package2.DupedType var2
  ...
\end{lstlisting}
\longline

In contrast, here is an example of what \emph{not} to do:

\longline

\code{MyModel.bl}\vspace*{-10pt}

\longline
\begin{lstlisting}
import package1.DupedType
import package2.DupedType

model MyModel{
  random DupedType var1
  random DupedType var2
  ...
}
\end{lstlisting}
\longline

A related construct is the extension import mechanism, described in more detail in Section~\ref{subsec:xexp}.

\subsection[Automatic imports in Blang files]{Automatic imports in \proglang{Blang} files} \label{subsec:autoimports}

Any \proglang{Blang} file automatically imports:\footnote{The relevant Javadocs can be found at \url{https://www.stat.ubc.ca/~bouchard/blang/Javadoc.html}.}
\begin{itemize}
  \item all the types in the following packages: \\
    \code{blang.core}, \\
    \code{blang.distributions}, \\
    \code{blang.io}, \\
    \code{blang.types}, \\
    \code{blang.mcmc}, \\
    \code{java.util}, \\
    \code{xlinear}
  \item all the static functions in the following files: \\
    \code{xlinear.MatrixOperations}, \\ 
    \code{bayonet.math.SpecialFunctions}, \\
    \code{org.apache.commons.math3.util.CombinatoricsUtils}, \\
    \code{blang.types.StaticUtils}
  \item as static extensions all the static functions in the following files:  \\
    \code{xlinear.MatrixExtensions}, \\
    \code{blang.types.ExtensionUtils}, \\
    \code{blang.distributions.Generators}
\end{itemize}

\subsection{Model Variables} \label{subsec:modvars}

Model variables encompass all observed (fixed) variables, latent variables, unknown parameters, and known (constant) parameters in a statistical model.
Model variables are declared using one of two methods, declared with no default initialization:

\longline
\begin{lstlisting}
random Type1 name1
param Type2 name2
\end{lstlisting}
\longline

or with default initialization:

\longline
\begin{lstlisting}
random Type3 name3 ?: XExpression1
param Type4 name4 ?: XExpression2
\end{lstlisting}
\longline

\updated{
Observed and latent random variables are declared with \code{random}, while parameters are declared with \code{param} (see Section~\ref{sec:model-overview}).
The initialization blocks, denoted by \code{XExpression1} and \code{XExpression2}, are imperative blocks of code used to provide default values in the absence of CLI arguments.
For example:
}

\longline
\begin{lstlisting}
random Double abc ?: {
  val x = 123.0
  return exp(x)
}
\end{lstlisting}
\longline

The expressions in initialization blocks are constructed with so called \emph{XExpressions}. XExpressions are introduced in more detail in Section~\ref{subsec:xexp} and are used to construct several aspects of \proglang{Blang} programs.
For now, think about XExpressions as chunks of code (lists of statements or expressions) capable of performing arbitrary computations (loops, conditionals, creating temporary variables, calling other functions, etc), and returning one value. 
\updated{The statements or expressions in a block can be terminated by a new line or by a semicolon.}

\updated{
If the block contains only one expression, the brackets can be omitted:
}

\longline
\begin{lstlisting}
random Double abc ?: exp(123.0)
\end{lstlisting}
\longline

Initialization blocks can use values of previously listed variables.
If a CLI argument is provided, then the initialization block will be overridden by it.

\subsection{Laws block} \label{subsec:laws}

Laws blocks are used to \emph{declare} the (conditional) probability distribution associated with each random variable.
Note that unlike common programming languages used today for data analyses such as \proglang{Python} and \proglang{R}, the \code{laws} block is declarative.
In particular, the interpretation of a model is invariant to the order in which the individual laws are declared in the code.

\subsubsection{Composite laws} \label{subsubsec:complaw}

Described conceptually in Section~\ref{sec:interpretation-composite}, composite laws have the following syntax in \proglang{Blang}:

\longline
\begin{Code}
variableExpression1, variableExpression2, ...
  | conditioning1, conditioning2, ...
    ~ MyDistributionName(argumentExpression1, argumentExpression2, ...)
\end{Code}
\longline

For example:

\longline
\begin{Code}
y | mu, variance  ~ Normal(mu + 123,  variance)
\end{Code}
\longline

Where \code{variableExpression1}, \code{conditioning1}, \code{conditioning2}, \code{argumentExpression1}, and \code{argumentExpression2} correspond to  \code{y}, \code{mu}, \code{variance}, \code{mu + 123}, and \code{variance} respectively.

\code{MyDistributionName} refers to another \proglang{Blang} model.
Each element in \code{argumentExpression1, argumentExpression2, ...}  is matched from left to right in the same order as the \code{param} variables are declared in the model \code{MyDistributionName}.

The list \code{(argumentExpression1, argumentExpression2, ...)} corresponds to the transformation $t : T \times \Theta \to \Theta'$ in the notation used in Section~\ref{sec:interpretation-composite}.
This is implemented by allowing each element in \code{argumentExpression1, argumentExpression2, ...} to be an XExpression which is recomputed each time the value of the density $f_\theta(x)$ is queried;
\updated{
the expressions \code{argumentExpression1, argumentExpression2, ...} are compiled to lambda expressions.
Continuing with the example above, \code{mu + 123} will be computed at every iteration of an MCMC algorithm (when factors dependent on $\mu$ are required).
In other probabilistic programming languages, these expressions are often referred to as deterministic nodes/variables (in \proglang{RevBayes} and \proglang{BUGS} for example).\footnote{In contrast to these other languages, these deterministic nodes cannot be straightforwardly named and traced at the moment. We are investigating ways to incorporate this feature in future releases. }
}

Each element in \code{variableExpression1, variableExpression2, ...} is matched from left to right in the same order as the \code{random} variables are declared in model \code{MyDistributionName}.
To relate this to Section~\ref{sec:interpretation-composite}, the list \code{variableExpression1, variableExpression2, ...} corresponds to the output of the selection function $s : T \to T'$. This is implemented by allowing each \code{variableExpression} to be an XExpression which is executed only once, at initialization time. Often this XExpression is only a variable name, but it could also be an expression selecting an entry in a list or vector.

The conditioning block, \code{conditioning1, conditioning2, ...} is used to restrict what can be accessed by the transformation $t$. This is called the \emph{scope} of the transformation $t$. It is useful to restrict the scope as much as possible since this restriction induces sparsity patterns in the model. Sparsity is then exploited by our efficient inference algorithms.

Specification of the scope is implemented as follows. Each item within \code{conditioning1, conditioning2, ...} can take one of two possible forms.
First, it can be one of the variable names declared via the keyword \code{random} or \code{param}. For example, this first method is used in all conditionings of the Doomsday model (see Figure~\ref{fig:doomsDayBlangDefn}).

The second method to specify a conditioning is as follows:

\longline
\begin{Code}
variableExpression1, variableExpression2, ...
  | MyType myConditioningVariable = XExpression1, ...
    MyDistributionName(argumentExpression1, argumentExpression2, ...)
\end{Code}
\longline

where \code{MyType} is a type, \code{myConditioningVariable} is a local variable that exists only for the declaration of \code{variableExpression1, variableExpression2, ...}'s law.
The code in \code{XExpression1} has access to all model variables.
For example:

\longline
\begin{Code}
y | RealVar mu = manyMus.get(0) ~ Normal(mu, 1)
\end{Code}
\longline

The \code{XExpression1} code is executed only once at initialization.
We show in Section~\ref{sec:markovChainExample} an example of the typical use case for this initialization process, where in a model for a Markov Chain, this initialization is simply to select, in a list of random variables, the variable corresponding to the previous time step.

\subsubsection{Atomic laws} \label{subsubsec:atomlaw}

Informally, atomic laws are used to compute factors, and are the building blocks for composite laws.
Described conceptually in Section~\ref{sec:interpretation-atomic}, atomic laws have the following syntax in \proglang{Blang}:

\longline
\begin{lstlisting}
laws {
  logf(expression1, expression2, ...) { XExpression }
}
\end{lstlisting}
\longline

For example, $x \sim \text{Normal}(\mu,\sigma^2)$ ($\text{realization} \sim \text{Normal}(\text{mean}, \text{variance})$) would have the following encoding:

\longline

\code{Normal.bl} \vspace*{-10pt}

\longline
\begin{lstlisting}
laws {
  logf(mean, variance, realization) {
    if (variance < 0.0) return NEGATIVE_INFINITY
    return (- log(2*PI) / 2.0
            - 0.5 * log(variance)
            - 0.5 * pow(mean - realization, 2) / variance)
  }
}
\end{lstlisting}
\longline

It is recommended to separate factors with as few arguments together as possible, as this will help the runtime architecture determine dependencies and avoid redundant computation.
For example, the \code{Normal.bl} implementation is recommended to be factorized as:

\longline

\code{Normal.bl} \vspace*{-10pt}\footnote{\url{https://github.com/UBC-Stat-ML/blangSDK/blob/master/src/main/java/blang/distributions/Normal.bl}}

\longline
\begin{lstlisting}
laws {  
  logf() {
    - log(2*PI) / 2.0
  }
  logf(variance) {
    if (variance < 0.0) return NEGATIVE_INFINITY
    return - 0.5 * log(variance)
  }
  logf(mean, variance, realization)  {
    if (variance < 0.0) return NEGATIVE_INFINITY
    return - 0.5 * pow(mean - realization, 2) / variance
  }
}
\end{lstlisting}
\longline

Recall that in Section~\ref{sec:interpretation-atomic}, each atomic law was denoted as $\log\left(f^{(k)}(x, \theta)\right)$.
Here the list \code{expression1, expression2, ...} is used to restrict the scope of $f^{(k)}$, with the same motivation and mechanism as for composite laws, described in the last section.
Each item in the list \code{expression1, expression2, ...} follows the same syntax as the items in \code{conditioning1, conditioning2, ...} also described in the last section.

The \code{XExpression} is responsible for computing the numerical value of $\log\left(f^{(k)}(x, \theta)\right)$, and as such, should return a value of type \code{Double}. The \code{XExpression} is recomputed each time the value of the density $f_\theta(x)$ is queried.

\subsubsection{Declarative loops} \label{subsubsec:declarative}

In practice, the factorization in Equation~(\ref{eq:factorization}) may have a large number of factors. To assist the user in declaring these factors, we provide a ``declarative loop''  construct:

\longline
\begin{lstlisting}
for (MyIteratorType myIteratorName : XExpression) { ... }
\end{lstlisting}
\longline

This will repeat all the declarations inside \code{{ ... }} be they atomic or composite. Loops can be nested with the expected cross product behaviour.

The \code{XExpression} should return an object of type \code{java.lang.Iterable}. Some important loop idioms:

\begin{itemize}
  \item Simple loop from 0 (inclusively) to 10 (exclusively):\\ \code{for (Integer i : 0 ..< 10) \{ ...\ \}},
  \item Loops based on a \code{Collection}, which offer a wide choice of data structures via \proglang{Java}'s SDK\footnote{\url{https://docs.oracle.com/javase/tutorial/collections/index.html}} or Google's Guava project\footnote{\url{https://github.com/google/guava/wiki/CollectionUtilitiesExplained}}.
  \\ An example iterating over a power set\footnote{This requires the import line \code{import static extension com.google.common.collect.Sets.powerSet}}: \\\code{for (Set<Integer> s : (0 ..< 5).powerSet) \{ ...\ \}}
  \item Loops based on \proglang{Xtend}'s or \proglang{Java}'s utilities.\footnote{Documentation for \proglang{Xtend}'s utilities available at \url{https://www.eclipse.org/xtend/documentation/203_xtend_expressions.html} and documentation for \proglang{Java}'s streams is available at \url{https://docs.oracle.com/javase/tutorial/collections/streams/index.html}.} An example iterating over the first four even integers: \\
  \code{for (Integer i : (0 ..< 10).filter[it \% 2 == 0]) \{ ...\ \}} \\
  The keyword \code{it} is explained in section~\ref{sec:implicitIt}.
  \item The \code{Plate} data structure supplied by the \proglang{Blang} SDK is described in more detail in Section~\ref{sec:input}.
\end{itemize}

The current runtime infrastructure assumes that the \code{XExpression} specifying the range should not be random, in particular, it should not change during sampling.
As such it is only computed at initialization.
Therefore, declarative loops, which \emph{surround} atomic and composite laws, are different than the loops \emph{within} \code{XExpressions}.
Although \proglang{Blang} does not currently have built-in data types for sampling of infinite dimensional objects, they can be handled by creating dedicated types and/or using \code{XExpression} loops \emph{inside} a \code{logf} block. 

\subsection{Generate block} \label{subsubsec:generate}

The generate block is responsible for the forward generating mechanism of a model.
This is optional in that it is only required when more sophisticated inference algorithms are desired, as discussed in Section~\ref{subsec:interpretation-generate}.
An important distinction from \code{laws} blocks is that \code{generate} blocks are imperative. 
Furthermore, they are not referentially transparent as random variables will be modified in-place.
We formalize the syntax used to encode the generate block introduced conceptually in Section~\ref{subsec:interpretation-generate}:

\longline
\begin{lstlisting}
generate(myRandomSeed) {
  XExpression
}
\end{lstlisting}
\longline

The argument \code{myRandomSeed} is the name of an input object of type \code{java.util.Random} (the type declaration for this input is skipped since this is the only possible type allowed). To connect this syntax with its interpretation described in Section~\ref{subsec:interpretation-generate}, the input argument can be thought as an outcome $\omega \in \Omega$, from which the \code{XExpression} should form the realization $X(\omega)$. 

If the \code{model} has exactly one \code{random} variable of type \code{IntVar} or \code{RealVar}, then the \code{generate} block should return an \code{int} or \code{double} respectively, corresponding to the new realization.
Otherwise, the generate block should modify the \code{random} variable(s) in-place.
\updated{
The special case for univariate \code{IntVar} and \code{RealVar} is just syntactic sugar: under the hood, generated code uses the returned realization to modify the single variable to be sampled in-place.
}

\subsection{Latent random variables and their reference measures} \label{subsubsec:constraints}

Each type of \code{random} variable which we would like to be latent is required to declare one or more sampling algorithms.
This is achieved by adding the following \emph{type annotation} in the \proglang{Xtend} or \proglang{Java} class for that data type:

\longline
\begin{lstlisting}
@Samplers(MySampler1, MySampler2, ...)
class MyDataType {
  ...
}
\end{lstlisting}
\longline

Here each item in the list \code{MySampler1, MySampler2, ...} should be subtypes of the interface \code{Sampler}.\footnote{\api{sdk}{blang/mcmc/Sampler}{blang.mcmc.Sampler}}

Implicitly, the samplers associate a default reference measure to the latent \code{random} variables.
In some models, it may be necessary to overwrite these default reference measures for a particular \code{random} variable.
In such cases, \proglang{Blang} provides a mechanism to change them by adding in the laws block, a line of the following form:
\newline

\longline
\begin{lstlisting}
laws {
  ...
  myVariableName is Constrained
}
\end{lstlisting}
\longline

\updated{
In the above, \code{myVariableName} refers to the \code{random} variable name for which the default reference measure is to be changed, and \code{Constrained} can also be any class which implements \code{Factor}.\footnote{\api{dsl}{blang/core/Constrained}{blang.core.Constrained}, \api{dsl}{blang/core/Factor}{blang.core.Factor}}
Effectively, the intended behaviour is to disable samplers which would be inoperative with the alternate choice of reference measure.

To illustrate this necessity, consider a $K$-dimensional  Dirichlet distributed random variable  (i.e., $p = (p_1, p_2, \dots, p_K)$). 
By default, \proglang{Blang} would automatically designate slice samplers for each of the coordinates $p_1, p_2, \dots, p_K$, as they are of type \code{RealVar} variables.
However, because of the simplex constraint requiring $\sum_{i=1}^K p_i=1$, this would lead to proposal rejections almost surely.
The keyword \code{Constrained} is used to prevent this automatic assignment of ineffective or incorrect samplers.
\footnote{Technically this could be achieved by creating a simplex type that is constructed without referencing an \code{RealVar} types, but this is cumbersome as it is natural to use standard matrix objects within a simplex implementation.}

Thus for the Dirichlet distribution, we require the line \code{realization is Constrained} to disable each coordinate's default sampler, as seen below:
}

\longline

\code{Dirichlet.bl}\vspace*{-10pt}\footnote{\url{https://github.com/UBC-Stat-ML/blangSDK/blob/master/src/main/java/blang/distributions/Dirichlet.bl}}

\longline
\begin{lstlisting}
...

model Dirichlet {
  random Simplex realization 
  param  Matrix concentrations 
  
  laws {
    logf(concentrations, realization) {
      ...
    }
    realization is Constrained  
  }
  
  generate(rand) { 
    ...
  }
}
\end{lstlisting}
\longline

\updated{
Having disabled default samplers, the next logical step is to ensure the variable (of type \code{Simplex}) has an appropriate sampler of its own.
The reader is recommended to return to this section after building familiarity with Sections~\ref{sec:customSamplers} and \ref{subsec:mcmcKernels}, where the details of creating custom samplers are discussed.
A very high-level introduction to creating samplers is discussed here only to highlight the role of the constraint mechanism (if and when required) within samplers.

Consider \proglang{Blang}'s implementation of a simplex sampler:
}

\longline

\code{SimplexSampler.xtend}\vspace*{-10pt}\footnote{\url{https://github.com/UBC-Stat-ML/blangSDK/blob/master/src/main/java/blang/mcmc/SimplexSampler.xtend}}

\longline
\begin{lstlisting}
...
class SimplexSampler implements Sampler {
  @SampledVariable DenseSimplex simplex
  @ConnectedFactor List<LogScaleFactor> numericFactors
  @ConnectedFactor Constrained constrained
  
  override void execute(Random rand) {
    ...
  }
  
  ...
}
\end{lstlisting}
\longline

\updated{
The annotation \code{@SampledVariable} informs \proglang{Blang} that the \code{simplex} field is the variable to be sampled in-place.
Here we focus on the line \code{@ConnectedFactor Constrained constrained} which signals that it is appropriate to use this sampler, even in the presence of a constrained factor connected to the sampled variable in the factor graph. 
In contrast, the default sampler for real variables, \code{RealSliceSampler}, does not have \code{@ConnectedFactor Constrained constrained} stating that it is not able to accommodate sampling of the variable when it is connected to such factor.
Again refer to Section~\ref{subsec:mcmcKernels} for a more detailed discussion.

The function \code{execute} samples \code{simplex} in-place, or in other words, mutates the variable as an update.
The implementation details of \code{execute} are not important for the discussion of the constraint mechanism, and are thus hidden.

In short, to change a reference measure for a variable, a user should first disable the sampler for a variable by declaring \code{myVariableName is Constrained} in the \code{laws} block of a \proglang{Blang} model.
Then create a sampler, and annotate a field of type \code{Constrained} with \code{@ConnectedFactor} to signal that it can handle this type of constraint. 

A more refined typology of constraints can be built by the user, simply by creating subtypes of \code{blang.core.Factor}. Those used in the standard library might also be refined in future releases. 
}

\subsection{XExpressions} \label{subsec:xexp}

Syntax for XExpressions is provided by the \proglang{Xtext} language engineering framework. 
XExpressions are imperative expressions.
Thus the \code{logf}, \code{generate}, and variable initialization blocks for example are imperative, while \code{laws} blocks are declarative.

Here we highlight key aspects commonly used in \proglang{Blang} programs.
We refer the reader to the \proglang{Xtext} documentation for more information.\footnote{Documentation page can be found at \url{https://www.eclipse.org/Xtext/documentation/index.html}}

XExpressions can be either a single instruction as in the \emph{argument} of the following Exponential composite law:

\longline
\begin{Code}
y | a, b, x ~ Exponential(exp(a * x + b))
\end{Code}
\longline

or there can be several instructions nested in braces, with the last one providing the return value, as in this equivalent version of the above code

\longline
\begin{Code}
y | a, b, x ~ Exponential({
  val product = a * x
  exp(product + b)
})
\end{Code}
\longline

\subsubsection{Types}

We classify types into three main categories: primitives, object references, and array references. The most common primitives are \code{boolean}, \code{int}, and \code{double}.\footnote{They have the same characteristics as in \proglang{Java}, see \url{https://docs.oracle.com/javase/tutorial/java/nutsandbolts/datatypes.html} for technical details.}
Object references can be thought of as an annotated address to a memory location, possibly \code{null}. Lastly, array references are rarely used directly in \proglang{Blang}. Instead, arrays are typically encapsulated in more convenient data structures.

\subsubsection{Literals}

Examples of expressions that create constants of type\dots

\begin{itemize}
  \item \code{boolean}: \code{true}, \code{false}
  \item \code{int}: \code{42}, \code{12000}
  \item \code{double}: \code{1.0}, \code{1.3e2}, making sure to include the decimal suffix or to use scientific notation.
  \item String literals: either via \code{"A"},\\ or \textquotesingle \textquotesingle  \textquotesingle\code{This version allows "quotes inside" and more}\textquotesingle \textquotesingle  \textquotesingle 
  \item type literals: \code{MyType}, which is equivalent to \proglang{Java}'s \code{MyType.class}.
  \item \code{List}: \code{\#[true,false]} (note the hash symbol \code{\#} is not a comment as in other languages, it is used to construct lists, sets, and maps)
  \item \code{Set}: \code{\#\{"A","C","G","T"\}}
  \item \code{Map}: \code{\#\{"key1" -> 1, "key2" -> 2\} }
  \item \code{Pair}: \code{"likelihood" -> 1.43} (this example returns type \code{Pair<String, Double>}; this syntax can be used with arbitrary key and value types).
\end{itemize}

\subsubsection{Declaring variables with XExpressions}

Local variables have to be declared at their first occurrence. The main syntax variant to do so are:

\longline
\begin{lstlisting}
var int myModifiableInt = 17 
var typeInferred = #[1,2,3]
val int myConstantInt = 17
\end{lstlisting}
\longline

In the example, \code{var} encodes a variable that is mutable whereas \code{val} encodes a variable that is immutable.
The meaning of immutability is simple to understand in the case of a primitive, but it should be interpreted carefully in the context of references. In the latter, it means that the reference will always point to the same object in the heap, however the internal state of that object might change over time.

In the above, \code{typeInferred} illustrates that the type can be inferred automatically, in this example a \code{List<Integer>}. 

\subsubsection{Conditionals}

Conditional expressions have the following form:

\longline
\begin{lstlisting}
val String variable = if (condition) value1 else value2
\end{lstlisting}
\longline

Conditional expressions return values depending on a condition, where \code{condition} evaluates to a boolean.
When \code{(condition)} is \code{true}, \code{value1} is returned, otherwise \code{value2} is returned. 
The shorthand notation without \code{else}

\longline
\begin{lstlisting}
if (condition) value
\end{lstlisting}
\longline

or

\longline
\begin{lstlisting}
if (condition) {
   (value)
}
\end{lstlisting}
\longline

is equivalent to \code{if (condition) value else null}.

\subsubsection{Scope}

The \emph{scope} of a variable is defined as the portion of code in which the variable can be accessed.
Scoping in \proglang{Blang} is similar to most languages where in order to find the scope of a variable we identify the parent braces and determine the region of the code where the variable can be accessed.
For example, a local variable declared within the body of a \code{for} loop (the regions between curly braces) cannot be accessed outside of the body.
If one variable reference is in the scope of several variables declared with the same name, then the innermost braces have priority. 

The only exception is the arguments of the atomic and composite laws.
Recall our example in Section~\ref{subsubsec:complaw} (repeated below),

\longline
\begin{Code}
variableExpression1, variableExpression2, ...
  | conditioning1, conditioning2, ...
    ~ MyDistributionName(argumentExpression1, argumentExpression2, ...)
\end{Code}
\longline

These laws require explicit declaration of the variables to include in the scope, where these variables should be identified at the right of the \code{|} symbol.
This design choice is primarily motivated by its flexibility in handling complex dependencies, to be demonstrated in Sections~\ref{sec:undirected} and \ref{sec:markovChainExample}.

\subsubsection{XExpression loops}

In addition to allowing loops following the declarative loop syntax, loops within XExpressions allow the number of iterations to be random as well as a few syntactic alternatives:

\begin{enumerate}
  \item Basic, C-like for loops: \\ \code{for (var IteratorType iteratorName = init; condition; update) \{...\}} \\ An example of which would be \\ \code{for (var int i = 0; i <= 10; i++) \{...\}}.
  \item While loops: \\ \code{while (condition) \{...\}}.
\end{enumerate}

\subsubsection{Function calls}

Functions are called as one would expect: \code{nameOfFunction{(expression1, expression2)}} where each element in \code{expression1} and \code{expression2} are XExpressions.
\updated{
These expressions are evaluated prior to being passed into the function (i.e.,  a form of ``eager/greedy evaluation''), in order from left to right.
}

The only exceptions are composite laws, where the evaluation of an argument is delayed at initialization and instead repeated each time the density is evaluated during  sampling (i.e., a form of ``lazy evaluation''). 
To see why this is needed, consider a factor declaration of the form \code{y | x ~ Normal(2 * x, 1)}. Each time this factor is computed during inference, we would like the mean parameter \code{2 * x} to be recomputed. One way to think about lazy evaluation in this context is that when the factor graph is created, \code{2 * x} is converted into a lambda expression which is computed each time we are computing the value of the normal factor.

In all cases, the actual function call only involves copying a constant size register making these calls very cheap. For primitives, the value of the primitive is copied and therefore the original primitive can never suffer side effects from the call.
For object references, the memory address in the reference is copied and hence the original reference cannot be changed, although the object it points to might have its state changed by the function call.

\subsubsection{User defined functions}

To create supporting functions, the user can create a separate \proglang{Xtend} or \proglang{Java} file. In \proglang{Xtend}, use the following template for the separate file, say \code{MyFunctions.xtend}:

\longline

\code{MyFunctions.xtend} \vspace*{-10pt}

\longline
\begin{lstlisting}
package my.pack
class MyFunctions {
  def static ReturnType myFunction(ArgumentType1 arg1, ArgumentType2 arg2) {
    // some computation
    return result
  }
}
\end{lstlisting}
\longline

Back to the \proglang{Blang} file being developed, the user can then import the functions into the \proglang{Blang} file using \code{import static my.pack.MyFunctions.*} allowing us to call \code{myFunction(arg1, arg2)}.

\subsubsection{Extensions}\label{sec:extensions}

Extension methods provide a kind of lightweight trait, i.e., adding methods to existing classes on demand.

Continuing the same example in the last section, this is done by adding an extension import statement:

\longline
\begin{lstlisting}
import static extension my.pack.MyFunctions.myFunction
\end{lstlisting}
\longline

Provided a variable, say \code{myVar}, of type \code{ArgumentType1} (the type of the first input argument to the function \code{myFunction} defined in the previous section), the user can then invoke the function via \code{myVar.myfunction(arg2)}.

As a concrete example of how this is used to create more readable code, consider a typical \code{generate} snippet, showing here how a Yule Simon distributed variate can be generated as a mixture

\longline
\begin{lstlisting}
generate(rand) {
  val w = rand.exponential(rho)
  return rand.negativeBinomial(1.0, 1.0 - exp(-w)) 
}
\end{lstlisting}
\longline

This can be equivalently written, more explicitly, as

\longline
\begin{lstlisting}
generate(rand) {
  val w = Generators.exponential(rand, rho)
  return Generators.negativeBinomial(rand, 1.0, 1.0 - exp(-w))
}
\end{lstlisting}
\longline

The underpinning of this code is that since \proglang{Blang} automatically imports all functions in \code{Generators} as extension methods,\footnote{\api{sdk}{blang/distributions/Generators}{blang.distributions.Generators}}
which contains the function:

\code{def static double exponential(Random random, double rate)}

then we can call \code{rand.exponential(...)} on the variable \code{rand} of type \code{java.util.Random}.

\subsubsection{Creating objects}

An object of type \code{MyClass} is created by calling \code{new MyClass(argument1, ...)}. This can be shortened to \code{new NameOfClass} if there are no arguments. To find which argument(s) are necessary, look for the \emph{constructor} in \code{MyClass}, which uses the keyword \code{new} in \proglang{Xtend} and the name \code{MyClass(...)} in \proglang{Java}.

In some libraries, for example in the package we use for linear algebra, \pkg{xlinear}, the call to \code{new} is wrapped inside a static function. In this case, just call the function to instantiate the object.
For example, to create a new sparse matrix with $1\,000$ rows and $10\,000$ columns, use \code{sparse(1_000, 10_000)} (automatically imported from \code{xlinear.MatrixOperations}).\footnote{\api{xlinear}{xlinear/MatrixOperations}{xlinear.MatrixOperations}}

\subsubsection{Using objects}

Classes have \emph{instance variables} or \emph{fields}, which are variables associated with objects, as well as \emph{methods},  which are functions associated with the object having access to the object's instance variables. Collectively, fields and methods are called \emph{features}.

Features are accessed using the ``dot'' notation: \\ \code{myObject.myVariable}
and \code{myObject.myMethod(...)}. When a method has no argument, the call can be shortened to \code{myObject.myMethod}.

The ability to call a feature is subject to \proglang{Java} visibility constraints. In short, only public features can be called from outside the file declaring a class.

\subsubsection[Implicit variable it]{Implicit variable \code{it}} \label{sec:implicitIt}

The special variable \code{it} allows users to provide a default object for feature calls:\footnote{\url{https://www.eclipse.org/xtend/documentation/203_xtend_expressions.html}}

\longline
\begin{lstlisting}
val it = myObject
doSomething 
\end{lstlisting}
\longline

This is merely a shorthand notation for:

\longline
\begin{lstlisting}
myObject.doSomething 
\end{lstlisting}
\longline

and is used in lambda expressions which we discuss next.

\subsubsection{Lambda expressions}

A \emph{lambda expression} is a succinct way to write a function without having to give it a name. This construction makes it easy to call functions which take functions as argument (e.g., to apply a function to each item in a list). Since they are so useful, many syntactic shortcuts are available.

The explicit syntax for lambda expressions is:

\longline
\begin{Code}
[Type1 argument1, Type2 argument2, ... | functionBody ]
\end{Code}
\longline

For example, to capitalize words in a list, we can use the function \code{map(myFunction)} which applies \code{myFunction} to every entry in the list.
Here \code{map()} is the function that takes in another function \code{myFunction} as an argument.
More concretely, we have:

\longline
\begin{Code}
#["foo", "bar"].map([String s | s.toUpperCase])
\end{Code}
\longline

When there is a single input argument in the lambda expression (i.e., in the above case \code{String s}), you can skip declaring the argument, and instead the argument will be assigned to the implicit variable \code{it} (described in the previous section).
This allows us to write:

\longline
\begin{Code}
#["one", "two"].map([it.toUpperCase])
\end{Code}
\longline

which further simplifies to:

\longline
\begin{Code}
#["one", "two"].map([toUpperCase])
\end{Code}
\longline

Finally, when the last argument of a function (\code{map()} in this case) is a function, you can simply put the lambda after the parentheses of the function call (\code{map()}).
For example:

\longline
\begin{Code}
#["one", "two"].map()[toUpperCase]
\end{Code}
\longline

Which further simplifies to:

\longline
\begin{Code}
#["one", "two"].map[toUpperCase]
\end{Code}
\longline

\subsubsection{Boxing and unboxing}

Boxing refers to wrapping a primitive such as \code{int} or \code{double} into an object such as \code{Integer} or \code{Double}. Deboxing is the reverse process.
The \code{Integer} or \code{Double} objects are immutable data structures necessary as many data structures assume all their contents are references to objects rather than primitives. As in \proglang{Java}, the conversion between the two representations is automatic in the vast majority of the cases.
\proglang{Blang} adds boxing/deboxing to and from \code{IntVar} and \code{RealVar},\footnote{\api{dsl}{blang/core/IntVar}{blang.core.IntVar}, and \api{dsl}{blang/core/RealVar}{blang.core.RealVar}}
which are mutable versions of \code{Integer} or \code{Double}. See Appendix~\ref{app:samplerStateRep} for a discussion on why these mutable data structures are necessary in \proglang{Blang}.

\subsubsection{Operator overloading}

Operator overloading is permitted. When in the \proglang{Blang} IDE, command click on an operator to reveal its definition. One important case to be aware of is \code{==} which is overloaded to \code{.equals(...)}. For the low-level equality operator that checks if the two sides are identical (point to the same object or in the case of primitive, have the same value) use \code{===} (with the exception of \code{Double.NaN} which, following IEEE convention, is never \code{===} to anything). 


Some useful operators that are automatically imported:
\begin{itemize}
  \item ``\code{0..10}'', and ``\code{0..<11}'': These expressions are range operators and return integers 0, 1, 2, ..., 10.
  \item \code{object => lambdaExpression}: calls the lambda expression with the input given by object e.g., \code{new ArrayList => [add("to be added in list")]}
\end{itemize}

When overloading operators of custom type refer to \proglang{Xtend}'s official documentation \citep{XtendDoc2019}.

\subsubsection{Parameterized types}

Types can be parameterized as in \proglang{Java}'s \code{List} type. 
For example, we use \code{List<String>} to declare that a string will be stored, just as we would in \proglang{Java} and \proglang{Blang}. 
At the moment, models can use variables with type parameters but models themselves cannot have type parameters.

\subsubsection{Throwing exceptions}

Throw exceptions to signal abnormal behaviour and to terminate the \proglang{Blang} runtime with an informative message:

\longline
\begin{lstlisting}
throw new MyException("Some error message.")
\end{lstlisting}
\longline

Here \code{MyException} should be of type \code{java.lang.Throwable}. A reasonable default choice is \code{java.lang.RuntimeException}. 
To signal that the current factor has invalid parameters return the value \code{NEGATIVE\_INFINITY}.
If not possible due to a particular code structure, one can also return the value \code{invalidParameter}.\footnote{\api{sdk}{blang/types/StaticUtils}{blang.types.StaticUtils.invalidParameter}} 
This will be caught and interpreted as a factor having zero probability.
In contrast to \proglang{Java}, \proglang{Blang} exceptions are never required to be declared or caught. If an exception needs to be caught, the syntax is as follows:

\longline
\begin{lstlisting}
try {
  // code that might throw an exception
} catch (ExceptionType exceptionName) {
  // process exception
} // optionally:
finally {
  // code executed whether the exception is thrown or not
}
\end{lstlisting}
\longline

\section{Cheatsheet interlude} \label{sec:cheatsheet}
Learning a language can be a time-consuming task with new grammar and syntax to remember.
Before continuing with more examples, ideas, and patterns, we present a condensed summary of the concepts covered thus far in the form of a recipe.
We will draw connections to the GMM example from Section~\ref{sec:gmm} where appropriate.
For convenience, we repeat the model below:

\begin{align*}
  \text{concentration} && \alpha &= [1 , 1 ] \\
  \text{proportions} && \pi \mid \alpha & \sim \text{Dirichlet}(\alpha) \\
  \text{labels} && z_i \mid \pi & \sim \text{Categorical}(\pi) \\
  \text{means} && \mu_k & \sim \text{Normal}(0,10^2) \\
  \text{standard deviations} && \sigma_k \ & \sim \text{Uniform}(0, 10) \\
  \text{observations} && y_i \mid \mu, \sigma, z_i & \sim \text{Normal}(\mu_{z_i}, \sigma_{z_i}^2)
\end{align*}
for $i \in \{1, 2,\dots, n\}$ and $k\in \{1,2\}$.

\emph{The cheatsheet}
\begin{enumerate}
  \item Write down your package statement.\\
  \textbf{Example:} 
  \begin{lstlisting}
package jss.gmm
  \end{lstlisting}
  \item If it is already known which packages you will work with, then import them.
  We did not require additional packages for the GMM.\\
  \textbf{Example:} 
  \begin{lstlisting}
import some.other.pack
  \end{lstlisting}
  \item Name your model.\\
  \textbf{Example:} 
  \begin{lstlisting}
model MixtureModel { ... }
  \end{lstlisting}
  \item Identify all model variables: observed (constant) random variables (RV), latent RVs, unknown parameters, and known (constant) parameters.\\
  \textbf{Example:} observed RVs $y_i$, latent RVs $z_i$, unknown parameters $\pi, \mu_k, \sigma_k$, and known parameter $\alpha$.
  \item Identify model variables' types, and values of known parameters. \\
  \textbf{Example:} $y_i$ are real numbers, $z_i$ are integers, $\pi$ is a simplex, $\mu_k$ are real numbers, $\sigma_k$ are real numbers, and $\alpha=[1,1]$.  
  \item Declare all observed and latent RVs, and unknown parameters with the keyword \code{random}; declare all known parameters with the keyword \code{param}.\\
  \textbf{Example:} 
 \begin{lstlisting}
random List<RealVar> observations
random Simplex pi
param Matrix concentrations
  \end{lstlisting}
  \item Initialize latent RVs and unknown parameters with their latent types, and initialize known parameter values.
  Realization of observations will be delayed until inference. \\
  \textbf{Example:} 
  \begin{lstlisting}
random List<RealVar> observations
random Simplex pi ?: latentSimplex(2)
param Matrix concentrations ?: fixedVector(1.0, 1.0)
  \end{lstlisting}
  \item Next we declare their respective distributions in \code{laws\{ ... \}}. \\
  \textbf{Example:} 
  \begin{lstlisting}
pi | concentration ~ Dirichlet(concentration)
  \end{lstlisting}
  \item Use \code{for} loops to declare over a list of variables for cleaner code.\\
  \textbf{Example:} 
  \begin{lstlisting}
for (int k : 0 ..< means.size) {
  means.get(k) ~ Normal(0.0, pow(10.0, 2.0))
}
  \end{lstlisting}
  \item Perform inference by using the CLI. Append \code{-{}-help} to the CLI for model-specific input description. \\
  \textbf{Example:} 
  \begin{Code}
blang --model jss.gmm.MixtureModel \
      --model.observations file "path/to/line_separated_values.txt"
  \end{Code}
\end{enumerate}

\section{Custom samplers for custom data structures}\label{sec:customSamplers}
Although the focus of this section is on custom data structures and samplers, it will also provide insight to \proglang{Blang}'s underlying sampling mechanisms.

In examples we have demonstrated thus far, sampling variables could be handled via default samplers.
This luxury is typically unavailable when working with complex state spaces such as trees, partitions, permutation spaces, or more generally discrete, non-ordinal spaces.
In such situations \proglang{Blang} still assists the user in several ways described in more detail in Section~\ref{sec:sdk}.
Here we focus on how \proglang{Blang} helps implement a complete sampler for a model that consists of such custom data structures.

Consider a model with latent variables taking values in a set of permutations (perfect bipartite matching). 
For example, \emph{record linkage} problems \citep{Tancredi2011, Steorts2016} rely on this type of latent variable.
In short, record linkage is the process of matching de-identified noisy records from multiple data sources that reference the same entity or individual.
For example, consider a datum taking value $170$ in one dataset, and $170.1$ in another dataset.
The process of recognizing these two data reference the same entity is a case of record linkage.
In the following, we demonstrate how to implement a custom sampler for data of type permutation.

We will implement a data type, \code{Permutation}, equipped with its tailor-made sampler, then apply it in the context of a model.\footnote{Complete and commented implementations in this section are available in the reproduction materials located in the directory \code{reproduction\_materials/example}, or at \url{https://github.com/UBC-Stat-ML/JSSBlangCode/tree/master/reproduction_material/PermutationExample/src}.}
We begin by implementing a \emph{class} describing how permutations will be encoded. 
Note a permutation can be represented with or stored as a list of integers.
We present an implementation of the \code{Permutation} class below, and discuss each component individually.
\\
\\

\longline

\code{Permutation.xtend}\vspace*{-10pt}\footnote{\url{https://github.com/UBC-Stat-ML/JSSBlangCode/blob/master/reproduction_material/PermutationExample/src/main/java/jss/perm/Permutation.xtend}}

\longline
\begin{lstlisting}
package jss.perm

import org.eclipse.xtend.lib.annotations.Data
import blang.mcmc.Samplers
import java.util.Random
import static java.util.Collections.sort
import static java.util.Collections.shuffle
import java.util.List

@Samplers(PermutationSampler)
@Data class Permutation {

  val List<Integer> connections
  
  new (int componentSize) {
    connections = (0 ..< componentSize).toList
  }
  
  def int componentSize() { 
    return connections.size
  }

  def void sampleUniform(Random random) { 
    sort(connections)
    shuffle(connections, random)
  }

  override String toString() { 
    return connections.toString
  }
}
\end{lstlisting}
\longline

A first observation is the annotation \code{@Samplers(...)} which informs the runtime engine to sample \code{Permutation} objects with an instance of \code{PermutationSampler}.\footnote{More than one sampler can be specified as a comma-separated list, more on this in Section~\ref{subsec:mcmcKernels}.}
We will discuss \code{PermutationSampler} later.

A second observation is the annotation \code{@Data}.\footnote{Complete documentation available at \url{http://archive.eclipse.org/modeling/tmf/xtext/javadoc/2.9/org/eclipse/xtend/lib/Data.html}.}
Briefly, this annotation should be interpreted as a \emph{data class}, a terminology in object oriented programming (and an unfortunate clash with the conventional use of ``data'' in statistics), meaning the class can only declare final fields, and that \code{.equals}, \code{.hashcode} are automatically implemented in addition to other defaults.

Moving on to the main code block, as noted previously, we can represent the mathematical permutation object with a list of integers, where each element in the list is the permuted value of the element's index.
Thus we encoded the permutation object with the field \code{connections}, and a constructor as repeated below:
\\
\\

\longline

\begin{lstlisting}
val List<Integer> connections

new (int componentSize) {
  connections = (0 ..< componentSize).toList
}
\end{lstlisting}

\longline

This permutation constructor is to related to  Section~\ref{sec:gmm}'s \code{latentSimplex(K)} constructor for simplex variables. We will see its use later to construct a latent permutation to be sampled.
Technically, this is all that is required to represent the permutation type.
However, it will be convenient to define a few more helper functions, in particular a function \code{sampleUniform} to uniformly draw a realization of a permutation. 

\longline

\begin{lstlisting}
def void sampleUniform(Random random) { 
  sort(connections)
  shuffle(connections, random)
}
\end{lstlisting}

\longline

Notice \code{sampleUniform} sorts connections, then shuffles connections in-place.
The sorting is required from a computational perspective to ensure the sampling is not affected by the \code{connection}'s current state, thus uniform when shuffled. In other words, it enforces the contract that for a given random seed encoded in the \code{rand} object, the behaviour of the \code{generate} block is fully deterministic and not affected by the current state of the object. 
This behaviour is exploited to design test cases, as in \code{TestCompositeModel.xtend} described shortly. 
The sampling performed in-place is a technical requirement for the inference engine, detailed in Section~\ref{sec:inference}.

Finally the last piece of the puzzle, the \code{toString} function.
Its purpose is best illustrated by an example followed by an explanation.
Here is what our sample output would read without overriding \code{toString}:

\longline

\code{permutation.csv}\vspace*{-10pt}

\longline
\begin{Code}
sample,value
0,"Permutation [
  connections = ArrayList (
    2,
    0,
    1
  )
]"
1,"Permutation [
  connections = ArrayList (
    1,
    2,
    0
  )
]"
...
\end{Code}
\longline

With the \code{toString} function, we have:

\longline

\code{permutation.csv}\vspace*{-10pt}

\longline

\begin{Code}
sample,value
0,"[2, 0, 1]"
1,"[1, 2, 0]"
...
\end{Code}

\longline

Thus we see that by overriding the default string output of our object, we enable the engine to output something more legible.
One can customize this output to respect the Tidy philosophy, details of which we leave to Appendix~\ref{sec:tidily}.

With all the pieces in place for our \code{Permutation} class, we are now ready to discuss samplers.

To perform posterior inference on permutation spaces, we need an invariant sampler designed specifically for the object \code{Permutation}.
In this example, we assume familiarity with the Metropolis algorithm \citep{Metropolis1953}, and begin by presenting the full code below, followed by a breakdown of its main components:

\longline

\code{PermutationSampler.xtend}\vspace*{-10pt}\footnote{\url{https://github.com/UBC-Stat-ML/JSSBlangCode/blob/master/reproduction_material/PermutationExample/src/main/java/jss/perm/PermutationSampler.xtend}}

\longline
\begin{lstlisting}
package jss.perm

import java.util.List
import bayonet.distributions.Random
import blang.core.LogScaleFactor
import blang.mcmc.ConnectedFactor
import blang.mcmc.SampledVariable
import blang.mcmc.Sampler
import blang.distributions.Generators
import static java.lang.Math.exp
import static java.lang.Math.min
import static extension java.util.Collections.swap

class PermutationSampler implements Sampler {
  @SampledVariable Permutation permutation
  @ConnectedFactor List<LogScaleFactor> numericFactors
  
  override void execute(Random rand) {
    val n = permutation.componentSize
    val i = Generators.discreteUniform(rand, 0, n)
    val j = Generators.discreteUniform(rand, 0, n)
    
    val currentLogDensity = logDensity()
    permutation.connections.swap(i,j)
    val newLogDensity = logDensity()
    
    val acceptProb = min(1.0, exp(newLogDensity - currentLogDensity))
    val accept = Generators.bernoulli(rand, acceptProb)
    if (!accept) {
      permutation.connections.swap(i, j)
    }
  }
  
  def double logDensity() {
    var double sum=0.0 
    for (LogScaleFactor f : numericFactors) sum += f.logDensity() 
    return sum 
  }
}
\end{lstlisting}
\longline

A first observation is the \emph{implementation} of the \code{Sampler} \emph{interface}, and the two annotations \code{@SampledVariable} and \code{@ConnectedFactor}:\footnote{For more on interfaces, see \url{https://docs.oracle.com/javase/tutorial/java/concepts/interface.html}.}

\longline

\begin{lstlisting}
class PermutationSampler implements Sampler {
  @SampledVariable Permutation permutation
  @ConnectedFactor List<LogScaleFactor> numericFactors
\end{lstlisting}

\longline

Briefly, this implies the \code{PermutationSampler} class necessarily implements methods specified in the interface \code{Sampler}, namely \code{execute}.
The \code{execute} method is invoked with each iteration of the inference algorithm (Section~\ref{sec:inferenceEngines}), and updates our variable of interest in-place.
The field annotated with \code{@SampledVariable} will automatically be populated with an instance of the object to be sampled, in this example, an instance of \code{Permutation}.
This annotation in tandem with \code{@Samplers} enables the linkage of variables and samplers.\footnote{Sampling of multiple variables can also be performed. For example, the SDK incorporates an elliptic slice algorithm which samples many real variables at once, see  \url{https://github.com/UBC-Stat-ML/blangSDK/blob/master/src/main/java/blang/distributions/NormalField.bl} and \url{https://github.com/UBC-Stat-ML/blangSDK/blob/master/src/main/java/blang/mcmc/EllipticalSliceSampler.xtend}}
Similarly, the field annotated with \code{@ConnectedFactor} will automatically be populated with factors dependent on the sampled object, which is inferred automatically via a factor graph built from scope analysis (described in detail in Section~\ref{sec:inference}). 
By default once a type and its sampler have been implemented, variables of such type will be sampled with this sampler.
We discussed how this default is altered in Section~\ref{subsubsec:constraints}, and from another perspective in Section~\ref{subsec:mcmcKernels}.

With this setup, we are ready to implement the Metropolis algorithm for permutations:

\longline

\begin{lstlisting}
override void execute(Random rand) {
  val n = permutation.componentSize
  val i = rand.nextInt(n)
  val j = rand.nextInt(n)
  
  val currentLogDensity = logDensity()
  permutation.connections.swap(i,j)
  val newLogDensity = logDensity()
  
  val acceptProb = min(1.0, exp(newLogDensity - currentLogDensity))
  val accept = Generators.bernoulli(rand, acceptProb)
  if (!accept) {
    permutation.connections.swap(i, j)
  }
}

def double logDensity() {
  var double sum=0.0 
  for (LogScaleFactor f : numericFactors) sum += f.logDensity() 
  return sum 
  }
}
\end{lstlisting}

\longline

The implementation of \code{execute()} shown above is a standard Metropolis algorithm \citep{Metropolis1953}, which invokes \code{logDensity} when density evaluation is required.
Since the field \code{numericFactors} is a list of all log factors dependent on our variable, the \code{logDensity} method merely returns the sum of log factors.

Notice the syntax \code{Generators.bernoulli(rand, acceptProb)} is used to determine acceptance of the proposal.
This syntax equivalent to \code{Bernoulli.distribution(p).sample(rand)} (in fact the latter calls the former).
However the second variant creates an intermediate object of type \code{IntDistribution} which could be a performance issue as the body of the sampling algorithm is in the inner loop of inference.
In other contexts, having this intermediate object is useful, e.g., if one would like to provide a distribution as input parameter to another distribution, as in Section~\ref{sec:distributions-as-parameters}.
As for the relationship with \code{... ~ Bernoulli(...)},  recall that the difference is that \code{... ~ Bernoulli(...)} is used in a declarative context, while the first two syntaxes are for imperative blocks such as MCMC samplers (of course, in all three variants there is no code duplications in the SDK, i.e., higher-level functions such as the declarative syntax call lower level implementations).

With our custom \code{Permutation} type and \code{PermutationSampler} in place, we are ready to apply them in models.

An example of a uniform distribution over the permutation space is implemented as follows:

\longline

\code{UniformPermutation.bl}\vspace*{-10pt}\footnote{\url{https://github.com/UBC-Stat-ML/JSSBlangCode/blob/master/reproduction_material/PermutationExample/src/main/java/jss/perm/UniformPermutation.bl}}

\longline
\begin{lstlisting}
model UniformPermutation {
  random Permutation permutation 
  
  laws {
    logf(permutation) {
      - logFactorial(permutation.componentSize)
    }
  }

  generate(rand) {
    permutation.sampleUniform(rand)
  }
}
\end{lstlisting}
\longline

As we have seen before, the \code{logf} block provides a method for evaluating log densities, while the \code{generate} block provides a method for sampling permutations in place.
In this case, \code{logf} returns the log density of a permutation with uniform distribution, and \code{generate} samples a permutation uniformly.
As for \code{logFactorial}, which computes $\log(n!)$, it is part of the automatically imported functions described in Section~\ref{subsec:autoimports} (a list of the most commonly used automatically imported functions can also be found in Appendix~\ref{app:functions}).

As with any distribution, model \code{UniformPermutation} can be used in composition with other models.
Here we present a minimal, illustrative example:

\longline

\code{CompositeModel.bl}\vspace*{-10pt}\footnote{\url{https://github.com/UBC-Stat-ML/JSSBlangCode/blob/master/reproduction_material/PermutationExample/src/main/java/jss/perm/CompositeModel.bl}}

\longline
\begin{lstlisting}
package jss.perm

model CompositeModel {
  random List<RealVar> y ?: fixedRealList(2.1,-0.3,0.8)
  random Permutation permutation ?: new Permutation(y.size)

  laws {
    permutation ~ UniformPermutation
    for (int i : 0 ..< y.size) {
      y.get(i) | permutation, i 
        ~ Normal(permutation.getConnections.get(i), 0.3)	
    }
  }
}
\end{lstlisting}
\longline

This should look rather similar to other models, with the exception of the use of a custom constructor \code{new Permutation(y.size)} to instantiate the latent permutation variable.
\code{CompositeModel} provides a toy example of how one can incorporate \code{UniformPermutation} into larger models.
An example of custom data types with emphasis on a practical application using a spike and slab model \citep{Mitchell1988} is presented in Appendix~\ref{subsec:tutorial}.

This concludes our tutorial on creating custom data types and samplers. 
We dedicate the remainder of this section to showcasing some available resources that assist users in testing the correctness of samplers.

A first test utility provided by the SDK is \code{DiscreteMCTest}, which is specialized to fully-discrete spaces.
The idea behind \code{DiscreteMCTest} is that, for small discrete spaces, we can explicitly form a sparse transition matrix and numerically check properties such as invariance and irreducibility.
In our experience, many software defects can be found in problems just large enough to achieve code coverage.

To run tests, we need to setup a project directory, i.e., \code{create-blang-gradle-project -{}-name PermutationExample}.\footnote{Commented implementations on testing are available in the reproduction materials located in the directory \code{PermutationExample}.}
This will create a directory named ``PermutationExample'', with directory structure \code{src/main/java}.
Place our implementations in this directory, and create additional directories \code{PermutationExample/src/test/java/}.
Making sure the package names are matching, place \code{TestCompositeModel.xtend} in the testing directory \code{src/test/java/jss/perm/} with implementations as follows:

\longline

\code{TestCompositeModel.xtend}\vspace*{-10pt}\footnote{\url{https://github.com/UBC-Stat-ML/JSSBlangCode/blob/master/reproduction_material/PermutationExample/src/test/java/TestCompositeModel.xtend}}

\longline
\begin{lstlisting}
package jss.perm

import static blang.types.StaticUtils.*
import blang.runtime.SampledModel
import blang.validation.DiscreteMCTest
import com.rits.cloning.Cloner
import org.junit.Test
import static org.apache.commons.math3.util.CombinatoricsUtils.factorial
import static java.lang.Math.pow
import blang.runtime.internals.objectgraph.GraphAnalysis
import blang.runtime.Observations
import blang.types.ExtensionUtils

class TestCompositeModel {
  
  val static y = fixedRealList(2.1, -0.3, 0.8)
  val static CompositeModel compositeModel = new CompositeModel.Builder()
    .setY(y)
    .setPermutation(new Permutation(y.size))
    .build

  val static observations = {
    val Observations result = new Observations
    result.markAsObserved(y)
    result
  }

  val static DiscreteMCTest test = 
    new DiscreteMCTest(
      new SampledModel(new GraphAnalysis(compositeModel, observations)),
      [
        val CompositeModel cm = model as CompositeModel
        return new Cloner().deepClone(cm.permutation)
      ]
    ) 

  @Test 
  def void stateSize() {
    test.verbose = true
    test.checkStateSpaceSize(factorial(y.size) as int)
  }
  
  @Test
  def void invariance() {
    test.verbose = true
    test.checkInvariance
  }
  
  @Test
  def void irreducibility() {
    test.verbose = true
    test.checkIrreducibility
  }
}
\end{lstlisting}
\longline

The test can be performed using \code{./gradlew test} while in the \code{PermutationExample} directory, or via the Eclipse IDE.

\updated{
The \code{DiscreteMCTest} object takes in two arguments.
The first argument is a small, discrete model.
The construction of this discrete model is achieved using the compiled builder in \code{CompositeModel.java} (or in general, \code{ModelName.java}).
The second argument is a lambda function (denoted by square brackets) that accepts a model and creates a new object encoding the identity of the current configuration, with identity being mediated by the \code{.equals()} function of the returned object. 

As \code{DiscreteMCTest} is created (i.e., handled in the construction of the object), the samplers involved in the input model are automatically translated into explicit sparse transition matrices, via a type of non-standard evaluation of the sampling code.\footnote{More information is available at \url{https://www.stat.ubc.ca/~bouchard/blang/Testing_Blang_models.html} under ``Exhaustive tests.''}  
Given these inputs, irreducibility and invariance tests boil down to an application of linear algebra and graph algorithms.

Detailed testing resources are discussed in Section~\ref{subsec:testing}, including tests for models defined on continuous spaces.
}

\section{Tools and software development kit} \label{sec:sdk}

\proglang{Blang} comes with ``batteries included'': more than just a language, it is a suite of tools and libraries supporting common tasks in Bayesian data analysis.
In this section, we present an overview of these libraries.
Briefly, we start with a description of the \proglang{Blang} integrated development environment (IDE), followed by a discussion on how input of data is handled in \proglang{Blang}.
This includes an introduction to implementing plate and plated variables for plate notation used in traditional graphical models.
Next, we discuss how samplers, distributions, and other components fit into the core inference algorithms' architecture.
Finally, we conclude with brief discussions on post-processing options, monitoring logs, testing frameworks, and additional packages and dependencies.

\subsection{Integrated development environment (IDE)} \label{sec:IDE}

Integrated development environments are software applications built for software construction.
They are typically equipped with features such as syntax highlighting, code completion, refactoring, debugging and other tools that assist programmers in software development.

\subsubsection{Desktop IDE}\label{sec:desktop-ide}

We provide an IDE for \proglang{Blang} built on the Eclipse. The only requirement is that \proglang{Java} 11, 13, or 15 should be installed. 

\updated{There are two ways to install it: one pre-packaged, the other by adding a plug-in to an already installed Eclipse instance. The former method is more straightforward but currently we only distribute the pre-packaged \proglang{Blang} Eclipse for Mac OS X (tested with Mac OS 10.11.6, 10.14.6, 10.15.7). The latter method supports Mac OS X, Windows and Linux. 
	
For the Mac-specific method, download the IDE at \url{https://www.stat.ubc.ca/~bouchard/blang/downloads/blang-mac-4.0.7.zip}. Unzip the downloaded file and copy the contents to a directory of your choice. The folder contains both the IDE, a template for your own projects, and some command line tools. The first time you try to launch \proglang{BlangIDE}, depending on the version of Mac OS X and/or security settings, you may get a message saying the ``app is not registered with Apple by an identified developer.'' To work around this, follow these instructions (from Apple) the first time you open the BlangIDE (then Mac OS will remember your decision for subsequent launches): \url{https://support.apple.com/en-ca/guide/mac-help/mh40616/mac}. }

The second installation process for the \proglang{Blang} IDE, which is the most portable across platforms, is the following:

\begin{enumerate}
  \item \updated{Install \emph{DSL tools for Eclipse}, which can be downloaded from the Eclipse website.\footnote{\url{https://www.eclipse.org/}} At the time of writing, the supported version is Eclipse IDE for \proglang{Java} and DSL (domain specific language) Developers, Release 2020-12 R.\footnote{Linked at \url{https://www.eclipse.org/downloads/packages/release/2020-12/r/eclipse-ide-java-and-dsl-developers}.}
  Using the standard version of Eclipse (i.e., not the DSL version) and/or a different version is unlikely to work. }
  \item From Eclipse: select \textit{Install New Software} from the \textit{Help} menu.
  \item Click \textit{Add} and enter \\  \url{https://www.stat.ubc.ca/~bouchard/maven/blang-eclipse-plugin-4.0.7/} \\ in the location field.
  \item Click \textit{Select All}, \textit{Next}, then follow instructions as prompted.
\end{enumerate}

\begin{figure}[!h]
  \centering
  \includegraphics[width=1.0\linewidth]{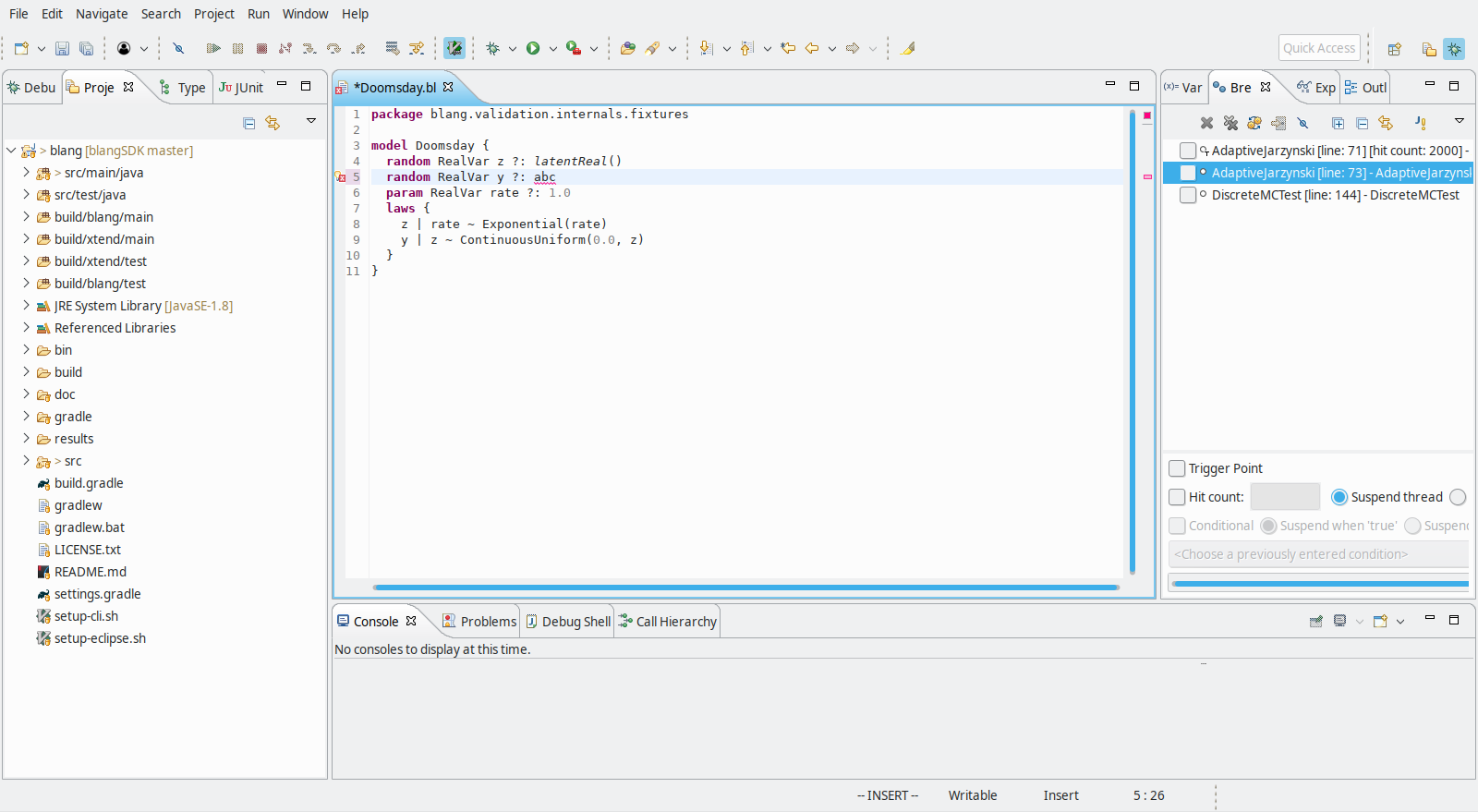}
  \caption{A preview of \proglang{Blang} IDE; Warning of syntactical error is underlined in red.}
  \label{fig:images/IDE}
\end{figure}

\updated{
To create or open a new project, follow these instructions:

\begin{enumerate}
	\item (Skip this step if you want to open an existing project) To create a template for a new project, if you have the \proglang{Blang} CLI installed, type \code{create-blang-gradle-project -{}-githubOrganization myOrg -{}-name myProject} where you should replace ``myOrg'' by the name of your organization, and ``myProject'' by the required name for the project. You can also find a template project at \url{https://github.com/UBC-Stat-ML/blangExample} (the method based on \code{create-blang-gradle-project} as it guarantees the library versions will be in sync with the version of \proglang{Blang} used by the CLI). 
	\item The next step is to generate configuration files suitable for Eclipse. This can be done by using the command \code{bash setup-eclipse.sh} which can be found at the root of freshly the generated project, or, if  \code{./gradlew assemble eclipse} if an older template project is used.   
	\item Starting from Eclipse's menus, select \code{File > Import > General > Existing project into Workspace}. Select the root of the project you created in the previous step. 
	\item The \proglang{Blang} project is ready. In the left tool bar in Eclipse, the project is in the file explorer. Right click on \code{src/main/java/[package name]/} and select the contextual menu \code{New > File}. Name the file \code{MyModel.bl}. The extension choice must always be \code{.bl}.
\end{enumerate}
}

The key IDE features useful for development include:

\begin{itemize}
  \item Ability to navigate a \proglang{Blang} code base and the \proglang{Blang} SDK by holding command while clicking on any symbol to jump to its definition, or hovering on it to see documentation. This and other related features are possible thanks to the static type system used by \proglang{Blang}.
  \item Incremental compilation in parallel in the background, which implies little time is spent waiting for compilation on modern multicore architectures. It also means that error messages appear interactively as the user types. See example in Figure~\ref{fig:images/IDE}.
  \item Quickly viewing the generated \proglang{Java} files, by right clicking anywhere in a \proglang{Blang} editor and selecting ``Open Generated File''.
  \item From any generated file, inference on the model can be launched by right-clicking ``Run As... Java Application". After doing this the first time, a shortcut is accessible via the menu ``Run > Run Configurations...'' Run Configurations allow setting the command-line arguments being passed to \proglang{Blang}.
  \updated{When using this method to launch a \proglang{Blang} execution, note that the argument \code{-{}-model [name of model]} should be skipped.}
  \item A full-feature debugger is built-in. Double clicking on the left margin of a \proglang{Blang} or \proglang{Xtend} file sets a break point. Use the menu ``Debug > Debug Configurations" to start the debugger.
  \item Being built on Eclipse, the IDE also inherits Eclipse's comprehensive set of features, such as utilities for unit testing, code coverage analysis, git integration, visualization of call and type hierarchies among others. 
\end{itemize}

More information on the \proglang{Blang} IDE is available from the \proglang{Blang} documentation page, \url{https://www.stat.ubc.ca/~bouchard/blang/Blang_IDE.html}.

\subsubsection{Web IDE}\label{sec:web-ide}

To facilitate deployment on large number of cores on the cloud, for example in a teaching or reproducible research context,  \proglang{Blang} is also available on the Web scientific platform Silico (\url{https://silico.io/}).

To setup a \proglang{Blang} project in Silico, create a \code{Model} from the user profile page, and create a file with \code{.bl} extension. Command-line arguments can be passed in by pasting them in a file called \code{configuration.txt}.

\subsection{Data types provided in the SDK} \label{subsec:types}

The interfaces \code{RealVar} and \code{IntVar} are automatically imported.\footnote{\api{dsl}{blang/core/IntVar}{blang.core.IntVar}, and \api{dsl}{blang/core/RealVar}{blang.core.RealVar}}
They can be either latent (unobserved, sampled), or fixed (conditioned upon). See Table~\ref{app:rvarfuncs} for commonly used functions to provide default initializations to these basic random variables.

\proglang{Blang}'s linear algebra is based on \pkg{xlinear} (for more information see Appendix~\ref{app:xlinear}) which is in turn based on a portfolio of established libraries.

The basic classes available are \code{Matrix}, \code{DenseMatrix}, and \code{SparseMatrix}. \proglang{Blang}/\proglang{XBase} allows operator overloading, so it is possible to write expressions of the type \code{matrix1 * matrix2}, \code{2.0 * matrix}, and so on.
Vectors do not have a distinct type, they are just $1 \times n$ or $n \times 1$ matrices.
Standard operations are supported using unsurprising syntax, e.g., \code{identity(100\_000)} (underscore delimited 100,000), \code{ones(3,3)}, \code{matrix.norm}, \code{matrix.sum}, \code{matrix.readOnlyView}, \code{matrix.slice(1, 3, 0, 2)}, \code{matrix.cholesky}, etc.\footnote{See \url{https://github.com/UBC-Stat-ML/xlinear} for more information on \pkg{xlinear}.}

\proglang{Blang} augments \pkg{xlinear} with two specialized types of matrices: \code{Simplex}, vector of positive numbers summing to one, and \code{TransitionMatrix}. Refer to Table~\ref{app:linalgfuncs} for key functions related to these specialized types of matrices.

\subsection{Distributions} \label{sec:distributions}

A range of distributions are included in the SDK. See Appendix~\ref{app:list-distributions} for the current list. 
These distributions are themselves written in \proglang{Blang}.
The SDK also contains tests covering all the included distributions. Our development workflow performs all the unit tests each time a commit is made in the \proglang{Blang} GitHub repository.

The implementation of the random number generators used in forward simulation of the SDK distributions are all grouped in the file \code{Generators}.\footnote{\api{sdk}{blang/distributions/Generators}{blang.distributions.Generators}}

\subsection{Input} \label{sec:input}

Inputs are parsed and managed via the \pkg{inits} package's injection framework.
Model variables can be provided a default initialization in the model's \code{.bl} file, or they can be initialized with arguments through the CLI.
Should both methods exist, the latter takes precedence; an example exposing only the pertinent snippets of code is shown below:

\longline

\code{Name.bl}\vspace*{-10pt}

\longline
\begin{lstlisting}
model Name {
  param IntVar h ?: 3
  param IntVar a 
  random Type1 p
  ...
}
\end{lstlisting}
\longline

The field \code{h} is initialized to $3$ by default, but can be overridden by the command-line argument to, say $5$, \code{-{}-model.h 5}.
The field \code{a} must be assigned a value, say 9, via the CLI through \code{-{}-model.a 9}.
For observations or custom data types such as \code{Type1}, annotations can be easily added to control parsing. \updated{The following is an example of a constructor that can parse command-line arguments such as \code{-{}-model.p.file abc.csv -{}-model.p.option 2}.}

\longline

\code{Type1.xtend}\vspace*{-10pt}

\longline
\begin{lstlisting}
import blang.inits.ConstructorArg;
import blang.inits.DesignatedConstructor;
import blang.inits.GlobalArg;
import blang.runtime.Observations;

class Type1
{
  ...
  @DesignatedConstructor
  def static Type1 loadObservedData( 
      @ConstructorArg(value = "file") File file,
      @ConstructorArg(value = "option") Integer x,
      @GlobalArg Observations observations)
  {
    val Type1 result = doSomethingWith(file, x) // Parse the file
    observations.markAsObserved(result)
    return result
  }
  ...
}
\end{lstlisting}
\longline

\updated{
In the code above, \code{doSomethingWith(file, x)} is any user-defined function that returns the parsed result as desired.
}

Additional information is provided at the relevant \proglang{Blang} documentation pages.\footnote{\url{https://www.stat.ubc.ca/~bouchard/blang/Javadoc.html}.}

\subsubsection{Plate notation} \label{subsec:plates}

Simple collections of random variables can be handled via \proglang{Java}'s built-in \code{List} and related data structures.
However, this can quickly become cumbersome and error-prone when working with sophisticated hierarchical Bayesian models.
To address this problem, \proglang{Blang} provides a specialized data structure based on the \emph{plate notation}, a method for representing repeated random variables in a graphical model.
A concise encoding of these models can be achieved with built-in types \code{Plate} and \code{Plated}.

Consider the following rocket launching data in Tidy format:

\begin{center}
\begin{tabular}{llll}
  \hline

  \textbf{Country} & \textbf{Rocket} & \textbf{nLaunches} & \textbf{nFails} \\
  \hline  \hline
  CHN & Chang Zheng 1 &$ 2 $& $0$ \\
  ESA & Ariane 44P &$ 15$ &$ 0$ \\
  RUS & Molniya 8K78M & $272$ &$ 13$ \\
  USA & Delta 2914 &$ 30$ & $2$ \\
  \vdots & \vdots & \vdots & \vdots \\
  \hline
\end{tabular}
\end{center}

\begin{figure}[h]
  \begin{minipage}{0.5\textwidth}
    \begin{align*}
    \alpha_c                            &\sim \text{Gamma}(1,1) \\
    \beta_c                            &\sim \text{Gamma}(1,1) \\
    p_{r,c} \mid \alpha_c, \beta_c          &\sim \text{Beta}(\alpha_c,\beta_c) \\
    f_{r,c} \mid p_{r,c}, l_{r,c}  &\sim \text{Binomial}(l_{r,c}, p_{r,c}) \\
    \end{align*}
  \end{minipage}
  \begin{minipage}{0.5\textwidth}
    \includegraphics[width=1.0\linewidth]{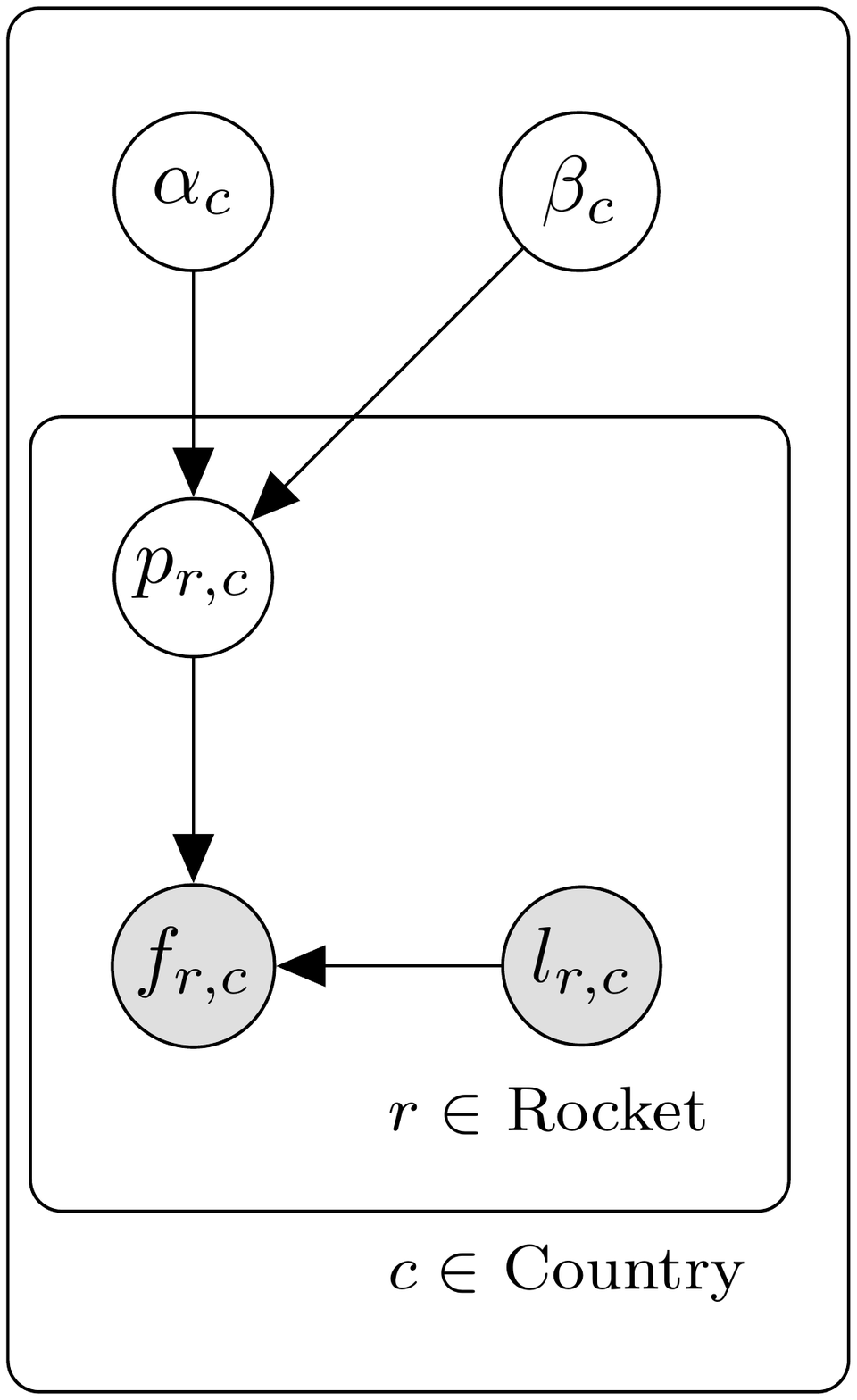}
  \end{minipage}
  \caption{
  Left: a toy hierarchical model for the rocket launching data set.
  Indices  $c, r$ index countries and rockets respectively.
  Observations $f_{r,c}, l_{r,c}$ are the number of failures and launches for rocket $r$ in country $c$ respectively.
  Latent variables $p_{r,c}, \alpha_c, \beta_c$ are parameters of interest.
  Right: A graphical representation of the hierarchical rocket model with plate notation.
  }
  \label{fig:graphicalModel}
\end{figure}

Where ``Country'' is the origin of the rocket, ``Rocket'' is the name of rockets, ``nLaunches'' is the number of launches, and ``nFails'' is the number of failed launches.
A toy model for this data set is the hierarchical model in Figure~\ref{fig:graphicalModel}.

Its corresponding \proglang{Blang} encoding is as follows:\footnote{Complete code and data can be found in the prepackaged repository of examples.}
\newpage
\longline

\code{Rocket.bl} \vspace*{-10pt}\footnote{\url{https://github.com/UBC-Stat-ML/JSSBlangCode/blob/master/reproduction_material/example/jss/hier/Rocket.bl}}

\longline

\begin{lstlisting}
package jss.hier

model Rocket {

  param GlobalDataSource data

  param Plate<String> countries
  param Plate<String> rockets

  random Plated<RealVar> alpha
  random Plated<RealVar> beta
  random Plated<RealVar> prob
  random Plated<IntVar> nFails
  param Plated<Integer> nLaunches

  laws {
    for(Index<String> c : countries.indices()){
      alpha.get(c) ~ Gamma(1,1)
      beta.get(c) ~ Gamma(1,1)

      for(Index<String> r : rockets.indices(c)){
        prob.get(r, c) | 
          RealVar a = alpha.get(c),
          RealVar b = beta.get(c) 
          ~ Beta(a, b)
        nFails.get(c, r) | 
          RealVar p = prob.get(r, c),
          Integer n = nLaunches.get(r, c) 
          ~ Binomial(n, p)
      }
    }
  }
}
\end{lstlisting}

\longline

A first observation is the additional \code{param GlobalDataSource data}, which we have not seen in our previous models.
We will discuss its function in more details shortly. 
At a high level, it is used to specify a CSV file from which many variables will be parsed.

A second observation is the use of \code{Plate<...>} and \code{Plated<...>} types.

\longline

\begin{lstlisting}
param Plate<String> countries
param Plate<String> rockets

random Plated<RealVar> alpha
random Plated<RealVar> beta
random Plated<RealVar> prob
random Plated<IntVar> nFails
param Plated<IntVar> nLaunches
\end{lstlisting}

\longline

A \code{Plate} is a collection of indices, such as Country and Rocket indices (columns one and two of our example data set above).
As they are non-random known indices, we declare plates with \code{param}.
On the other hand \code{Plated} types, as its name suggests, are variables within plates.
The usual rules for selection between \code{param} or \code{random} apply to plated variables (see Section~\ref{sec:model-overview}).

With this setup, we are ready to examine the laws block.

\longline

\begin{lstlisting}
for(Index<String> c : countries.indices()){
  alpha.get(c) ~ Gamma(1,1)
  beta.get(c) ~ Gamma(1,1)

  for(Index<String> r : rockets.indices(c)){
    prob.get(r, c) | 
      RealVar a = alpha.get(c),
      RealVar b = beta.get(c) 
      ~ Beta(a, b)
    nFails.get(c, r) | 
      RealVar p = prob.get(r, c),
      Integer n = nLaunches.get(r, c) 
      ~ Binomial(n, p)
  }
}
\end{lstlisting}

\longline

This should look rather similar to code we have presented thus far.
We highlight the key differences:
first, the set of index values is obtained by appending \code{.indices} to a \code{Plate} variable.
Each index is of type \code{Index<T>}, where \code{T} is the same type as the corresponding \code{Plate<T>}.
Plated variables can subsequently be retrieved by using \code{.get()}.
Second, notice the syntax of the second \code{for} loop over rocket indices, in particular \code{rockets.indices(c)}.
This syntax retrieves the set of rocket indices such that its country index is \code{c}.
Lastly, we note the ordering of indices within \code{.get()} is exchangeable, for example, \code{nFails.get(c, r)} is equivalent to \code{nFails.get(r, c)}. This is possible since \code{Index<...>} objects keep track of which plate they belong to.

Additional methods available for types \code{Index<>} are described in Figure~\ref{app:platefuncs} of Appendix~\ref{app:functions}.

With variables, parameters, and laws declared, we tie these concepts back to the promised discussion of \code{param GlobalDataSource data}.
Its purpose becomes clear when we invoke \code{blang} and its corresponding arguments:

\longline
\begin{CodeInput}
> git clone https://github.com/UBC-Stat-ML/JSSBlangCode.git
> cd JSSBlangCode/reproduction_material/example/
> blang --model jss.hier.Rocket \
    --model.data data/rockets.csv \
    --model.countries.name Country \
    --model.rockets.name Rocket \
\end{CodeInput}
\longline

Variable of type \code{Plate} and \code{Plated} can be put in correspondence with a column in a Tidy CSV file.
Command-line arguments can be used to set the CSV file for \emph{each} variable individually. 
\emph{Alternatively} by declaring a dummy variable of type \code{GlobalDataSource}, here called \code{data}, we can set a default CSV file that will be used by default by all  \code{Plate} and \code{Plated} variables.
In our example, specifying the default CSV is achieved via \code{-{}-model.data pathToData/data.csv}. 
For each \code{Plate} and \code{Plated} variable, the data input algorithm will attempt to  find a column in the CSV file matching with the variable name.  \updated{The algorithm does not require all columns in the CSV file to be matched to \code{Plate} or \code{Plated} variables.}

By default, matching is done by using the same string for the column header as the variable name, but this can be overridden via CLI arguments. 
In our example, this is achieved via e.g., \code{-{}-model.rockets.name Rocket}. Notice we did not require this argument for \code{nFails}, as the column name in the CSV file is also \code{nFails}. 

When a plated variable is not found in the CSV file, it is assumed to be latent \updated{(a message is displayed to standard out when this happens)}. 
Should a plate not correspond to a column in the CSV, then its \code{maxSize} should be set via for e.g., \code{-{}-model.varName.maxSize 3}, or initialized in the model. 
An example using the CLI is presented in the advanced tutorial in Appendix~\ref{subsec:tutorial},\footnote{Use the argument \code{-{}-model Rocket -{}-help} for full documentation.} and an example using default initializations is shown below.

Recall the Gaussian mixture model from Section~\ref{sec:gmm}.
We can implement the same model with plate syntax:

\code{MixtureModelPlated.bl}\vspace*{-10pt}\footnote{\url{https://github.com/UBC-Stat-ML/JSSBlangCode/blob/master/reproduction_material/example/jss/gmm/MixtureModelPlated.bl}}

\longline

\begin{lstlisting}
package jss.gmm

model MixtureModelPlated {
    
  param GlobalDataSource data

  param  Integer        K  ?: 2
  param  Plate<Integer> N
  param  Plate<Integer> components ?: Plate.ofIntegers("components", K)

  random Plated<IntVar>  z 
  random Plated<RealVar> y
  random Plated<RealVar> mu 
  random Plated<RealVar> sd 
  random Simplex         pi ?: latentSimplex(K)

  laws {
    pi | K ~ SymmetricDirichlet(K, 1.0)

    for (Index<Integer> k : components.indices) {
      mu.get(k) ~ Normal(0.0, 100.0)
      sd.get(k) ~ ContinuousUniform(0.0 ,10.0)
    }

    for (Index<Integer> i : N.indices) {
      z.get(i) | pi ~ Categorical(pi)
      y.get(i) | List<RealVar> muList = mu.asList(components),
                 List<RealVar> sdList = sd.asList(components),
                 IntVar k = z.get(i)
        ~ Normal(muList.get(k) , pow(sdList.get(k), 2.0))
    }
  }
}
\end{lstlisting}

\longline

A first observation is the use of \code{Plate.ofIntegers()} to initialize a plate with a predetermined size.
The function \code{ofIntegers()} takes in two arguments: a column name, and a maximum size.
For other related functions, see Appendix~\ref{app:functions}.

A second observation is the \code{asList()} function, which returns the given plate (\code{components}) as a list.

\longline

\begin{lstlisting}
y.get(i) | List<RealVar> muList = mu.asList(components),
           List<RealVar> sdList = sd.asList(components),
           IntVar k = z.get(i)
  ~ Normal(muList.get(k) , pow(sdList.get(k), 2.0))
\end{lstlisting}

\longline

Similar conversion utilities automatically imported from \code{ExtensionUtils} are documented in Figure~\ref{app:extenfuncs} of Appendix~\ref{app:functions}.\footnote{\api{sdk}{blang/types/ExtensionUtils}{blang.types.ExtensionUtils}.}
With this setup, our command line arguments for inference can be written succinctly:

\longline
\begin{CodeInput}
> git clone https://github.com/UBC-Stat-ML/JSSBlangCode.git
> cd JSSBlangCode/reproduction_material/example/
> blang --model jss.gmm.MixtureModelPlated \
    --model.data data/obs1Plated.csv 
\end{CodeInput}
\longline

\subsubsection{PlatedMatrix}
One special case of a \code{Plated} variable is the type \code{PlatedMatrix},\footnote{\api{sdk}{blang/types/PlatedMatrix}{blang.blang.types.PlatedMatrix}}
which is a built-in type that facilitates easy representation of higher dimensional random variables such as random vectors or matrices, as well as lists and arrays of vectors and matrices.

\code{PlatedMatrix} can be used to represent both random vectors and matrices that are enclosed within a \code{Plate}.
\code{PlatedMatrix} generally works in the same way as \code{Plated}, but provide specialized mechanisms to access vectors, matrices, simplices, etc. For example, to access a dense vector, use the method \code{myPlatedMatrix.getDenseVector(myRowPlate, myParentIndex1, myParentIndex2, ...)}. Here \code{myRowPlate} refers to the plate from which row indices will be constructed. The indices can be either of the form $0, 1, 2, \dots$, or, if they are of other types, say strings or non-consecutive integers, in which case a fixed correspondence with $0, 1, 2, \dots$ is maintained internally.

The other arguments, \code{myParentIndex1, myParentIndex2, ...} consist in the indices for the plates in which this vector belongs to. For example, a set of vectors can be obtained as follows:

\longline

\code{PlatedMatrixExample.bl}\vspace*{-10pt}

\longline
\begin{lstlisting}
model PlatedMatrixExample {
  param Plate<String> dims
  param Plate<String> replicates 
  random PlatedMatrix vectors
  laws {
    for (Index<String> n : replicates.indices) {
      vectors.getDenseVector(dims, n) | 
        int size = dims.indices.size 
      ~ MultivariateNormal(dense(size), identity(size).cholesky)
    }
  }
}
\end{lstlisting}
\longline

\subsection{Testing framework} \label{subsec:testing}
\updated{
There is considerable emphasis in the Markov chain Monte Carlo (MCMC) literature on efficiency, but much less on correctness, in the sense of the implementation being ergodic with respect to the distribution of interest.
Consider a Monte Carlo procedure producing samples $X_1, X_2, \dots$, targeting some distribution $\pi$.
We say the procedure is correct if its ergodic averages for any integrable function $f$ admits a law of large numbers converging to the posterior expectation of $f$ under the target $\pi$,
\begin{align}
  \frac{1}{N}\sum_{i=1}^Nf(X_i) \to \int f(x)\pi(x)\mathrm{d}x,
\end{align}
almost surely.

Two common defects of an incorrect Monte Carlo procedure include erroneous mathematical derivations of algorithms and software implementation bugs.
This section discusses the tools available to test and detect both types of problems.
}

\subsubsection{Exhaustive random objects}

\updated{
We provide in \code{bayonet.distributions.ExhaustiveDebugRandom} a non-standard replacement implementation of \code{bayonet.distributions.Random} which enumerates all the possible realizations of an arbitrary finite random process along with the probability of each realization. 
}

\subsubsection{Testing unbiasedness}

We use \code{ExhaustiveRandom} to test the unbiasedness of the normalization constant estimate provided by our sequential Monte Carlo (SMC) implementation.
The code forming the basis of this test is shown below:

\longline

\code{UnbiasednessTest.xtend}\vspace*{-10pt}

\longline
\begin{lstlisting}
package blang.validation

import java.util.function.Supplier
import bayonet.distributions.ExhaustiveDebugRandom

class UnbiasednessTest {
  def static double expectedZEstimate(Supplier<Double> logZEstimator, 
                                      ExhaustiveDebugRandom exhaustiveRand) {
    var expectation = 0.0
    var nProgramTraces = 0
    while (exhaustiveRand.hasNext) {
      val logZ = logZEstimator.get
      expectation += Math.exp(logZ) * exhaustiveRand.lastProbability
      nProgramTraces++
    }
    println("nProgramTraces = " + nProgramTraces)
    return expectation
  }
}
\end{lstlisting}
\longline

\updated{
  The above code defines a function, \code{expectedZEstimate} which 
  takes as input an estimator \code{logZEstimator} and an 
  \code{ExhaustiveRandom} object. The estimator is assumed to 
  use internally the \code{ExhaustiveRandom} object to provide a randomized estimate. Assuming that the estimator is defined on 
  a finite probability space, the above code can therefore compute 
  the exact value of the expectation of \code{logZEstimator}.
  
  Note that SMC executed on a small finite model, e.g., a short hidden Markov model, has a finite number of possible execution 
  traces (defined as all possible intermediate particle, i.e., 
  possible proposal and resampling vectors).  
  
  The above code is called in our continuous integration 
  test suite to verify unbiasedness of our SMC implementation on a model small enough for the number of possible execution traces to be manageable while achieving code coverage.\footnote{\url{https://github.com/UBC-Stat-ML/blangSDK/blob/master/src/test/java/blang/TestSMCUnbiasness.xtend}}
}

The output of a test based on the unbiasedness test has the form:

\longline
\begin{CodeChunk}
  \begin{CodeOutput}
nProgramTraces = 23868
true normalization constant Z: 0.345
expected Z estimate over all traces: 0.34500000000000164
  \end{CodeOutput}
\end{CodeChunk}
\longline

\updated{
  where \code{true normalization constant Z} is computed by explicitly enumerating all states in the space of the posterior (SMC code is not needed since the state space is finite), and \code{expected Z estimate over all traces} is computed by enumerating all SMC execution traces along with their probabilities.
}

\subsubsection{Tests based on linear algebra}

\updated{
  We can leverage the exhaustive random object to assert the invariance and irreducibility of a transition kernel, in a similar flavour to the test for unbiasedness.
  As such, the aim of this short section is to highlight the key ideas of such tests, with details deferred to reference code.

The \code{DiscreteMCTest} contains algorithms that use \code{ExhaustiveDebugRandom} to check via linear algebra whether Markov kernels on small discrete (finite state space) models are invariant and irreducible.\footnote{\api{sdk}{blang/validation/DiscreteMCTest}{blang.validation.DiscreteMCTest}}
}

Algorithmically, \code{DiscreteMCTest} takes a model and a kernel, and constructs the corresponding sparse transition matrix.
From this matrix it is then trivial to check, numerically, irreducibility and invariance via linear algebra and graph algorithms.
See \code{TestDiscreteModels} for an example.\footnote{\url{https://github.com/UBC-Stat-ML/blangSDK/blob/master/src/test/java/blang/TestDiscreteModels.xtend}}

\subsubsection{Exact invariance test}

\updated{
  Tests discussed thus far focused around the idea of exhaustively enumerating outcomes and probabilities.
  Although these tests are attractive due to their deterministic property, they are only applicable to a small set of models, namely models with finite state spaces.
  These tests are not applicable to models with continuous state spaces.

For continuous models, we provide a modified form of the Geweke test \citep{Geweke2004}, which we call the exact invariance test.
Consider the goal of testing the invariance of a kernel $T$ with respect to some target distribution $\pi(\theta \mid y) \propto p_\theta(\theta)p_{y \mid \theta}(y \mid \theta)$.
Assume $T$ is a combination of individual kernels $T_i$ for $i=1,2,\dots,Q$.
Briefly, the Geweke test examines the correctness of an MCMC procedure by comparing two sets of simulated random variables, $F=\{F_1, F_2, \dots, F_{M_1}\}$ and $G=\{G_1, G_2, \dots, G_{M_2}\}$ using an approximate test based on an asymptotic result.

The set $F$ is generated by the marginal-conditional simulator defined by iterating the three steps:
\begin{enumerate}
  \item $\theta_{m} \sim p_\theta(\cdot)$,
  \item $y_{m} \mid \theta_m \sim p_{y\mid \theta}(\cdot\mid \theta_m)$,
  \item $F_m = f(\theta_m, y_m)$,
\end{enumerate}
for $m=1,2,\dots,M_1$ and some integrable, real-valued test function $f$.
Similarly, $G$ is generated by the successive-conditional simulator defined by the following steps:
\begin{enumerate}
  \item $\theta_1 \sim p_\theta(\cdot)$,
  \item $y_1 \mid \theta_1 \sim p_{y\mid \theta}(\cdot \mid \theta_1)$,
  \item $G_1 = f(\theta_1, y_1)$,
  \item iterate for $m \geq 2$:
        \begin{enumerate}
          \item $\theta_m \mid \theta_{m-1}, y_{m-1} \sim T(\cdot \mid \theta_{m-1}, y_{m-1})$
          \item $y_m \mid \theta_m \sim p_{y \mid \theta}(\cdot \mid \theta_m)$
          \item $G_m = f(\theta_m, y_m)$.
        \end{enumerate}
\end{enumerate}
The Geweke test is based on the observation that both $F$ and $G$ can be used to approximate the expectation of $f$ under the joint distribution of $p_\theta$ and $p_{y\mid \theta}$.
However, there are several limitations to the original Geweke approach:
\begin{enumerate}
  \item The validity of the approximate test relies on $T$ being irreducible.
        As a consequence, individual kernels $T_i$'s often cannot be tested in isolation.
  \item The test is an approximate test relying on asymptotics.
        It is difficult to verify the accuracy of this asymptotic result in practice.
        Furthermore, the problem is compounded when several such tests need to be combined using a multiple-testing framework.
  \item The validity of the approximate test also relies on a central limit theorem for Markov chains to hold, which typically involves establishing geometric ergodicity.
  The task of proving geometric ergodicity is model-dependent and rather involved.
\end{enumerate}

To address the problems above, \proglang{Blang} employs a modified version of the Geweke test, the exact invariance test (EIT).
The EIT does not rely on irreducibility of $T_i$'s, thus allowing individual tests.
Furthermore, it does not rely on establishing geometric ergodicity, and as its name suggests, it is an exact test independent of asymptotics.

Similar to the Geweke test, it compares two sets of samples $F$ and $H=\{H_1, H_2, \dots, H_{M_3}\}$.
The samples $H$ are generated from the exact invariant simulator defined by the steps:
\begin{enumerate}
  \item $\theta_{1,m} \sim p_\theta(\cdot)$,
  \item $y_{1, m} \mid \theta_{1,m} \sim p_{y\mid\theta}(\cdot\mid\theta_{1,m})$
  \item For $k=2,3,\dots,K$ 
        \begin{enumerate}
          \item $\theta_{k, m} \mid \theta_{k-1, m}, y_{k-1, m} \sim T_i (\cdot \mid \theta_{k-1, m}, y_{k-1,m})$ 
        \end{enumerate}
  \item $H_m = f(\theta_{K, m}, y_{K, m})$.
\end{enumerate}

By construction, for any $K\geq 1$, $j \in \{1, 2, \dots, M_1 \}, l \in \{1, 2, \dots, M_3\}$, $F_j$ and $H_l$ are equal in distribution if and only if $T_i$ is $\pi$-invariant.
Thus the appropriate exact tests (e.g., Fisher's exact test), or well-understood asymptotic tests (e.g., Kolmogorov-Smirnov) may be employed.
Note $H_m$'s are also independent, thus the asymptotics do not rely on irreducibility nor geometric ergodicity; standard IID tests can be employed.

An example of how EIT is used in \proglang{Blang} to automatically test all distributions in the SDK can be found in \url{https://github.com/UBC-Stat-ML/blangSDK/blob/master/src/test/java/blang/TestSDKDistributions.xtend}.
}
A complete example of an EIT for our permutation model (Section~\ref{sec:customSamplers}) is provided in the reproduction materials.

\subsection{Package distribution and injection} \label{sec:injectingDepedencies}

Distributing and reusing packages is standard practice in software development.
Any user can create a model and publish it in a versioned fashion via GitHub.\footnote{Under the hood, the mechanism for dependency management is \pkg{Maven}. However, GitHub repositories are seamlessly imported via \pkg{JitPack}. See \url{https://jitpack.io/} for details.} 

To use a package developed by another user, 
\proglang{Blang} projects compiled via the CLI automatically handle dependencies hosted on GitHub by parsing a file called \code{dependencies.txt} placed in the project root directory.
For correct parsing, GitHub dependencies' format must be of the forms:

\longline

\code{dependencies.txt}\vspace*{-10pt}

\longline
\begin{CodeInput}
com.github.Username:Repository:Branch-CommitHash
com.github.Username:Repository:ReleaseTag
\end{CodeInput}
\longline

where \code{CommitHash} may be replaced by \code{SNAPSHOT} to automatically select latest commits.
For compilation through Eclipse IDE, users should manually input dependencies in the \code{build.gradle} file.

To distribute packages, users can create a \proglang{Blang} project with the command-line interface \code{create-blang-gradle-project}, and publish it in a GitHub repository.

\section{Design patterns}\label{sec:design}
This section discusses design patterns specific to programming in \proglang{Blang}.
The goal of these design patterns is to enable users to design models going beyond Bayes nets, improve computational efficiency, and improve code readability.

\subsection{Undirected graphical models}\label{sec:undirected}

The mechanisms in \proglang{Blang}'s default inference engine require the models to be in generative normal form. 
In some cases, in particular for users interested in undirected graphical models or Markov random fields (MRF), this may appear a stringent condition, since forward simulation in these models is computationally intractable.

We illustrate here a construction based on a type of ``pseudo-prior''.
Let $f_\theta(x) \propto \prod_{i\in I} \psi_\theta(x)$ denote an MRF, where $I$ denotes a set of cliques that factorizes the MRF.
We rewrite the model as $f_\theta(x) \propto f_0(x) \prod_{i\in I} \tilde \psi_\theta(x)$, where $f_0(x)$ is a ``tractable'' pseudo-prior.
By tractable, we mean that we can sample and compute the normalization constant of the pseudo-prior.
Annealing is then automatically performed on the factors $\prod_{i\in I} \tilde \psi_\theta(x)$ only, not on the pseudo-prior, ensuring finite marginalization for all interpolating distributions.

For example, consider the Ising model \citep{Ising1925} which is a type of MRF. 
In this case we use a product of independent Bernoulli random variables as a pseudo-prior.
Note, we will make use of an ``empty pipe symbol'', i.e., ``\code{| IntVar first = ...}'' which is explained after the example:

\longline

\code{Ising.bl}\vspace*{-10pt}\footnote{\url{https://github.com/UBC-Stat-ML/JSSBlangCode/blob/master/reproduction_material/example/jss/others/IsingExample.bl}}

\longline
\begin{lstlisting}
model Ising {
  param Double moment ?: 0.0
  param Double beta ?: log(1 + sqrt(2.0)) / 2.0 
  param Integer N ?: 5
  random List<IntVar> vertices ?: latentIntList(N*N)
  
  laws {
    for (UnorderedPair<Integer, Integer> pair : squareIsingEdges(N)) {
      | IntVar first  = vertices.get(pair.getFirst), 
        IntVar second = vertices.get(pair.getSecond),
        beta
      ~ LogPotential({
          if ((first < 0 || first > 1 || second < 0 || second > 1))
            return NEGATIVE_INFINITY
          else
            return beta*(2*first-1)*(2*second-1))
        })
    }
    for (IntVar vertex : vertices) {
      vertex | moment ~ Bernoulli(logistic(-2.0*moment))
    }
  }
}
\end{lstlisting}
\longline

Rather than using \code{logf} here for the likelihood, which would have violated the technical conditions for generative normal forms, we used the \code{LogPotential} utility in the SDK (shown below for reference).
Since \code{LogPotential} does not define random variables, when it is invoked there are no variables to the left of the conditioning symbol in \code{| IntVar first = ...}.
This follows naturally from the formal definition of composite laws. 
A second observation worthy of note is the conditioning of \code{| IntVar first = vertices.get(pair.getFirst)} as opposed to \code{| vertices}.
This prevents the runtime architecture from assuming that these factors depend on the full \code{vertices} object, hence improving computational efficiency by a scaling proportional to the size of \code{vertices}.
We emphasize this computational advantage in Section~\ref{sec:markovChainExample} with a Markov chain example.

\longline

\code{LogPotential.bl}\vspace*{-10pt}\footnote{\url{https://github.com/UBC-Stat-ML/JSSBlangCode/blob/master/reproduction_material/example/jss/others/LogPotentialExample.bl}}

\longline
\begin{lstlisting}
model LogPotential {
  param RealVar logPotential
  laws {
    logf(logPotential) {
      return logPotential
    }
  }
\end{lstlisting}
\longline

\subsection{Delayed graphical model construction} \label{sec:markovChainExample}

The runtime engine is able to decrease computational expense when it can detect sparsity patterns in models.
This is handled automatically for simple objects but requires user input for complex objects.
For an example with a complex object \code{chain} consider the following Markov Chain:

\longline

\code{MarkovChain.bl}\vspace*{-10pt}\footnote{\url{https://github.com/UBC-Stat-ML/JSSBlangCode/blob/master/reproduction_material/example/jss/others/MarkovChainExample.bl}}

\longline
\begin{lstlisting}
model MarkovChain {

  param Simplex initialDistribution
  param TransitionMatrix transitionProbabilities
  random List<IntVar> chain

  laws {
    chain.get(0) | initialDistribution ~ Categorical(initialDistribution)

    for (int step : 1 ..< chain.size) {
      chain.get(step) | IntVar previous = chain.get(step - 1), 
                        transitionProbabilities
        ~ Categorical({
            if (previous >= 0 && previous < transitionProbabilities.nRows)
              transitionProbabilities.row(previous)
            else
              invalidParameter
          })
    }
  }
} 
\end{lstlisting}
\longline

We condition on the previous step instead of the whole chain, using \code{chain.get(step) | IntVar previous = chain.get(step - 1)} as opposed to \code{chain.get(step) | chain}.
This prevents the runtime architecture from computing factors involving the full \code{chain} object, potentially improving computational efficiency by a scaling proportional to the chain size.
In other words, here the exact specification of the graphical model is delayed until the data is available. 

In general, it is optimal to condition on the smallest possible scope.
For example, suppose we have \code{SomeObject x} with conditional distribution on \code{ConditionalObject y}, where \code{y} has two \code{IntVar} fields \code{a} and \code{b}.
If the distribution on \code{x} only requires the first field of \code{y}, \code{a}, then we should condition only on \code{a}. 
Hence, we use \code{x | IntVar v = y.a ~ Distribution(v)} as opposed to \code{x | y ~ Distribution(y.a)}.
For a detailed understanding of this efficiency gain, we refer readers to Section~\ref{sec:sparsity}.

\subsection{Model reparameterization} \label{subsec:modelreparam}

It is often the case that distribution families can be written using different parameterizations, or that a family can be expressed as a special case of another family.
Following ``Don't Repeat Yourself'' (DRY) coding principles, the following pattern shows what is the best practice to express such reparameterizations. 

To illustrate the pattern, consider how the Exponential distribution is coded in the \proglang{Blang} SDK as a special case of the Gamma distribution:

\longline

\code{Exponential.bl}\vspace*{-10pt}\footnote{\url{https://github.com/UBC-Stat-ML/JSSBlangCode/blob/master/reproduction_material/example/jss/others/ExponentialExample.bl}}

\longline
\begin{lstlisting}
model Exponential {
  random RealVar realization
  param  RealVar rate
  laws {
    realization | rate ~ Gamma(1.0, rate)
  }
}
\end{lstlisting}
\longline

\subsection{Distributions as parameters}\label{sec:distributions-as-parameters}

In many situations, it is useful to have one or several parameters of a model to be themselves distributions. Consider for example a mixture model: it takes as input a list of distributions as well as mixture proportions, and creates a new distribution from it. Here is an example of how this is implemented for mixtures of integer-valued distributions in \proglang{Blang}:

\longline

\code{IntMixture.bl}\vspace*{-10pt}\footnote{\url{https://github.com/UBC-Stat-ML/JSSBlangCode/blob/master/reproduction_material/example/jss/others/IntMixtureExample.bl}}

\longline
\begin{lstlisting}
model IntMixture {
  param Simplex proportions
  param List<IntDistribution> components
  random IntVar realization
  
  laws {
    logf(proportions, components, realization) {
      var sum = 0.0
      if (components.size !== proportions.nEntries) {
        throw new RuntimeException
      }
      for (i : 0 ..< components.size) {
        val prop = proportions.get(i)
        if (prop < 0.0 || prop > 1.0) return NEGATIVE_INFINITY
        sum += prop * exp(components.get(i).logDensity(realization))
      }
      return log(sum)
    }
  }
  
  generate (rand) {
    val category = rand.categorical(proportions.vectorToArray)
    return components.get(category).sample(rand)
  }
}
\end{lstlisting}
\longline

And this is invoked from another model as follows, and in this case, to create a mixture of two Poisson distributions:

\longline

\code{PoissonPoissonMixtureExample.bl}\vspace*{-10pt}\footnote{\url{https://github.com/UBC-Stat-ML/JSSBlangCode/blob/master/reproduction_material/example/jss/others/PoissonPoissonMixtureExample.bl}}

\longline
\begin{lstlisting}
x | lambda1, lambda2, pi 
  ~ IntMixture(
    pi, 
    #[Poisson::distribution(lambda1), Poisson::distribution(lambda2)]
  )
\end{lstlisting}
\longline

Here \code{Poisson::distribution(...)} is a convenient shortcut generated automatically: any model with only one random variable is automatically endowed with a \code{distribution(...)} function taking as input the model's parameters. 
The \code{distribution(...)} function returns a simplified application programming interface (API) for models having only one random variable. 
If that single random variable is of type \code{RealVar} (respectively, \code{IntVar}), the returned value of \code{distribution(...)} is of type   \code{RealDistribution}\footnote{\api{dsl}{blang/core/RealDistribution}{blang.core.RealDistribution}}
(respectively, \code{IntDistribution}\footnote{\api{dsl}{blang/core/IntDistribution}{blang.core.IntDistribution}}).
If the type of the single random variable is neither \code{RealVar} nor \code{IntVar}, the returned value of \code{distribution(...)} is of the type \code{Distribution}.\footnote{\api{dsl}{blang/core/Distribution}{blang.core.Distribution}}

\section{Inference} \label{sec:inference}

\proglang{Blang} efficiently samples from posterior distributions by detecting sparsity patterns in the model, matching variable types with their associated roles in inference, then sample using state-of-the-art Monte Carlo methods.

In the following sections, we detail intermediate steps in the process described above. 
We first assume that a continuum of probability distributions is available.
On one end of the spectrum, we have the posterior distribution, and the prior on the other.
The prior is a distribution from which we can sample from assuming the model is in generative normal form.
Then we describe the technical details used to automatically construct this continuum of interpolating probability distributions, along with invariant Markov chain kernels for each distribution in the interpolation.

\subsection{Inference algorithms} \label{sec:inferenceEngines}

\proglang{Blang} currently focuses on two complementary inference algorithms: sequential change of measure (SCM), and non-reversible parallel tempering (PT).
SCM infers the exact posterior distribution asymptotically in memory, while PT infers the exact posterior distribution asymptotically in time.
The former is an SMC algorithm and the latter a parallel MCMC algorithm.

A core concept present in both algorithms is the use of an adaptive sequence of tempered distributions extracted from a continuum interpolating from the prior to the posterior distribution. 
Through these tempering schemes, we are able to explore complex, multimodal distributions without the need for automatic differentiation; as such, these techniques are not limited to Euclidean spaces.
For example, the default sampler for real and integer data types are their respective slice samplers \citep{Neal2003},\footnote{More precisely, a doubling and shrinking procedure is used as an adaptive scheme, whose details and validity are described and proved by \cite{Neal2003}.} which when used in a naive MCMC algorithm could perform poorly in highly correlated models.
However in the context of SCM or PT, it is frequently the case that simple MCMC algorithms perform better than using specialized moves in a single chain \citep{ballnus_comprehensive_2017}.

Furthermore, due to the inherent characteristics of these algorithms, they are trivially parallelized for efficient computing, and provide computation of model evidence at negligible cost.
For these reasons, SCM and PT are good candidates for automatic inference on generalized state spaces.
These two algorithms can be used individually, but by default the SCM is used to initialize PT.
This combination is motivated by the fact that SCM appears to often be better suited to quickly find a crude approximation.
In particular SCM is able to find configurations of positive probability even in the presence of deterministic constraints (i.e., configurations having zero posterior probability).
However, to obtain high quality samples, SCM may require a number of particles larger than what can be fitted in memory.
PT on the other hand can provide approximations of arbitrary high quality without asymptotically infinite memory consumption.

\subsubsection{Constructing a sequence of measures} \label{sec:sequenceOfMeasures}

Both SCM and PT inference algorithms require a continuum of measures.
To retain theoretical guarantees, we must ensure each measure in this sequence has a finite normalization constant.
To achieve this, we factorize our joint density into what we call likelihood $l_i(x)$ and prior $p_j(x)$ factors.
Assuming a \proglang{Blang} model in generative normal form, the construction of such a continuum of probability measures begins with an exhaustive unrolling of composite laws to identify all atomic laws, or log factors. 
Each factor belongs to a model and as such each of its dependencies can be classified as either corresponding to \code{random} or \code{param}.
If its dependency is \code{random}, we direct the corresponding edge in the factor graph as out-going.
Otherwise, if it is a \code{param}, we direct the edge as in-coming.
Likelihood factors are then defined as factors whose outgoing edges, if any, all connect to an observed variable; factors are classified as priors otherwise.

Suppose we have factorized our posterior as follows
$$\pi(x) \propto \prod_{i=1}^I l_i(x) \prod_{j=1}^J p_j(x)$$
where $l_i(x)$, $p_j(x)$ denote likelihood and prior factors respectively.
As opposed to raising the product of likelihood and prior factors to some 
$t \in [0,1]$, which may not yield a probability distribution, 
it is preferable to exponentiate the likelihood factors.

Additionally, it is common to have configurations of zero probability when performing inference over discrete combinatorial objects.
In some scenarios, for example in pedigree analysis, these zero-valued likelihood evaluations can create difficulties in building irreducible samplers, thus invalidating convergence guarantees.
We alleviate this restriction using the annealing scheme shown below,
\begin{align}\label{eq:sequence}
	\pi_{t}(x) = \frac{\gamma_t(x)}{Z_t} = \frac{\Big(\prod_{i \in I}[(l_i(x))^t + \mathbb{I}(l_{i}(x) = 0) \epsilon_{t}]\Big)p(x) }{Z_t} 
\end{align}
where $\ \epsilon_t = \text{exp}(-10^{100}t)\mathbb{I}(t < 1) \ $, $\ p(x) = \prod_{j=1}^J p_j(x)$ and we use the convention $0^0 = 0$ so that $\pi_0(x) = p(x)$.
The conditions and effects of $\epsilon_t$ on the performance of algorithms have yet to be explored and is part of our future work.
By design, the interpolating chains have a wide support (i.e., $p(x) >0 \Rightarrow \pi_t(x) > 0$ for $t < 1$), while maintaining the guarantee of having a finite normalization constant for all annealing parameters:
\begin{align}
	\begin{split}
		\int \pi_{t}(x) dx ={}& \int p(x) \prod_{i \in I}[(l_i(x))^t + \mathbb{I}(l_{i}(x) = 0) \epsilon_{t}]dx\\
		& \leq \sum_{K:K\subset I} \epsilon_{t}^{|I| - |K|} \int p(x)(\prod_{i \in K}l_i (x))^t dx\\
		& = \sum_{K:K\subset I}\epsilon_{t}^{|I| - |K|} \int p(x)(\prod_{i \in K}l_i (x))^t[I(\prod_{i\in K} l_i (x) \geq 1) + I(\prod_{i \in K} l_i (x) < 1)]dx\\
		& \leq \sum_{K:K\subset I} \epsilon_{t}^{|I| - |K|} [\int p(x) \prod_{i \in K} l_i (x)dx + \int p(x)dx]\\
		& < \infty 
	\end{split}
\end{align}

This proposed annealing scheme allows our sampler to traverse across multimodal distributions, preserve the correct marginal posterior distribution at room temperature, and guarantee convergence of normalizing constant estimates.

In a given execution of the PT and SCM inference algorithms, the full continuum of distributions $\{ \pi_t : t \in [0, 1]\}$ is only instantiated on a finite grid $0 = t_0 < t_1 < \dots < t_N = 1$, called an \emph{annealing schedule}.
Since the performance of both PT and  SCM are sensitive to the choice of annealing schedules, they each use a specialized algorithm to automatically optimize the  annealing schedule (described in the next sections).
To perform a continuous optimization over $(t_1, t_2, \dots, t_{N-1})$ with monotonicity constraints, the algorithms rely on the fact that the discrete sequence of distributions is embedded in a continuum of distributions.

\subsubsection{Sequential change of measure} \label{sec:scm}

Informally, \proglang{Blang}'s SCM inference engine initializes a population of particles from a prior distribution, and iteratively perturbs and reweighs the particles.  
The number of particles used is $1\,000$ by default and can be set using the \texttt{-{}-engine.nParticles} option. 
More precisely, SCM is a special case of the sequential Monte Carlo (SMC) sampler \citep[Section 3.3.2.3]{DelMoral2006} combined with an adaptive tempering schedule described by \cite{Zhou2016} (also called ``annealed SMC'', e.g., in  \cite{wang_annealed_2020}). 
SMC \emph{samplers} are an extension or generalization of SMC methods, which allow for sampling from a sequence of distributions defined on a fixed state space, as opposed to the more common SMC setup \citep{doucet2009tutorial} consisting of product spaces of increasing dimensionality. 

As described in \cite{DelMoral2006}, Section 3.3.2.3, the proposals we use consist in MCMC kernels targeting each of the intermediate distributions (see Section~\ref{sec:sparsity} for a detailed description on their construction).
\cite{DelMoral2006} also justify in this context the incremental weight updates given by 
\begin{align}\label{eq:weight}
	w_t(x_{t-1}, x_t) &= \frac{\gamma_t(x_{t-1})}{\gamma_{t-1}(x_{t-1})},
\end{align}
where $\gamma_t(x)$ is the numerator in the right-hand side of Equation~(\ref{eq:sequence}). Initialization is done using the prior sampler described in Section~\ref{sec:prior-samplers}. 

\begin{figure}
  \centering
  \begin{subfigure}[b]{0.45\linewidth}
    \includegraphics[width=\linewidth]{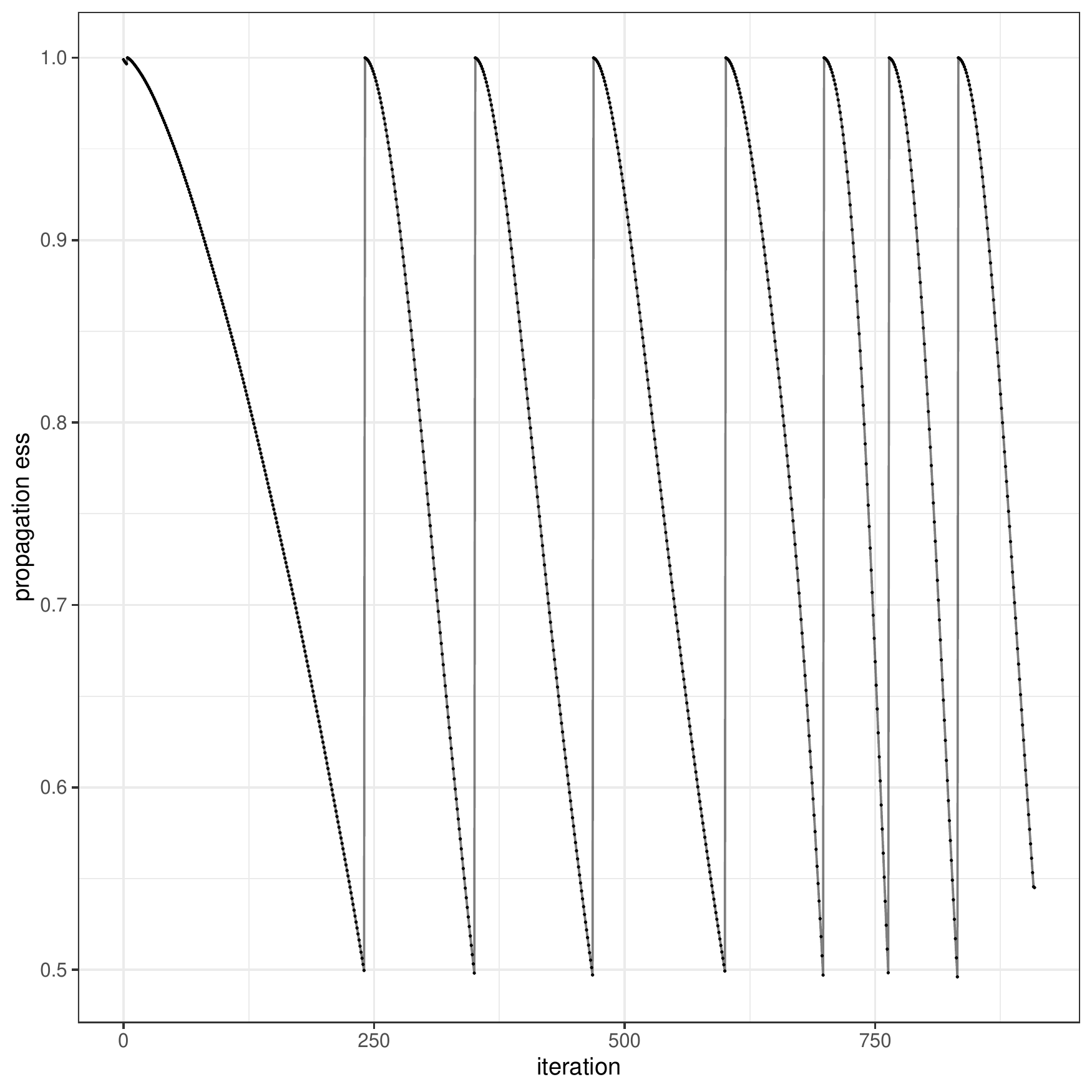}
  \end{subfigure}
  \begin{subfigure}[b]{0.45\linewidth}
    \includegraphics[width=\linewidth]{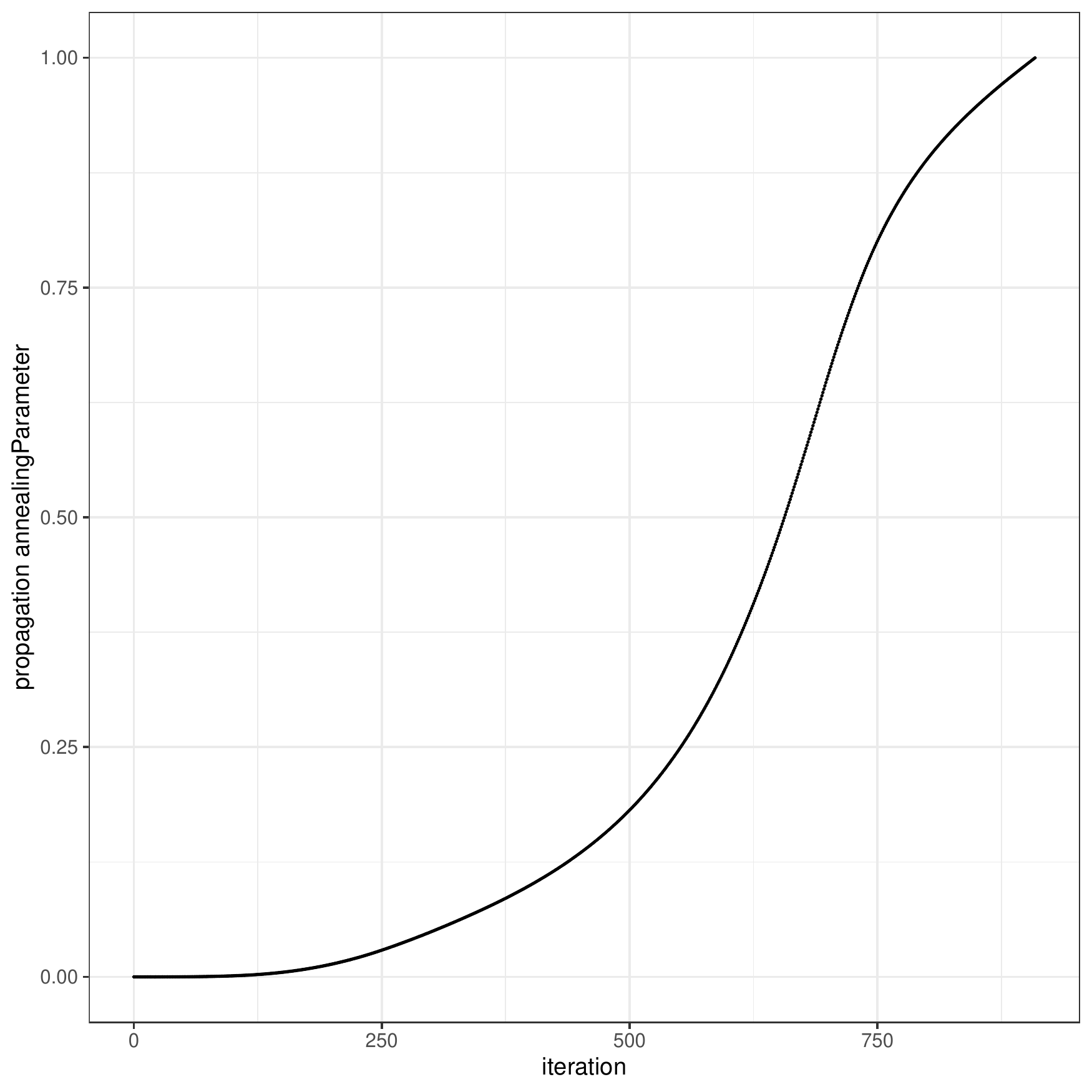}
  \end{subfigure}
  \caption{
  SCM monitoring plots automatically created when SCM is used and the default post-processing tool activated (\code{-{}-engine SCM -{}-postProcessor DefaultPostProcessor}). Here we show examples for the Gaussian mixture model in Section~\ref{sec:gmm}, using $10000$ particles.
  The adaptive resampling scheme based on ESS estimates is performed by default for SCM.
  Left: ESS plotted against iterations, automatically created  in \code{monitoringPlots/propagation-ess.pdf}.
        Each ``spike'' observed in this plot correspond to a resampling step taking place.
	When ESS falls beneath a predefined threshold, $0.5$ here, particles are resampled to prevent weight degeneracy.
	This resampling procedure ``refreshes'' the relative ESS to $1.0$.
  Right: The resulting adaptive annealing schedule, found in \code{monitoringPlots/propagation.pdf}.
          The y-axis corresponds annealing parameters, and x-axis the iteration number.
  }
  \label{fig:scmMonitoring}
\end{figure}

Due to the weight degeneracy problem reviewed in \cite{doucet2009tutorial}, a resampling procedure is required.
Resampling prevents the population of sample weights from collapsing into a point mass.
However, resampling injects additional noise into the sampling process.
Motivated by the need to balance these two factors, we use a standard procedure to adaptively determine when resampling should be performed. 
The effective sample size (ESS) is computed at each iteration \citep{kong_ess_1992}.  
If the relative ESS---ESS divided by population size---falls beneath a predetermined threshold, resampling is performed.
Figure~\ref{fig:scmMonitoring} (left) illustrates the resampling procedure's effect on ESS.
This threshold value defaults to $0.5$ in \proglang{Blang}, and can be set, for example, to $0.4$ using the command-line argument \code{-{}-engine.resamplingESSThreshold 0.4}. 
By default, the resampling scheme used is the stratified sampling of \cite{kitagawa_monte_1996} (use \texttt{-{}-engine.resamplingScheme MULTINOMIAL} for multinomial resampling). 

SMC samplers rely on a discrete set of interpolating distributions $0 = t_1 < t_2 < \dots < t_N = 1$.
As initially proposed in \cite{jasra_inference_2011} and improved in \cite{Zhou2016}, instead of building this sequence a priori, we construct it incrementally and adaptively.
At each step the next annealing parameter is determined so as to cause a fixed decay in the relative conditional ESS as defined in \cite{Zhou2016}.
Figure~\ref{fig:scmMonitoring} (right) shows an example of a resulting adaptive annealing schedule.
Finding the next annealing parameter is a simple univariate root finding problem. 
Since the weight update shown in Equation~(\ref{eq:weight}) does not depend on $x_t$, only the already available particles from the previous iteration, $x_{t-1}$, the computational cost of the root finding problem is negligible. By default, the targeted decay is set to $0.9999$ and can be controlled via \texttt{-{}-engine.temperatureSchedule.threshold}. Setting it to a lower value will speed up computation at the cost of a less accurate posterior distribution (and vice versa). 
One disadvantage of this adaptation scheme is that the running time of the method is random and may be hard to predict a priori. 
If the user requires a prespecified number of iterations, adaptive construction of the sequence of distribution can be turned off, for example to use a fixed number of 20 iterations and an equally spaced annealing schedule $0 = t_1 < t_2 < \dots < t_{20} = 1$, use  \texttt{-{}-engine.temperatureSchedule FixedTemperatureSchedule -{}-engine.temperatureSchedule.nTemperatures 20}. 
Custom mechanisms to control the schedule can be added by implementing the interface \code{TemperatureSchedule}.\footnote{\api{sdk}{blang/engines/internals/schedules/TemperatureSchedule}{blang.engines.internals.schedules.TemperatureSchedule}}
If for example the user implements their own algorithm in a class called \texttt{MySchedule} located in package \texttt{mypackage}, to enable its use during inference, add the command line arguments  \texttt{-{}-engine.temperatureSchedule mypackage.MySchedule}.

After SCM inference is performed, \blang performs one last round of resampling followed by 5 rounds of particle rejuvenation on each particle.
This results in a set of equally weighted particles.
\updated{The amount of rejuvenation to perform after the final resampling round can be controlled via \texttt{-{}-engine.nFinalRejuvenations}, for example use \texttt{-{}-engine.nFinalRejuvenations 10}, for 10 rejuvenation rounds.}

\subsubsection{Non-reversible parallel tempering} \label{sec:pt}

\proglang{Blang} incorporates a non-reversible, adaptive parallel tempering (PT) algorithm \citep{Syed2019}.
PT \citep{Geyer1991} is an MCMC method that operates on product spaces.
Informally, PT runs $N$ Markov chains in parallel, each targeting a distribution from a sequence of tempered (i.e., annealed) distributions indexed by $0 \le t \le 1$ (Section~\ref{sec:sequenceOfMeasures}). 
Each PT iteration consists of two phases:
a local exploration phase taking place within individual chains, and a communication phase taking place between chains. 

\begin{figure}
	\centering
	\includegraphics[width=\linewidth, trim={750 0 500 0}, clip=true]{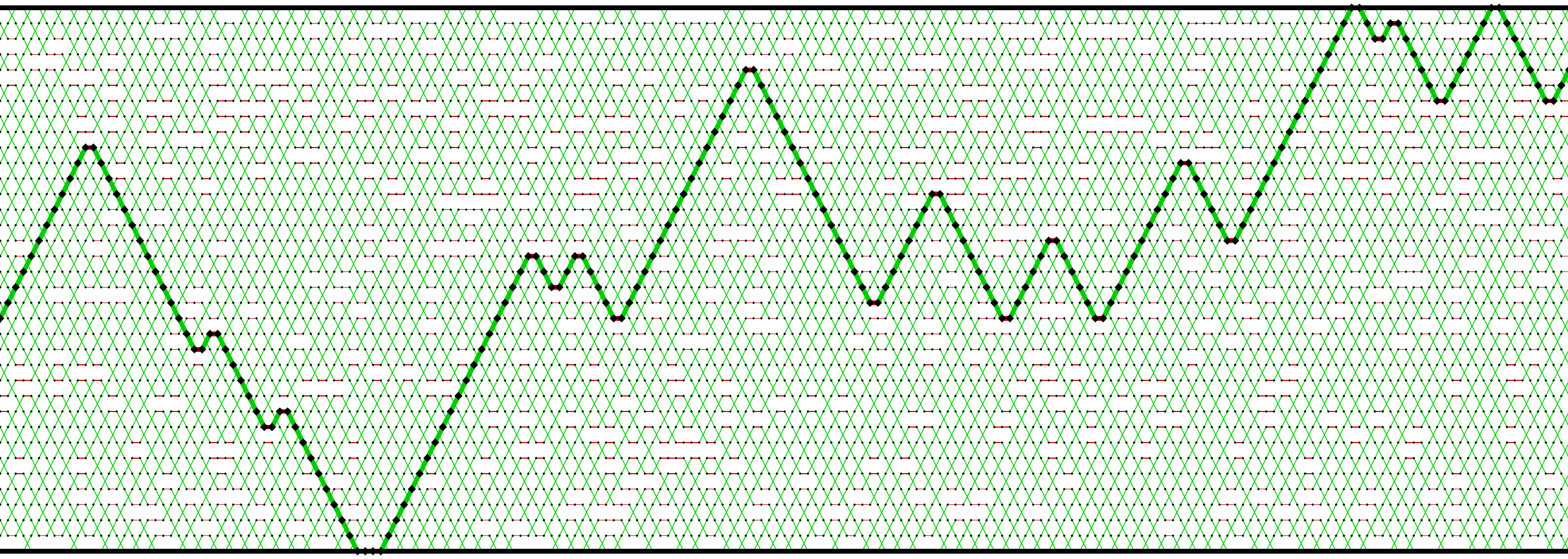}
	\caption{Visualization (cropped) of the chain swaps proposed while running non-reversible PT (add \code{-{}-postProcessor.runPxviz true -{}-postProcessor.boldTrajectory 1} to create this visualization).
		The x-axis corresponds to PT iterations, and y-axis corresponds to different parallel chains, with the one at the top corresponding the posterior distribution, the one at the bottom, to the prior distribution, and those in between interpolating between the two.
		When a swap is accepted (green line segments), two chains exchange their states, denoted by crossing lines.
		When a swap is rejected we use red line segments.
		An \emph{index process} is obtained by considering a path formed by these line segments (one index process is shown as a bold line for ease of interpretation). 
		An \emph{annealed restart} is defined as a path segment within an index process which starts at the prior and ends at the posterior. 
	}
	\label{fig:ptPaths}
\end{figure}

In the local exploration step, for chains with $t > 0$, the state is updated via samplers or MCMC kernels invariant with respect to the chain's target distribution (the construction of these kernels is detailed in Section~\ref{sec:sparsity}).
For the chain with $t = 0$, the local exploration step consists in an independent draw from the prior distribution (the construction of the independent sampler for the prior is detailed in Section~\ref{sec:prior-samplers}).
If the user requires using MCMC samplers for $t=0$ instead of prior sampling, the option \texttt{-{}-engine.usePriorSamples false} can be used. 
 
In the communication phase, swaps between neighbour chains are proposed and accepted/rejected according to the Metropolis-Hastings ratio.
Informally, this swapping procedure provides opportunities for states to traverse across modes, as the prior allows independent sampling and hence a form of regeneration. Even when sampling from the prior is not possible, annealing often yields MCMC kernels with better mixing rates.

In our implementation, both the exploration and communication phases are parallelized in the number of parallel chains $N$ (see \ref{sec:parallelize} for details). 
However to leverage this parallelism, following the theoretical analysis of  \cite{Syed2019},  special attention has been devoted (1) to the details of how the swap mechanism is performed, and (2) to the tuning of the annealing schedule $t_1, t_2, \dots, t_{N-1}$ introduced in the last section. 

Point (1) is motivated by a sharp contrast between the performance of reversible and non-reversible flavours of PT.
Performance in the following discussion is based on the notion of \emph{annealed restarts}, defined along with the related notion of the index process in Figure~\ref{fig:ptPaths}.
We define PT performance as the fraction of iterations where an annealed restart is just completed at the current iteration. This is called the restart rate, which we denote by $\tau$, and it is equivalent (up to an additive factor of 1) to the notion of round trip rate popular in the PT literature  \citep{katzgraber2006feedback,lingenheil2009efficiency}.

\newcommand\taurev{\tau_{\text{rev},N}}
\newcommand\taunonrev{\tau_{\text{non-rev},N}}

Previous theoretical work has focused on reversible PT where the groups of chains to swap are selected at random. 
In the reversible regime, several lines of work \citep{rathore2005optimal,atchade_towards_2011} have demonstrated that  even when a high number of cores is available, one still has to ensure that $N$, and hence the number of cores leveraged, is not too large. More precisely, the performance of reversible PT collapses as $N$ increases, even when communication and local exploration are fully parallel. For example the results in  \cite{atchade_towards_2011} imply that $\taurev \to 0$ as $N\to\infty$. 
Surprisingly, this performance collapse disappears when a non-reversible flavour of PT is used:  \cite{Syed2019} identified conditions where  $\taunonrev \to c$ as $N\to\infty$, where the model-dependent constant $c > 0$ is discussed further below. 
Even more surprising is that 
algorithmically, the distinction needed to make PT non-reversible is minimal: it is simply the use of a deterministic alternation of two specific types of swap kernels, those swapping $i, i+1$ with $i$ even, followed by similar swaps with $i$ odd. This algorithm can be traced back to \cite{okabe2001replica}, however, only recently its non-reversible dynamics have been identified and used to prove the existence of a qualitative gap  between the reversible and non-reversible flavours of PT.  
The gap can be established both non-asymptotically ($\taurev < \taunonrev$ for all $N$), and also asymptotically as $N \to \infty$, in which case the performance of non-reversible PT, $\taunonrev$, is furthermore guaranteed to be monotonically increasing for $N$ large enough.  

More importantly, non-reversibility opens the door for highly parallel algorithms to optimize over the annealing schedule, hence addressing point (2) above. By default, \proglang{Blang}'s PT engine uses the non-reversible schedule optimization from \cite{Syed2019} (labelled NRPT henceforth). In contrast, at the time of writing, mainstream probabilistic programming languages either lack support for parallel tempering \citep{Plummer2003,Lunn2012,Salvatier2015,Carpenter2017}, or require manual input of the annealing parameters \citep{foreman-mackey_emcee_2013}. 

\begin{figure}[H]
	\centering
	\begin{subfigure}[b]{0.3\linewidth}
		\includegraphics[width=\linewidth]{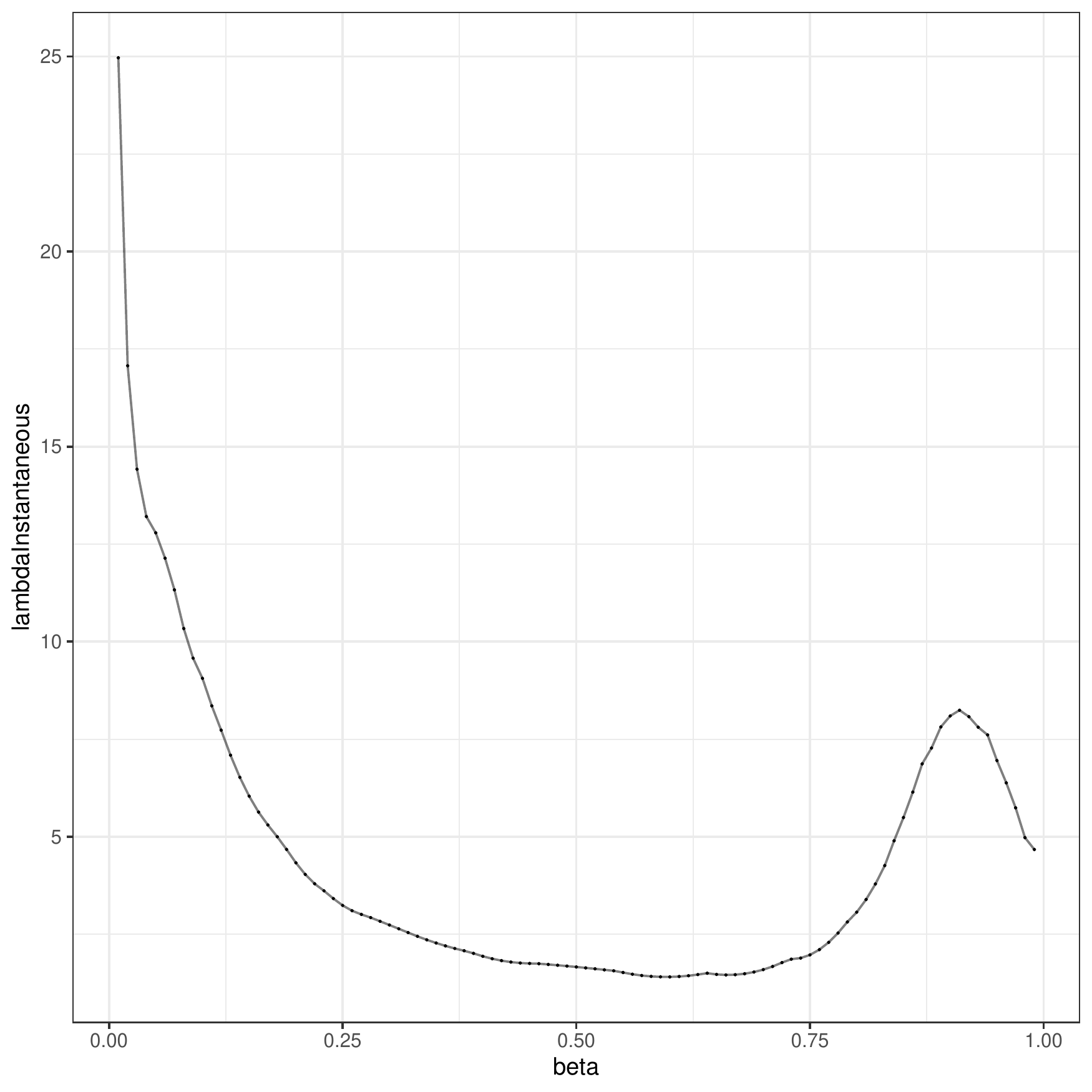}
	\end{subfigure}
	\begin{subfigure}[b]{0.3\linewidth}
		\includegraphics[width=\linewidth]{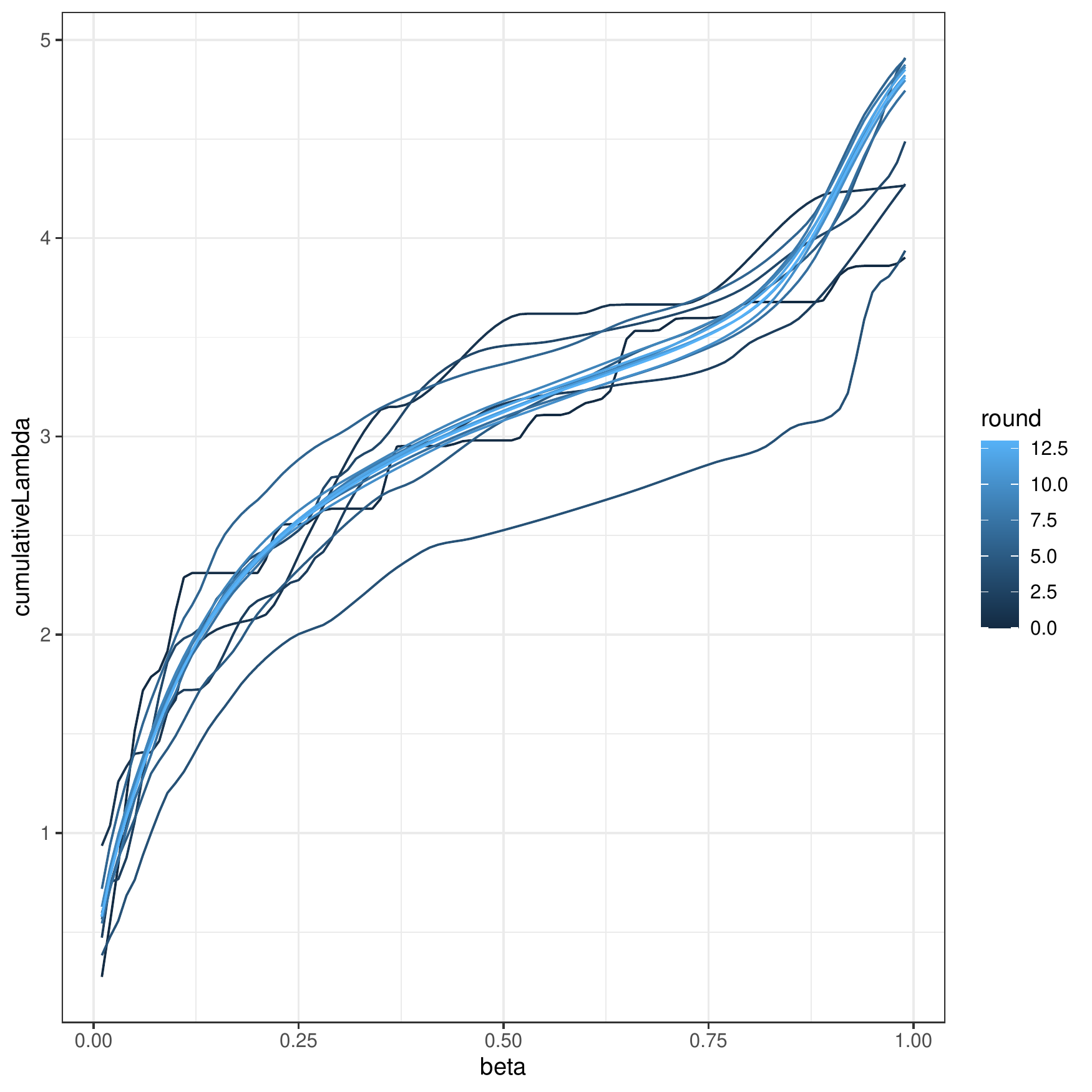}
	\end{subfigure}
	\begin{subfigure}[b]{0.3\linewidth}
		\includegraphics[width=\linewidth]{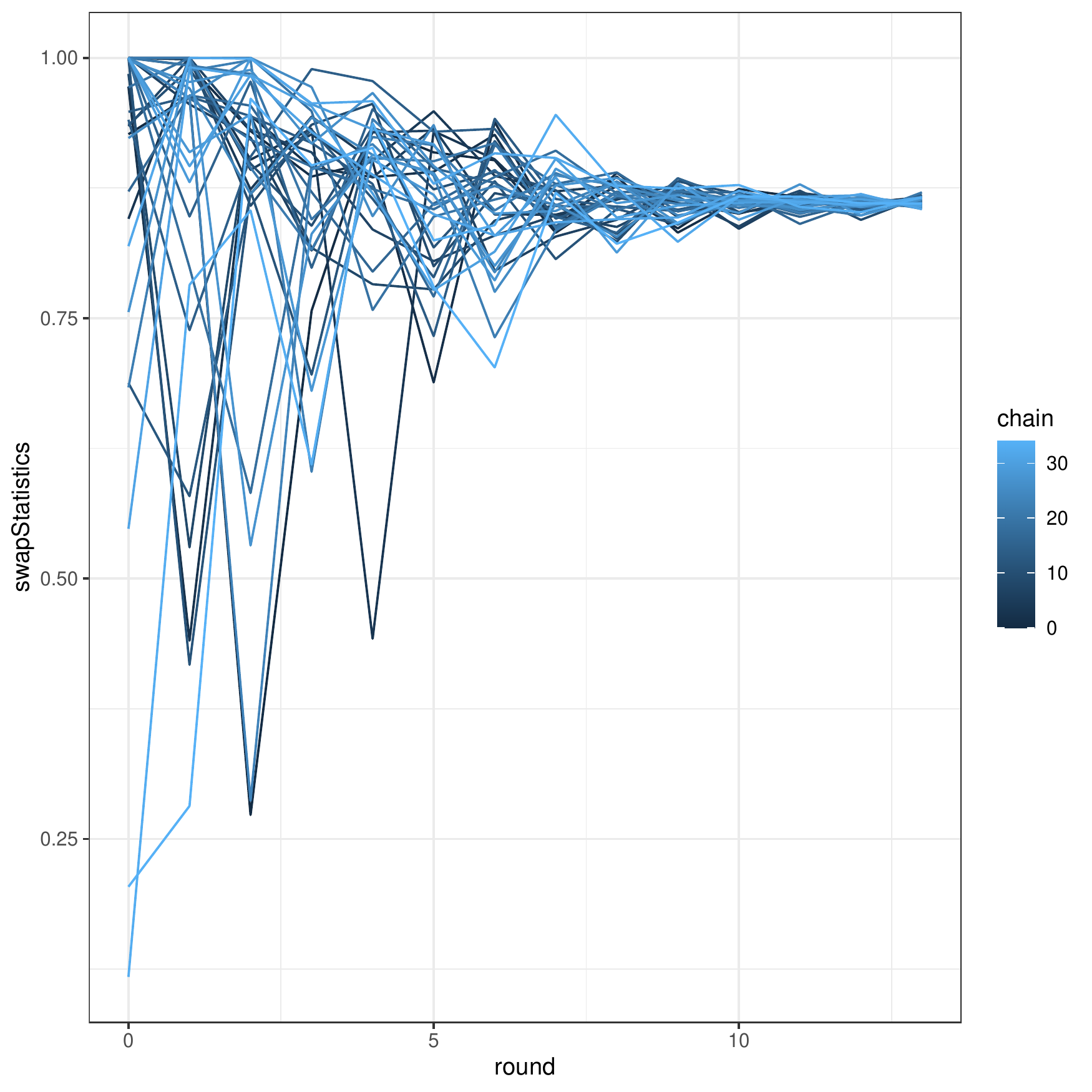}
	\end{subfigure}
	\caption{Left: \updated{final estimate of the local communication barrier $\hat \lambda$ versus the annealing parameter $t$ (labelled ``beta'' in the \proglang{Blang} output).
	The spiking phenomenon around $t=0.9$ is indicative of a phase transition.
	This corresponds to the mixture indicator variables going from a disorganized configuration (the side of the peak at $t=0.9$ closer to the prior on the left) to a clustered configuration (the side of the peak closer to the posterior).}
	Middle: estimates of the cumulative communication barrier, $\hat \Lambda(t)$, with each colour corresponding to a different iterative round of the annealing schedule optimization algorithm.
	Right: here each line (colour) is one of the $N$ chains, and the line tracks the average acceptance probability (ordinate) between that chain and its neighbour for each round of the schedule optimization  algorithm (abscissa).
	In contrast to reversible PT, NRPT does not need to restrict swap acceptance probability to low values such as the $23\%$ acceptance rule of \cite{atchade_towards_2011}.}
	\label{fig:monitoringPlots2}
\end{figure}

To outline how the NRPT algorithm works, we first outline the asymptotic distribution of a single index process (for example, the bold line in Figure~\ref{fig:ptPaths}) as $N \to \infty$. While for reversible PT this distribution converges to a diffusion, for non-reversible PT, it converges to a piecewise-deterministic Markov process (PDMP). See \cite{davis1993markov} for background on PDMPs. The rate parameter of this limiting PDMPs, $\lambda$, a positive function taking as input an annealing parameter $t \in [0, 1]$, can be interpreted as being proportional to the expected rejection rate for a swap between $\pi_t$ and $\pi_{t+\epsilon}$. See Figure~\ref{fig:monitoringPlots2} (left) for an example of an estimate $\hat \lambda$ from the model in Section~\ref{sec:gmm}. Moreover, the constant $c$ introduced earlier as the asymptotic non-reversible performance, $\taunonrev \to c$ can be written as $c = (2+\Lambda)^{-1}$, where $\Lambda(t) = \int_0^t \lambda(t') \ud t'$. We call $\lambda$, $\Lambda(t)$, and $\Lambda = \Lambda(1)$ the local, cumulative and global communication barriers respectively.

Importantly, all three communication barriers can be estimated from the MCMC output, and used as the basis for tuning PT as described in \citep{Syed2019}. 
First, $\Lambda$ can be used as a measure of the difficulty of PT-based inference for a given model: this is supported by its relation with the model-specific  constant $c$ described earlier.
We recommend to use a number of chains proportional to $\Lambda$.
Using at least $2\Lambda$ appears to provide a good starting point empirically. 
Second, NRPT uses $\Lambda(t)$ to optimize the annealing parameters using the following strategy.
The algorithm iteratively estimates $\hat \Lambda(t)$, using a simple and asymptotically consistent rule $\hat \Lambda(t_i) = \sum_{j=1}^i \hat r^{(j-1,j)}$, and $\hat \Lambda(\cdot)$ interpolated using a monotone cubic spline between the $t_i$'s, where $\{t_i\}$ are annealing parameters from the previous iterations, and $\hat r^{(j-1,j)}$ is the empirical swap rejection rates, also obtained from the previous iteration.
The algorithm then computes univariate quantile of $t \mapsto \hat \Lambda(t) / \hat \Lambda(1)$ to update the annealing schedule.
This is repeated using a doubling scheme, where the first round uses 1 iteration to estimate $\hat \Lambda(t)$, followed by annealing parameters update, the second round uses 2 iterations based on the updated schedule, followed by an update of the annealing parameters, then 4, 8, etc.
See Figure~\ref{fig:monitoringPlots2} (middle) for an example of how estimates of $\hat \Lambda$ progress as the number of rounds increases. 
As a byproduct of the NRPT algorithm we obtain a burn-in mechanism: by default, all post-processing uses only the samples produced by the last optimization round which is equivalent to a $50\%$ burn-in.  
The only exception is for the trace plot, which is shown for both the whole MCMC trace in the output folder \texttt{tracePlotsFull}, and for the post burn-in phase,  \texttt{tracePlots}. The default of $50\%$ burn-in can be customized via \texttt{-{}-postProcessor.burnInFraction}. 

Alternative mechanisms can be used to control the annealing parameters. By default, the initial annealing schedule is uniform, other initial values can be used, see \texttt{-{}-engine PT -{}-help} for various options.
Optimization of the annealing parameters can be disabled with the argument \texttt{-{}-engine.adaptFraction 0.0}. 
Custom mechanisms to control the initial schedule can be added by the user by implementing the interface \code{TemperatureLadder}.\footnote{\api{sdk}{blang/engines/internals/ladders/TemperatureLadder}{blang.engines.internals.ladders.TemperatureLadder}}
If for example the user implements their own algorithm in a class called \texttt{MyLadder} located in package \texttt{mypackage}, to enable it, use \texttt{-{}-engine.ladder mypackage.MyLadder}.

One tuning parameter that can be used to speed-up the execution of NRPT is the expected number of times each local exploration kernel should be used between two rounds of swap attempts, \texttt{-{}-engine.nPassesPerScan} (fractional values are accepted).
By default, this is set to 3, so that a theoretical assumption called Effective Local Exploration (ELE) \cite{Syed2019}, is well approximated.
However, we observed that performance was robust to this choice so if the local exploration kernels are reasonably efficient, lower values will lead to similar behaviour of the index processes for a lower computational budget.
Conversely, if the local exploration kernels perform very poorly, it may be useful to explore higher values for the argument \texttt{-{}-engine.nPassesPerScan}.

If the hard-drive space required to store the samples produced by PT becomes prohibitive, one option is to enable thinning by providing an input  \texttt{-{}-engine.thinning} greater than one. For example, \texttt{-{}-engine.thinning 2} will store samples only once every two PT iterations. An alternative (available for all engines), is to compress the samples in \texttt{.gz} format, which is enabled using \texttt{-{}-experimentConfigs.tabularWriter.compressed true}. All post-processing is compatible with the compressed samples format. 

Initialization of PT is by default performed by first running SCM using an annealing schedule containing all annealing parameters in PT's initial schedule. The SCM initialization can be configured using the same arguments as those described in Section~\ref{sec:scm} but with the prefix \texttt{engine.scmInit}. For example, to increase the number of particles (set to 100 for the initialization run), use \texttt{-{}-engine.scmInit.nParticles 200}.

\subsubsection{Other inference engines} \label{sec:other-engines}

When a model is not in generative normal form, the PT and SCM engines cannot be used. In such case, the user can still run a basic, single chain MCMC via \texttt{-{}-engine MCMC}.
This option is essentially a shortcut for setting the PT engine to use a single chain, to avoid using SCM for initialization, and to avoid other checks that assume a generative normal form.
The PT command line arguments from Section~\ref{sec:pt} that are relevant to single-chain MCMC can still be used, in particular \texttt{-{}-engine.nScans} and \texttt{-{}-engine.thinning}. 

Another convenient shortcut is \texttt{-{}-engine AIS} which uses SCM but with resampling disabled. This is known as the Annealed Importance Sampling algorithm \cite{Neal2001}.
The SCM arguments relevant to AIS can still be used with this engine, namely \texttt{-{}-engine.nParticles}, \texttt{-{}-engine.nFinalRejuvenation}, and \texttt{-{}-engine.temperatureSchedule.threshold}.

In cases where the user would like to sample independent and identically distributed realizations from a model where no observation is present, the engine \texttt{-{}-engine Forward} (for forward sampling) with option \code{-{}-engine.nSamples 1} can be used. 

When all random variables in a small model are discrete, the argument \texttt{-{}-engine Exact} will enumerate all possible scenarios. Note that the \texttt{DefaultPostprocessor} should not be used to analyze the output of the exact engine. This is because the output in the folder \texttt{samples} have a different interpretation than with the other engines: instead of representing equally weighted samples, they represent weighted samples with weight indicated in a row called \texttt{logProbability}.

Finally, the inference engine can be customized. This is achieved by implementing the interface \code{PosteriorInferenceEngine}.\footnote{\api{sdk}{blang/engines/internals/PosteriorInferenceEngine}{blang.engines.internals.PosteriorInferenceEngine}}
If for example the user implements their own algorithm in a class called \texttt{MyEngine} located in package \texttt{mypackage}, to enable its use during inference, add the command line arguments  \texttt{-{}-engine mypackage.MyEngine}.

\subsection{Pseudo-random generator}\label{sec:random-gen}

The current pseudo-random generator is the Mersenne Twister \cite{Matsumoto1998} as implemented in the \pkg{MathCommons} package.
By default, the seed $1$ is used. For inference engines based on randomized algorithms (all current algorithms except \texttt{Exact}), this can be changed using the command line argument \texttt{-{}-engine.random} followed by an integer.

\subsection{Parallelization} \label{sec:parallelize}
Due to the nature of PT and SCM algorithms, parallelization can be used to obtain significant performance improvements.
In both PT and SCM, transition MCMC kernels are applied in parallel across particles/chains.
In addition to parallelization of transition kernels, PT also performs its swap operations in parallel.

\proglang{Blang} uses lightweight threads to parallelize these operations \citep{Friesen2015}.
Specifically, it uses the algorithm described in \cite{Leiserson2012} as implemented in \cite{Steele2013}.
This implementation allows each chain to pertain to its own random stream, consequently avoiding any blocking between threads.
Furthermore, this implementation implies any numerical output will not be altered by the number of threads utilized given fixed random seeds.

For controlling multi-threading, use \texttt{-{}-engine.nThreads Max} to take advantage of as many threads as there are cores in the host machine, \texttt{-{}-engine.nThreads Dynamic} to dynamically allocate threads based on the overall system usage (the default behaviour, which ensures analysts can smoothly carry other tasks while inference is running in the background), \texttt{-{}-engine.nThreads Single} to force single-thread mode, and \texttt{-{}-engine.nThreads Fixed -{}-engine.nThreads.number 2} to fix a specific number of threads to use.

\subsection{Marginal likelihood computation} \label{sec:evidence}

Standard Bayesian model selection requires computing the marginal likelihood, also known as the evidence.
The marginal likelihood is conceptually simple: it is the probability or the density of the observed data. 
However computing or approximating this single scalar is often challenging. 
Fortunately, both PT and SCM automatically compute the marginal likelihood with no extra computational cost.

Our PT engine supports estimation of the marginal likelihood through two methods: thermodynamic integration \citep{Ogata1989}, and the stepping stone estimator \citep{xie2011improving}. 
For models with hard constraints (i.e., models whose likelihood is equal to zero for particular configurations of states proposed by the sampling algorithm), the technical conditions underlying thermodynamic integration may not be satisfied, and that estimator is automatically omitted in such cases.
The stepping stone estimator can still be used in these cases. 
In SCM, the evidence comes as a by-product of the weights computed by the algorithm, see e.g.,\ \cite{DelMoral2006}.

In contrast to \proglang{Blang}, other mainstream probabilistic programming languages require additional packages and external dependencies  to approximate the  marginal likelihood.
For example in \proglang{Stan}, one would require additional post-processing with bridge sampling \citep{Meng1996} using packages such as \pkg{bridgesampling} \citep{bridgesamplingR, Gronau2017}.

\subsection{Diagnostics}

We summarize here some diagnostic strategies that can be used to assess the quality of the posterior distribution approximation. With the PT inference engine, the key diagnostics are the ESS estimates (Section~\ref{sec:overview}) and the number of annealed restarts (see Figure~\ref{fig:ptPaths} for the definition, and \texttt{monitoring/actualTemperedRestarts.csv} for the estimates). Each annealed restart incorporates a unique independent draw from the prior chain successfully propagated to the posterior chain. This can be complemented with inspection of the trace plots (Section~\ref{sec:overview}). Finally, another strategy is to monitor the marginal likelihood: a separate estimate is provided for each adaptation round in the PT engine (in \texttt{monitoring/logNormalizationContantProgress.csv}), so its convergence can be readily monitored, and moreover one can check the agreement of PT's estimate with the orthogonal marginal likelihood  estimator used by SCM (either based on the automatic SCM initialization, or from a separate run; see \texttt{logNormalizationEstimate.csv}).

\subsection{Construction of prior samplers}\label{sec:prior-samplers}

Consider the sequence of distribution in Equation~(\ref{eq:sequence}) at $t = 0$, where we recover the prior distribution $p(x)$. 
When a model is in generative normal form (Section~\ref{sec:normalform}), Blang automatically constructs an efficient algorithm to sample from the prior distribution $p(x)$.
Briefly, the normal form property guarantees that we can orient the factor graph over the latent variables into a directed graphical model.
The generative normal form property enables the enumeration of forward samplers provided by the \texttt{generate} blocks.
Finally, as a preprocessing step, we order these \texttt{generate} blocks according to a linearization of the directed graphical model.

\subsection{Construction of invariant samplers} \label{sec:sparsity}

We first describe how a \proglang{Blang} model $m$ is transformed into an efficient representation aware of $m$'s sparsity patterns.
The transformed representation is an instance of \code{SampledModel},\footnote{\api{sdk}{blang/runtime/SampledModel}{blang.runtime.SampledModel}}
a mutable object keeping track of the state space and offering methods to:
1) change the annealing parameter of the model,
2) apply a transition kernel in place targeting the current annealing parameter,
3) perform forward simulation in place,
4) obtain the joint log density of the current configuration,
and 5) duplicate the state via a deep cloning library.

\subsubsection{Preprocessing} \label{subsec:preproc}

The process of translating $m$ into a \code{SampledModel} begins with the instantiation of model variables.
After this is done, a list $l$ of factors is recursively constructed.
That is, we recursively search through $m$ for sub-models, and terminate when we have identified and added all atomic laws to $l$.

\updated{
The next phase of initialization consists of building an \emph{accessibility graph} between all objects in a model,\footnote{Here objects refer to the same objects as defined in \proglang{Java}, i.e., dynamically allocated class instances or arrays.} defined as follows:
}
the set of vertices is the set of objects defined by a model, starting at the root model, and of the constituents of these objects recursively.
Constituents are fields in the case of objects and integer indices in the case of arrays.
Constituents can also be customized, for example, in order to index entries of matrices.
The directed edges of the accessibility graph connect objects to their constituents, and constituents to the object they resolve to, if any.
We say that object $o_2$ is accessible from $o_1$ if there is a directed path from $o_1$ to $o_2$ in the accessibility graph.

Once the accessibility graph has been constructed, the latent variables in $m$ are extracted from the vertex set of the accessibility graph.
These variables are the intersection of objects of a type annotated with \code{@Samplers} and objects that are mutable, or have accessible mutable children.
Mutability here corresponds to the class having fields that are either arrays or that are \emph{non-final} (the latter being the \proglang{Java} terminology, or equivalently, in \proglang{Xtend}, fields not marked by \code{val}).
In other words, latent variables are objects that have a designated sampler, and are or have access to non-final fields.
Immutability is therefore the main mechanism used to define observed (fixed) values. 
Additionally, we can mark indices in matrices and arrays as observed.
This is accomplished by the \code{Observations} object.\footnote{\api{sdk}{blang/runtime/Observations}{blang.runtime.Observations}}
For example, \code{observationsObject.markAsObserved(mtx.getRealVar(i, j))} marks entry $(i,j)$ of matrix \code{mtx} as observed.
In scenarios where objects or fields are accessible but unused in factors, the exploration of such objects and fields can be skipped.
This can be handled in the construction of the accessibility graph by using the annotation \code{@SkipDependency}.
With our accessibility graph constructed and latent variables identified, we can now exploit the model's sparsity patterns by constructing a factor graph.

\subsubsection{Exploiting sparsity}

Samplers can be made more efficient by avoiding unnecessary computation of model components;
we exploit a model's sparsity pattern by building factor graphs via linear time graph algorithms on our accessibility graph.

Given a latent variable $v$ and factor $f$, we can determine whether the application of a sampling operator on $v$ can change the numerical value of the factor $f$.
This is accomplished by assessing $v$ and $f$'s co-accessibility.
Two objects $o_1$ and $o_2$ are said to be co-accessible if there is a mutable object (as defined previously) $o_3$ such that $o_3$ is accessible from both $o_1$ and $o_2$.

Through this awareness of sparsity patterns, we can now perform sampling operations on variables without computing every factor involved in the model.
The cost of the entire preprocessing procedure has negligible cost in comparison with the performance to be gained from its implications.

For a concrete example, we refer readers to the Markov chain example (Section~\ref{sec:markovChainExample}).

\subsubsection{Matching transition kernels (samplers)}
\label{subsec:mcmcKernels}

Once sparsity patterns have been identified, samplers are matched to latent variables through the \code{@Samplers} annotation.
We have seen examples of this annotation in the permutation example of Section~\ref{sec:customSamplers}. Here we provide more details on this process.

\updated{
Given a certain model we would like to sample from, the first step is to identify which types of samplers are needed. 
To do so, recall that each latent variable is by definition of a type annotated with the \code{@Samplers} annotation. Now the \code{@Samplers} annotation is required to include as arguments a list of types responsible for sampling that type. 
For example, the class \code{DenseSimplex} is annotated with \code{@Samplers(SimplexSampler)}.\footnote{It is also possible to add or exclude samplers from the command line, using the options \code{-{}-samplers.additional} and \code{-{}-samplers.excluded} respectively, followed by fully qualified \proglang{Java} types pointing to sampler implementations. If the user wants to only rely on the set of samplers specified in the command line and not the ones obtained from the annotation, the option \code{-{}-useAnnotation false} can be used.} 
Repeating this process for each type of latent variable, this gives us a pool of sampler types. This pool of sampler types is summarized at the top of the standard output when sampling is performed, for example,}

\longline
\begin{CodeChunk}
	\begin{CodeOutput}
2 samplers constructed with following prototypes:
  RealScalar sampled via: [RealSliceSampler]
  IntScalar sampled via: [IntSliceSampler]
	\end{CodeOutput}
\end{CodeChunk}
\longline

The next step is to attempt to instantiate one Sampler object for each latent variable. We will walk through this process based on the example of instantiating a \code{SimplexSampler}, shown below.

\longline

\code{SimplexSampler.xtend}\vspace*{-10pt}\footnote{\url{https://github.com/UBC-Stat-ML/blangSDK/blob/master/src/main/java/blang/mcmc/SimplexSampler.xtend}}

\longline
\begin{lstlisting}
...
class SimplexSampler implements Sampler {
  @SampledVariable DenseSimplex simplex
  @ConnectedFactor List<LogScaleFactor> numericFactors
  @ConnectedFactor Constrained constrained
  
  override void execute(Random rand) { ... }
  override boolean setup(SamplerBuilderContext context) { ... }
}
\end{lstlisting}
\longline

\updated{
At this point of the process, we have on one hand a specific instance of a simplex random variable $s$ within a factor graph, and on the other hand, the \code{SimplexSampler} type. 
We first look at all the factors connected to $s$ that can be assigned to fields annotated by \code{@ConnectedFactor} in \code{SimplexSampler}.
For example, if any number of factors of type \code{LogScaleFactor} are encountered, it can be assigned by adding it to the list \code{numericFactors}.
Similarly, if $s$ is connected to no more than one \code{Constrained} factor, that factor can be assigned to the field \code{constrained}.
If all neighbours of $s$ can be matched in this fashion, then all these annotated fields will be automatically populated by the  neighbour factors of $s$.
If this is not the case, i.e., if some neighbour factor cannot be assigned to one of the fields marked by \code{@ConnectedFactor}, then the current sampler type will not be matched with $s$. 

Provided we have a match, then the field \code{simplex} annotated with \code{@SampledVariable} is automatically initialized with the sampled variable.
After this is done, the \code{setup} method is invoked to (1) perform any required pre-computation, and (2) to provide a chance to reject matching the sampler based on more complex criteria (for example, if deciding whether it is possible for this sampler to handle sampling $s$ based on information only available at runtime; the return value of \code{setup} determines if the sampler will be instantiated in the given context). 
The function \code{setup} is provided as input an object of type \code{SamplerBuilderContext} which makes it possible to use more fine grain information on the factor graph.\footnote{\url{https://www.stat.ubc.ca/~bouchard/blang/javadoc-sdk/blang/mcmc/internals/SamplerBuilderContext.html}}
For example, if $s$ is itself composed of several objects $s1$ and $s2$ to be sampled one after the other, it would be possible to cache a more precise decomposition of the list of factors connected to each by using \code{context.connectedFactors(new ObjectNode(s1))}.
}

This complete the setup phase. 
Next, during Monte Carlo sampling, the variable being sampled is updated in place through the invocation of \code{execute}. 

Going back to the simplex example, the 
\code{Constrained} factor is used here to indicate that the sampler being constructed is aware of the constraints posed by simplex variables.

\section*{Computational Details} \label{compdetails}
All the programs in this paper were run using \pkg{blangSDK} 2.13.1 on a Mac OS X version 10.14.5. The device used is a MacBook Pro (15-inch, 2018) with a 2.2 GHz 6-core Intel Core i7 processor and a 2.2 GHz Radeon Pro 555X graphics card and 32 GB of 2400 MHz DDR4 memory.

\section*{Funding}

This research was supported by a Discovery Grant from the National Science and Engineering Research Council, a Canadian Statistical Sciences Institute Collaborative Research Team Project, and Michael Smith Foundation RRF Grant. 



\bibliography{refs}

\begin{thebibliography}{69}
\newcommand{\enquote}[1]{``#1''}
\providecommand{\natexlab}[1]{#1}
\providecommand{\url}[1]{\texttt{#1}}
\providecommand{\urlprefix}{URL }
\expandafter\ifx\csname urlstyle\endcsname\relax
  \providecommand{\doi}[1]{doi:\discretionary{}{}{}#1}\else
  \providecommand{\doi}{doi:\discretionary{}{}{}\begingroup
  \urlstyle{rm}\Url}\fi
\providecommand{\eprint}[2][]{\url{#2}}

\bibitem[{Ackerman \emph{et~al.}(2017)Ackerman, Freer, and Roy}]{Roy2017}
Ackerman NL, Freer CE, Roy DM (2017).
\newblock \enquote{On Computability and Disintegration.}
\newblock \emph{Mathematical Structures in Computer Science}, \textbf{27}(8),
  1287--1314.

\bibitem[{Atchad\'e \emph{et~al.}(2011)Atchad\'e, Roberts, and
  Rosenthal}]{atchade_towards_2011}
Atchad\'e YF, Roberts GO, Rosenthal JS (2011).
\newblock \enquote{Towards Optimal Scaling of {M}etropolis-Coupled {Markov}
  Chain {Monte} {Carlo}.}
\newblock \emph{Statistics and Computing}, \textbf{21}(4), 555--568.
\newblock ISSN 0960-3174, 1573-1375.
\newblock \doi{10.1007/s11222-010-9192-1}.

\bibitem[{Ballnus \emph{et~al.}(2017)Ballnus, Hug, Hatz, G\"{o}rlitz,
  Hasenauer, and Theis}]{ballnus_comprehensive_2017}
Ballnus B, Hug S, Hatz K, G\"{o}rlitz L, Hasenauer J, Theis FJ (2017).
\newblock \enquote{Comprehensive Benchmarking of {Markov} Chain {Monte} {Carlo}
  Methods for Dynamical Systems.}
\newblock \emph{BMC Systems Biology}, \textbf{11}.
\newblock ISSN 1752-0509.

\bibitem[{Bingham \emph{et~al.}(2018)Bingham, Chen, Jankowiak, Obermeyer,
  Pradhan, Karaletsos, Singh, Szerlip, Horsfall, and Goodman}]{bingham2018pyro}
Bingham E, Chen JP, Jankowiak M, Obermeyer F, Pradhan N, Karaletsos T, Singh R,
  Szerlip P, Horsfall P, Goodman ND (2018).
\newblock \enquote{\proglang{Pyro}: Deep Universal Probabilistic Programming.}
\newblock \emph{Journal of Machine Learning Research}.

\bibitem[{Burkner and Vuorre(2019)}]{Paul2019}
Burkner PC, Vuorre M (2019).
\newblock \enquote{Ordinal Regression Models in Psychology: A Tutorial.}
\newblock \emph{Advances in Methods and Practices in Psychological Science},
  \textbf{2}(1), 77--101.

\bibitem[{Carpenter \emph{et~al.}(2017)Carpenter, Gelman, Hoffman, Lee,
  Goodrich, Betancourt, Brubaker, Guo, Li, and Riddell}]{Carpenter2017}
Carpenter B, Gelman A, Hoffman MD, Lee D, Goodrich B, Betancourt M, Brubaker M,
  Guo J, Li P, Riddell A (2017).
\newblock \enquote{\proglang{Stan} : A Probabilistic Programming Language.}
\newblock \emph{Journal of Statistical Software}, \textbf{76}(1).
\newblock ISSN 1548-7660.

\bibitem[{{Carter Brandon} and {McCrea W.
  H.}(1983)}]{carter_brandon_anthropic_1983}
{Carter Brandon}, {McCrea W H} (1983).
\newblock \enquote{The Anthropic Principle and its Implications for Biological
  Evolution.}
\newblock \emph{Philosophical Transactions of the Royal Society of London.
  Series A, Mathematical and Physical Sciences}, \textbf{310}(1512), 347--363.

\bibitem[{Chen and Shao(1999)}]{chen_monte_1999}
Chen MH, Shao QM (1999).
\newblock \enquote{Monte {Carlo} {Estimation} of {Bayesian} {Credible} and
  {HPD} {Intervals}.}
\newblock \emph{Journal of Computational and Graphical Statistics},
  \textbf{8}(1), 69--92.
\newblock ISSN 1061-8600.
\newblock \doi{10.2307/1390921}.

\bibitem[{Davis(1993)}]{davis1993markov}
Davis MH (1993).
\newblock \emph{Markov Models \& Optimization}.
\newblock Chapman and Hall.

\bibitem[{{Del Moral} \emph{et~al.}(2006){Del Moral}, Doucet, and
  Jasra}]{DelMoral2006}
{Del Moral} P, Doucet A, Jasra A (2006).
\newblock \enquote{Sequential Monte Carlo Samplers.}
\newblock \emph{Journal of the Royal Statistical Society B}, \textbf{68}(3),
  411--436.
\newblock ISSN 13697412.
\newblock \eprint{0212648}.

\bibitem[{Doucet and Johansen(2009)}]{doucet2009tutorial}
Doucet A, Johansen AM (2009).
\newblock \enquote{A Tutorial on Particle Filtering and Smoothing: Fifteen
  years later.}
\newblock \emph{Handbook of nonlinear filtering}, \textbf{12}(656-704), 3.

\bibitem[{Duane \emph{et~al.}(1987)Duane, Kennedy, Pendleton, and
  Roweth}]{Duane1987}
Duane S, Kennedy A, Pendleton BJ, Roweth D (1987).
\newblock \enquote{Hybrid Monte Carlo.}
\newblock \emph{Physics Letters B}, \textbf{195}(2), 216--222.
\newblock ISSN 03702693.

\bibitem[{Efftinge and V{\"{o}}lter(2006)}]{Efftinge2006}
Efftinge S, V{\"{o}}lter M (2006).
\newblock \enquote{oAW \proglang{xText}: A Framework for Textual DSLs.}
\newblock \emph{Proceedings of Workshop on Modeling Symposium at Eclipse
  Summit}.

\bibitem[{Flegal and Jones(2010)}]{Flegal2010}
Flegal JM, Jones GL (2010).
\newblock \enquote{Batch Means and Spectral Variance Estimators in Markov Chain
  Monte Carlo.}
\newblock \emph{The Annals of Statistics}, \textbf{38}(2), 1034--1070.
\newblock ISSN 00905364.

\bibitem[{Foreman-Mackey \emph{et~al.}(2013)Foreman-Mackey, Hogg, Lang, and
  Goodman}]{foreman-mackey_emcee_2013}
Foreman-Mackey D, Hogg DW, Lang D, Goodman J (2013).
\newblock \enquote{emcee: {The} {MCMC} {Hammer}.}
\newblock \emph{Publications of the Astronomical Society of the Pacific},
  \textbf{125}(925), 306.
\newblock ISSN 1538-3873.
\newblock \doi{10.1086/670067}.
\newblock Publisher: IOP Publishing.

\bibitem[{Friesen(2015)}]{Friesen2015}
Friesen J (2015).
\newblock \emph{{\proglang{Java} Threads and the Concurrency Utilities}}.
\newblock Apress, Berkeley, CA.
\newblock ISBN 978-1-4842-1699-6.

\bibitem[{Geweke(2004)}]{Geweke2004}
Geweke J (2004).
\newblock \enquote{Getting It Right: Joint Distribution Tests of Posterior
  Simulators.}
\newblock \emph{Journal of the American Statistical Association},
  \textbf{99}(467), 799--804.
\newblock ISSN 01621459.

\bibitem[{Geyer(1991)}]{Geyer1991}
Geyer CJ (1991).
\newblock \enquote{Markov Chain Monte Carlo Maximum Likelihood.}
\newblock \emph{Computing Science and Statistics: Proc. 23rd Symposium on the
  Interface, Interface Foundation, Fairfax Station, VA}, pp. 156--163.

\bibitem[{Goodman \emph{et~al.}(2012)Goodman, Mansinghka, Roy, Bonawitz, and
  Tenenbaum}]{church}
Goodman ND, Mansinghka VK, Roy DM, Bonawitz K, Tenenbaum JB (2012).
\newblock \enquote{\proglang{Church}: A Language for Generative Models.}
\newblock \emph{CoRR}, \textbf{abs/1206.3255}.
\newblock \eprint{1206.3255}.

\bibitem[{Greiner \emph{et~al.}(2016)Greiner, Burgess, Savchenko, and
  Yu}]{Greiner2016}
Greiner J, Burgess JM, Savchenko V, Yu HF (2016).
\newblock \enquote{On the Fermi-GBM event 0.4 s after {GW}150914.}
\newblock \emph{The Astrophysical Journal}, \textbf{827}(2), L38.

\bibitem[{Griffiths and Ghahramani(2011)}]{griffiths_indian_2011}
Griffiths TL, Ghahramani Z (2011).
\newblock \enquote{The {Indian} {Buffet} {Process}: {An} {Introduction} and
  {Review}.}
\newblock \emph{Journal of Machine Learning Research}, \textbf{12}(32),
  1185--1224.
\newblock ISSN 1533-7928.

\bibitem[{Gronau and Singmann(2018)}]{bridgesamplingR}
Gronau QF, Singmann H (2018).
\newblock \emph{\pkg{bridgesampling}: Bridge Sampling for Marginal Likelihoods
  and Bayes Factors}.
\newblock R package version 0.6-0,
  \urlprefix\url{https://CRAN.R-project.org/package=bridgesampling}.

\bibitem[{Gronau \emph{et~al.}(2017)Gronau, Singmann, and
  Wagenmakers}]{Gronau2017}
Gronau QF, Singmann H, Wagenmakers EJ (2017).
\newblock \enquote{\pkg{bridgesampling}: An R Package for Estimating
  Normalizing Constants.}
\newblock \eprint{1710.08162}.

\bibitem[{H{\"o}hna \emph{et~al.}(2014)H{\"o}hna, Heath, Boussau, Landis,
  Ronquist, and Huelsenbeck}]{hohna_probabilistic_2014}
H{\"o}hna S, Heath TA, Boussau B, Landis MJ, Ronquist F, Huelsenbeck JP (2014).
\newblock \enquote{Probabilistic {Graphical} {Model} {Representation} in
  {Phylogenetics}.}
\newblock \emph{Systematic Biology}, \textbf{63}(5), 753--771.
\newblock ISSN 1063-5157.
\newblock \doi{10.1093/sysbio/syu039}.

\bibitem[{H{\"o}hna \emph{et~al.}(2016)H{\"o}hna, Landis, Heath, Boussau,
  Lartillot, Moore, Huelsenbeck, and Ronquist}]{RevBayes}
H{\"o}hna S, Landis M, Heath T, Boussau B, Lartillot N, Moore B, Huelsenbeck J,
  Ronquist F (2016).
\newblock \enquote{\pkg{RevBayes}: Bayesian Phylogenetic Inference Using
  Graphical Models and an Interactive Model-Specification Language.}
\newblock \emph{Systematic Biology}, \textbf{65}, 1--11.

\bibitem[{Ising(1925)}]{Ising1925}
Ising E (1925).
\newblock \enquote{Beitrag zur Theorie des Ferromagnetismus.}
\newblock \emph{Zeitschrift f{\"u}r Physik}, \textbf{31}(1), 253--258.
\newblock ISSN 0044-3328.

\bibitem[{Jasra \emph{et~al.}(2005)Jasra, Holmes, and
  Stephens}]{jasra2005markov}
Jasra A, Holmes CC, Stephens DA (2005).
\newblock \enquote{Markov Chain Monte Carlo Methods and the Label Switching
  Problem in Bayesian Mixture Modeling.}
\newblock \emph{Statistical Science}, pp. 50--67.

\bibitem[{Jasra \emph{et~al.}(2011)Jasra, Stephens, Doucet, and
  Tsagaris}]{jasra_inference_2011}
Jasra A, Stephens DA, Doucet A, Tsagaris T (2011).
\newblock \enquote{Inference for {Lévy}-{Driven} {Stochastic} {Volatility}
  {Models} via {Adaptive} {Sequential} {Monte} {Carlo}.}
\newblock \emph{Scandinavian Journal of Statistics}, \textbf{38}(1), 1--22.
\newblock ISSN 1467-9469.
\newblock \doi{10.1111/j.1467-9469.2010.00723.x}.

\bibitem[{Katzgraber \emph{et~al.}(2006)Katzgraber, Trebst, Huse, and
  Troyer}]{katzgraber2006feedback}
Katzgraber HG, Trebst S, Huse DA, Troyer M (2006).
\newblock \enquote{Feedback-Optimized Parallel Tempering {M}onte {C}arlo.}
\newblock \emph{Journal of Statistical Mechanics: Theory and Experiment},
  \textbf{2006}(03), P03018.

\bibitem[{Kitagawa(1996)}]{kitagawa_monte_1996}
Kitagawa G (1996).
\newblock \enquote{Monte {Carlo} {Filter} and {Smoother} for {Non}-{Gaussian}
  {Nonlinear} {State} {Space} {Models}.}
\newblock \emph{Journal of Computational and Graphical Statistics},
  \textbf{5}(1), 1--25.
\newblock ISSN 1061-8600.
\newblock \doi{10.2307/1390750}.

\bibitem[{Kong(1992)}]{kong_ess_1992}
Kong A (1992).
\newblock \enquote{A Note on Importance Sampling Using Standardized Weights.}
\newblock \emph{Technical Report 348}, The University of Chicago.

\bibitem[{Lakner \emph{et~al.}(2008)Lakner, van~der Mark, Huelsenbeck, Larget,
  and Ronquist}]{Lakner2008}
Lakner C, van~der Mark P, Huelsenbeck JP, Larget B, Ronquist F (2008).
\newblock \enquote{Efficiency of Markov Chain Monte Carlo Tree Proposals in
  Bayesian Phylogenetics.}
\newblock \emph{Systematic Biology}, \textbf{57}(1), 86--103.
\newblock ISSN 1063-5157.

\bibitem[{Leiserson \emph{et~al.}(2012)Leiserson, Schardl, and
  Sukha}]{Leiserson2012}
Leiserson CE, Schardl TB, Sukha J (2012).
\newblock \enquote{Deterministic Parallel Random-Number Generation for
  Dynamic-Multithreading Platforms.}
\newblock \emph{ACM Sigplan Notices}, \textbf{47}(8), 193--204.

\bibitem[{Lingenheil \emph{et~al.}(2009)Lingenheil, Denschlag, Mathias, and
  Tavan}]{lingenheil2009efficiency}
Lingenheil M, Denschlag R, Mathias G, Tavan P (2009).
\newblock \enquote{Efficiency of Exchange Schemes in Replica Exchange.}
\newblock \emph{Chemical Physics Letters}, \textbf{478}(1-3), 80--84.

\bibitem[{Lunn \emph{et~al.}(2012)Lunn, Jackson, Best, Thomas, and
  Spiegelhalter}]{Lunn2012}
Lunn D, Jackson C, Best N, Thomas A, Spiegelhalter D (2012).
\newblock \emph{The \proglang{BUGS} Book: A Practical Introduction to Bayesian
  Analysis}.
\newblock First edition. Chapman and Hall/CRC.
\newblock ISBN 9781584888499.

\bibitem[{Lunn \emph{et~al.}(2009)Lunn, Spiegelhalter, Thomas, and
  Best}]{Lunn2009}
Lunn D, Spiegelhalter D, Thomas A, Best N (2009).
\newblock \enquote{The \proglang{BUGS} project: Evolution, Critique and Future
  Directions.}
\newblock \emph{Statistics in Medicine}, \textbf{28}(25), 3049--3067.
\newblock ISSN 0277-6715.

\bibitem[{Lunn \emph{et~al.}(2000)Lunn, Thomas, Best, and
  Spiegelhalter}]{Lunn2000}
Lunn DJ, Thomas A, Best N, Spiegelhalter D (2000).
\newblock \enquote{\pkg{WinBUGS} - A Bayesian Modelling Framework: Concepts,
  Structure, and Extensibility.}
\newblock \emph{Statistics and Computing}, \textbf{10}(4), 325--337.
\newblock ISSN 1573-1375.

\bibitem[{Matsumoto and Nishimura(1998)}]{Matsumoto1998}
Matsumoto M, Nishimura T (1998).
\newblock \enquote{Mersenne Twister: A 623-Dimensionally Equidistributed
  Uniform Pseudo-Random Number Generator.}
\newblock \emph{ACM Trans. Model. Comput. Simul.}, \textbf{8}(1), 3--30.
\newblock ISSN 1049-3301.

\bibitem[{Meng and Wong(1996)}]{Meng1996}
Meng XL, Wong WH (1996).
\newblock \enquote{Simulating Ratios of Normalizing Constants via a Simple
  Identity: A Theoretical Exploration.}
\newblock \emph{Statistica Sinica}, \textbf{6}, 831--860.

\bibitem[{Metropolis \emph{et~al.}(1953)Metropolis, Rosenbluth, Rosenbluth,
  Teller, and Teller}]{Metropolis1953}
Metropolis N, Rosenbluth AW, Rosenbluth MN, Teller AH, Teller E (1953).
\newblock \enquote{Equation of State Calculations by Fast Computing Machines.}
\newblock \emph{The Journal of Chemical Physics}, \textbf{21}(6), 1087--1092.
\newblock ISSN 0021-9606.

\bibitem[{Milch \emph{et~al.}(2005)Milch, Marthi, Russell, Sontag, Ong, and
  Kolobov}]{blog}
Milch B, Marthi B, Russell S, Sontag D, Ong DL, Kolobov A (2005).
\newblock \enquote{\proglang{BLOG}: Probabilistic Models with Unknown Objects.}
\newblock \emph{IJCAI: Proceedings of the 19th International Joint Conference
  on Artificial Intelligence}, pp. 1352--1359.

\bibitem[{Mitchell and Beauchamp(1988)}]{Mitchell1988}
Mitchell TJ, Beauchamp JJ (1988).
\newblock \enquote{Bayesian Variable Selection in Linear Regression.}
\newblock \emph{Journal of the American Statistical Association},
  \textbf{83}(404), 1023--1032.
\newblock ISSN 0162-1459.

\bibitem[{Murray and Sch{\"o}n(2018)}]{murray_automated_2018}
Murray LM, Sch{\"o}n TB (2018).
\newblock \enquote{Automated Learning with a Probabilistic Programming
  Language: {\proglang{Birch}}.}
\newblock \emph{Annual Reviews in Control}, \textbf{46}, 29--43.
\newblock ISSN 1367-5788.

\bibitem[{Neal(2001)}]{Neal2001}
Neal RM (2001).
\newblock \enquote{Annealed Importance Sampling.}
\newblock \emph{Statistics and Computing}, \textbf{11}(2), 125--139.
\newblock ISSN 09603174.
\newblock \eprint{9803008}.

\bibitem[{Neal(2003)}]{Neal2003}
Neal RM (2003).
\newblock \enquote{Slice Sampling.}
\newblock \emph{The Annals of Statistics}, \textbf{31}(3), 705--741.
\newblock ISSN 00905364.

\bibitem[{Neal(2011)}]{Neal2012}
Neal RM (2011).
\newblock \enquote{MCMC Using Hamiltonian dynamics.}
\newblock \emph{Handbook of Markov chain Monte Carlo}, \textbf{2}(11).

\bibitem[{Ogata(1989)}]{Ogata1989}
Ogata Y (1989).
\newblock \enquote{A Monte Carlo Method for High Dimensional Integration.}
\newblock \emph{Numerische Mathematik}, \textbf{55}(2), 137--157.
\newblock ISSN 0029-599X.

\bibitem[{Okabe \emph{et~al.}(2001)Okabe, Kawata, Okamoto, and
  Mikami}]{okabe2001replica}
Okabe T, Kawata M, Okamoto Y, Mikami M (2001).
\newblock \enquote{Replica-Exchange Monte Carlo Method for the
  Isobaric--Isothermal Ensemble.}
\newblock \emph{Chemical physics letters}, \textbf{335}(5-6), 435--439.

\bibitem[{Paige and Wood(2014)}]{probabilisticC}
Paige B, Wood F (2014).
\newblock \enquote{A Compilation Target for Probabilistic Programming
  Languages.}
\newblock \emph{arXiv preprint arXiv:1403.0504}.

\bibitem[{Paige and Wood(2016)}]{paige_inference_2016}
Paige B, Wood F (2016).
\newblock \enquote{Inference {Networks} for {Sequential} {Monte} {Carlo} in
  {Graphical} {Models}.}
\newblock In \emph{Proceedings of the 33rd {International} {Conference} on
  {International} {Conference} on {Machine} {Learning} - {Volume} 48},
  {ICML}'16, pp. 3040--3049.
\newblock Event-place: New York, NY, USA.

\bibitem[{Plummer(2003)}]{Plummer2003}
Plummer M (2003).
\newblock \enquote{\pkg{JAGS}: A Program for Analysis of Bayesian Graphical
  Models Using Gibbs Sampling.}
\newblock \emph{Proceedings of the 3rd International Workshop on Distributed
  Statistical Computing}.
\newblock ISSN 1609-395X.

\bibitem[{\proglang{Xtend}(2019)}]{XtendDoc2019}
\proglang{Xtend} (2019).
\newblock \enquote{\proglang{Xtend}.}
\newblock
  \urlprefix\url{https://www.eclipse.org/xtend/documentation/203_xtend_expressions.html}.

\bibitem[{Rathore \emph{et~al.}(2005)Rathore, Chopra, and
  de~Pablo}]{rathore2005optimal}
Rathore N, Chopra M, de~Pablo JJ (2005).
\newblock \enquote{Optimal Allocation of Replicas in Parallel Tempering
  Simulations.}
\newblock \emph{The Journal of Chemical Physics}, \textbf{122}(2), 024111.
\newblock \doi{10.1063/1.1831273}.

\bibitem[{Ronquist \emph{et~al.}(2020)Ronquist, Kudlicka, Senderov,
  Borgstr{\"o}m, Lartillot, Lund{\'e}n, Murray, Sch{\"o}n, and
  Broman}]{ronquist2020probabilistic}
Ronquist F, Kudlicka J, Senderov V, Borgstr{\"o}m J, Lartillot N, Lund{\'e}n D,
  Murray L, Sch{\"o}n TB, Broman D (2020).
\newblock \enquote{Probabilistic Programming: A Powerful New Approach to
  Statistical Phylogenetics.}
\newblock \emph{BioRxiv}.

\bibitem[{Salvatier \emph{et~al.}(2015)Salvatier, Wiecki, and
  Fonnesbeck}]{Salvatier2015}
Salvatier J, Wiecki T, Fonnesbeck C (2015).
\newblock \enquote{Probabilistic Programming in Python Using PyMC.}
\newblock \eprint{1507.08050}.

\bibitem[{Semmens \emph{et~al.}(2009)Semmens, Ward, Moore, and
  Darimont}]{Semmens2009}
Semmens BX, Ward EJ, Moore JW, Darimont CT (2009).
\newblock \enquote{Quantifying Inter- and Intra-Population Niche Variability
  Using Hierarchical Bayesian Stable Isotope Mixing Models.}
\newblock \emph{PLOS ONE}, \textbf{4}(7), 1--9.

\bibitem[{Steele and Lea(2013)}]{Steele2013}
Steele G, Lea D (2013).
\newblock \enquote{Splittable Random Application Programming Interface.}
\newblock
  \urlprefix\url{https://docs.oracle.com/javase/8/docs/api/java/util/SplittableRandom.html}.

\bibitem[{Steorts \emph{et~al.}(2016)Steorts, Hall, and Fienberg}]{Steorts2016}
Steorts RC, Hall R, Fienberg SE (2016).
\newblock \enquote{A Bayesian Approach to Graphical Record Linkage and
  Deduplication.}
\newblock \emph{Journal of the American Statistical Association},
  \textbf{111}(516), 1660--1672.
\newblock ISSN 0162-1459.

\bibitem[{Syed \emph{et~al.}(2019)Syed, Bouchard-C{\^{o}}t{\'{e}},
  Deligiannidis, and Doucet}]{Syed2019}
Syed S, Bouchard-C{\^{o}}t{\'{e}} A, Deligiannidis G, Doucet A (2019).
\newblock \enquote{Non-Reversible Parallel Tempering: an Embarassingly Parallel
  MCMC Scheme.}
\newblock \eprint{1905.02939}.

\bibitem[{Tancredi and Liseo(2011)}]{Tancredi2011}
Tancredi A, Liseo B (2011).
\newblock \enquote{A Hierarchical Bayesian Approach to Record Linkage and
  Population Size Problems.}
\newblock \emph{The Annals of Applied Statistics}, \textbf{5}(2B), 1553--1585.
\newblock ISSN 1932-6157.

\bibitem[{Teh \emph{et~al.}(2006)Teh, Jordan, Beal, and Blei}]{teh2006}
Teh YW, Jordan MI, Beal MJ, Blei DM (2006).
\newblock \enquote{Hierarchical Dirichlet Processes.}
\newblock \emph{Journal of the American Statistical Association},
  \textbf{101}(476), 1566--1581.
\newblock \doi{10.1198/016214506000000302}.

\bibitem[{van~de Meent \emph{et~al.}(2018)van~de Meent, Paige, Yang, and
  Wood}]{van_de_meent_introduction_2018}
van~de Meent JW, Paige B, Yang H, Wood F (2018).
\newblock \enquote{An {Introduction} to {Probabilistic} {Programming}.}
\newblock \emph{arXiv:1809.10756 [cs, stat]}.
\newblock ArXiv: 1809.10756.

\bibitem[{Wang \emph{et~al.}(2020)Wang, Wang, and
  Bouchard-Côté}]{wang_annealed_2020}
Wang L, Wang S, Bouchard-Côté A (2020).
\newblock \enquote{An {Annealed} {Sequential} {Monte} {Carlo} {Method} for
  {Bayesian} {Phylogenetics}.}
\newblock \emph{Systematic Biology}, \textbf{69}(1), 155--183.
\newblock ISSN 1063-5157.
\newblock \doi{10.1093/sysbio/syz028}.

\bibitem[{Wickham(2014)}]{Wickham2014}
Wickham H (2014).
\newblock \enquote{Tidy Data.}
\newblock \emph{Journal of Statistical Software, Articles}, \textbf{59}(10),
  1--23.
\newblock ISSN 1548-7660.

\bibitem[{Wood \emph{et~al.}(2014)Wood, van~de Meent, and
  Mansinghka}]{anglican}
Wood F, van~de Meent JW, Mansinghka V (2014).
\newblock \enquote{A New Approach to Probabilistic Programming Inference.}
\newblock In \emph{Proceedings of the 17th International conference on
  Artificial Intelligence and Statistics}, pp. 1024--1032.

\bibitem[{Xie \emph{et~al.}(2011)Xie, Lewis, Fan, Kuo, and
  Chen}]{xie2011improving}
Xie W, Lewis PO, Fan Y, Kuo L, Chen MH (2011).
\newblock \enquote{Improving Marginal Likelihood Estimation for Bayesian
  Phylogenetic Model Selection.}
\newblock \emph{Systematic biology}, \textbf{60}(2), 150--160.

\bibitem[{Zhao \emph{et~al.}(2015)Zhao, Cumberworth, Wang, Gsponer, de~Freitas,
  and Bouchard-C\^{o}t\'{e}}]{Zhao2015Bayesian}
Zhao T, Cumberworth A, Wang Z, Gsponer J, de~Freitas N, Bouchard-C\^{o}t\'{e} A
  (2015).
\newblock \enquote{Bayesian Analysis of Continuous Time {M}arkov Chains with
  Application to Phylogenetic Modelling.}
\newblock \emph{Bayesian Analysis}, \textbf{11}, 1203--1237.

\bibitem[{Zhou \emph{et~al.}(2019)Zhou, Gram{-}Hansen, Kohn, Rainforth, Yang,
  and Wood}]{Yuan19}
Zhou Y, Gram{-}Hansen BJ, Kohn T, Rainforth T, Yang H, Wood F (2019).
\newblock \enquote{{LF-PPL:} {A} Low-Level First Order Probabilistic
  Programming Language for Non-Differentiable Models.}
\newblock In K~Chaudhuri, M~Sugiyama (eds.), \emph{The 22nd International
  Conference on Artificial Intelligence and Statistics, {AISTATS} 2019, 16-18
  April 2019, Naha, Okinawa, Japan}, volume~89 of \emph{Proceedings of Machine
  Learning Research}, pp. 148--157. {PMLR}.

\bibitem[{Zhou \emph{et~al.}(2016)Zhou, Johansen, and Aston}]{Zhou2016}
Zhou Y, Johansen AM, Aston JA (2016).
\newblock \enquote{Toward Automatic Model Comparison: An Adaptive Sequential
  Monte Carlo Approach.}
\newblock \emph{Journal of Computational and Graphical Statistics},
  \textbf{25}(3), 701--726.
\newblock ISSN 15372715.
\newblock \eprint{1303.3123}.

\end{thebibliography}

\newpage
\begin{appendix}

\section{Advanced tutorials}\label{sec:tutorialAdvanced}
\subsection{Inference on a non-standard data structures via third-party libraries} \label{subsec:phylo}

Consider an inference problem where the data structure or parameter of interest is a phylogenetic tree.
A phylogenetic tree is a branching process encoding evolutionary relationships between organisms.
The following example illustrates how to perform inference on a phylogenetic tree model given sequence alignment data, using a third-party library.

The \proglang{Blang} language itself does not contain tree-valued random variables.
However, the language allows \emph{creating} custom types of random variables.
Moreover, these custom types can be packaged, published and imported.

First, we create a file called \code{dependencies.txt} at the root of the project directory.
Each line in \code{dependencies.txt} encodes a versioned third-party library to be imported (along with its transitive set of dependencies).
Here we use a \proglang{Blang} package providing phylogenetic-centric data types \citep{Zhao2015Bayesian}\footnote{\url{https://github.com/UBC-Stat-ML/conifer/tree/master/src/main/java/conifer}}:

\longline

\code{dependencies.txt} \vspace*{-10pt}

\longline
\begin{CodeChunk}
\begin{CodeInput}
com.github.UBC-Stat-ML:conifer:2.1.3
\end{CodeInput}
\end{CodeChunk}
\longline

We encode the \proglang{Blang} model, using the imported data type in the following code chunk

\longline

\code{PhylogeneticTree.bl} \vspace*{-10pt}

\longline
\begin{lstlisting}
package demo
import conifer.*
import static conifer.Utils.*

model PhylogeneticTree {

  random RealVar shape ?: latentReal()
  random RealVar rate ?: latentReal()
  random SequenceAlignment observations
  param EvolutionaryModel evoModel ?: kimura(observations.nSites)
  
  random UnrootedTree tree ?: unrootedTree(observations.observedTreeNodes) 
  
  laws {
    shape ~ Exponential(1.0)
    rate ~ Exponential(1.0)
    tree | shape, rate ~ NonClockTreePrior(Gamma.distribution(shape, rate))
    observations | tree, evoModel ~ UnrootedTreeLikelihood(tree, evoModel)
  }
}
\end{lstlisting}
\longline

In the first block of code, unobserved variables of the type \code{RealVar} and {IntVar} are declared using the functions \code{latentReal()} and \code{latentInt()}.\footnote{A summary of the most commonly used functions are listed in Figure~\ref{app:rvarfuncs} and Figure~\ref{app:linalgfuncs}.}
This is no different from our previous examples.
In the second block, \code{NonClockTreePrior} and \code{UnrootedTreeLikelihood} are themselves  \proglang{Blang} models defined in the imported \pkg{conifer} package.
\code{NonClockTreePrior} accepts a distribution as an argument, in the example an XExpression is used to pass in a Gamma distribution directly without the need to declare another variable in the model.

To run \code{PhylogeneticTree} we enter the following in the CLI:

\longline
\begin{CodeChunk}
%PARSESTART
\begin{CodeInput}
> git clone https://github.com/UBC-Stat-ML/JSSBlangCode.git
> cd JSSBlangCode/reproduction_material/example
> blang --model jss.phylo.PhylogeneticTree \
    --model.observations.file data/primates.fasta \
    --model.observations.encoding DNA
\end{CodeInput}
%PARSEEND
\begin{CodeOutput}
Preprocess {
  ...
  Initialization {
    ...
  } [ ... ]
} [ ... ]
Inference {
    ...
  Round(9/9) {
    ...
  } [ ... ]
} [ ... ]
Postprocess {
 ...
} [ ... ]
executionMilliseconds : ...
outputFolder :./JSSBlangCode/.../results/all/2019-06-18-09-42-15-sP.exec
\end{CodeOutput}
\end{CodeChunk}
\longline

Here \code{-{}-model.observations.file} specifies the data path, this is the standard \proglang{Blang} input method. 
\code{-{}-model.observations.encoding} is a model-specific option to parse our data, provided by the third-party library.
The usual outputs can be found in the \code{results} directory.

On the whole, to use third-party libraries or packages (not necessarily restricted to \proglang{Blang}), users just need to specify the dependencies in \code{dependencies.txt}, and include \code{import} statements as needed.
Running the model can be done via the usual CLI.
Inputs follow the same syntax, unless otherwise instructed by the third-party library (i.e., custom parsers).
Outputs are also placed in the usual directories.

\subsection{Spike and slab classification} \label{subsec:tutorial}

In this example, we focus on the implementation of a non-standard data type to handle a spike and slab model \citep{Mitchell1988}.
The spike and slab model is a mixture of prior distributions commonly used for coefficients in a regression model.
The non-standard data type \code{SpikedRealVar} is \proglang{Blang}'s representation of the type of the coefficients in a spike and slab model. 
The file \code{SpikedRealVar.xtend} shown in the code block below contains the implementation of the data type \code{SpikedRealVar} using \proglang{Xtend}. 

\longline

\code{SpikedRealVar.xtend}\vspace*{-10pt}

\longline
\begin{lstlisting}
package jss.glms

import blang.core.RealVar
import blang.core.IntVar
import blang.types.StaticUtils

class SpikedRealVar implements RealVar {
  public val IntVar selected = StaticUtils::latentInt()
  public val RealVar continuousPart = StaticUtils::latentReal()

  override doubleValue() {
    if (selected.intValue < 0 || selected.intValue > 1)
      StaticUtils::invalidParameter()
    if (selected.intValue == 0) return 0.0
    else return continuousPart.doubleValue
  }
  override toString() { "" + doubleValue }
}
\end{lstlisting}
\longline

Because we want to use \code{RealVar} and \code{IntVar} types in our \code{SpikedRealVar} type (\proglang{Xtend}), we require the import statements of core \proglang{Blang} types, as the usual automatic imports are only for \proglang{Blang} (\code{.bl}) files.
We declare its member variables, \code{selected} and \code{continuous}, as \code{IntVar} and \code{RealVar}.
These variables will encode the spike and slab component values for each explanatory variable.
Because these members are random, their values are initialized using \code{latentInt()} and \code{latentReal()}. 
We \code{override} \code{RealVar()}'s getter method \code{doubleValue()} to return the regression coefficient if the explanatory variable is selected.

We can now use this custom data type to build a simple classification model, this time using \proglang{Blang}. The code below is contained in \code{SpikeSlabClassification.bl}.

\longline

\code{SpikeSlabClassification.bl}\vspace*{-10pt}

\longline
\begin{lstlisting}
package glms

model SpikeSlabClassification {

  param GlobalDataSource data
  random RealVar activeProbability ?: latentReal
  random RealVar sigma ?: latentReal
  random RealVar intercept ?: latentReal
  
  param Plate<String> instances, features
  param Plated<Double> covariates
  random Plated<IntVar> labels
  random Plated<SpikedRealVar> parameters 
  
  laws {
    for (Index<String> instance : instances.indices) {
      labels.get(instance) | intercept, 
      DotProduct dotProduct 
      = DotProduct.of(features, parameters, covariates.slice(instance))
        ~ Bernoulli(logistic(intercept + dotProduct.compute))
    }

    for (Index<String> feature : features.indices) {
      parameters.get(feature).selected | activeProbability 
        ~ Bernoulli(activeProbability)
      parameters.get(feature).continuousPart | sigma 
        ~ StudentT(1.0, 0.0, sigma)
    }
    
    intercept | sigma ~ StudentT(1.0, 0.0, sigma)
    activeProbability ~ ContinuousUniform(0, 1)
    sigma ~ Exponential(1.0)
  }
}

\end{lstlisting}
\longline

In the above model, the random variable \code{parameters} is indexed by \code{instances} and \code{features}.
This relationship is encoded using built-in types \code{Plated} and \code{Plate} variables; where a \code{Plated} variable is indexed by one or more \code{Plate} variable.
Hence, \code{parameters} is of type \newline \code{Plated<SpikedRealVar>} and both \code{instances} and \code{features} are of type \code{Plate<String>}.
\code{Plate} and \code{Plated} variables are detailed in Section \ref{subsec:plates}.
Note this is not the only way to implement a spike and slab model.
For instance, a user could define a distribution and sampler for the \code{SpikedRealVar} type itself.
This example merely illustrates a  minimal implementation that takes advantage of \proglang{Blang}'s preexisting types, distributions, and samplers.

To perform inference on the model \code{SpikeSlabClassification}, we call the following using the CLI:

\longline
\begin{CodeChunk}
%PARSESTART
\begin{CodeInput}
> git clone https://github.com/UBC-Stat-ML/JSSBlangCode.git
> cd JSSBlangCode/reproduction_material/example
> blang --model jss.glms.SpikeSlabClassification \
    --model.data data/titanic/titanic-covariates.csv \
    --model.instances.name Name \
    --model.instances.maxSize 200 \
    --model.labels.dataSource data/titanic/titanic.csv \
    --model.labels.name Survived \
    --engine PT \
    --engine.nChains 20 \
    --engine.nScans 10000 \
    --postProcessor DefaultPostProcessor
\end{CodeInput}
%PARSEEND
\begin{CodeOutput}
Preprocess {
    ...
  Initialization {
    ...
  } [ ... ]
} [ ... ]
Inference {
    ...
  Round(9/9) {
    ... 
  } [ ... ]
} [ ... ]
Postprocess {
  Post-processing activeProbability
  Post-processing allLogDensities
  Post-processing energy
  Post-processing intercept
  Post-processing logDensity
  Post-processing parameters
  Post-processing sigma
  MC diagnostics
} [ ... ]
executionMilliseconds : ...
outputFolder :./JSSBlangCode/.../results/all/2019-06-18-10-05-29-ut.exec
\end{CodeOutput}
\end{CodeChunk}
\longline

The arguments \code{-{-}model.data} and \code{-{-}model.labels.dataSource} specify the data source, \\ \code{-{-}model.labels.name} and \code{-{-}model.instances.name} specify the column names that \code{labels} and \code{instances} correspond to, and \code{-{-}model.instances.maxSize} indicates the maximum size of the variable.
The \code{-{-}postProcessor} command creates additional summary statistics, posterior plots, trace plots, monitoring plots, and effective sample size in addition to the default outputs.

\begin{figure}[H]
  \begin{minipage}[c]{0.45\textwidth}
    \includegraphics[width=\textwidth, clip=true, trim={0 440 0 295}]{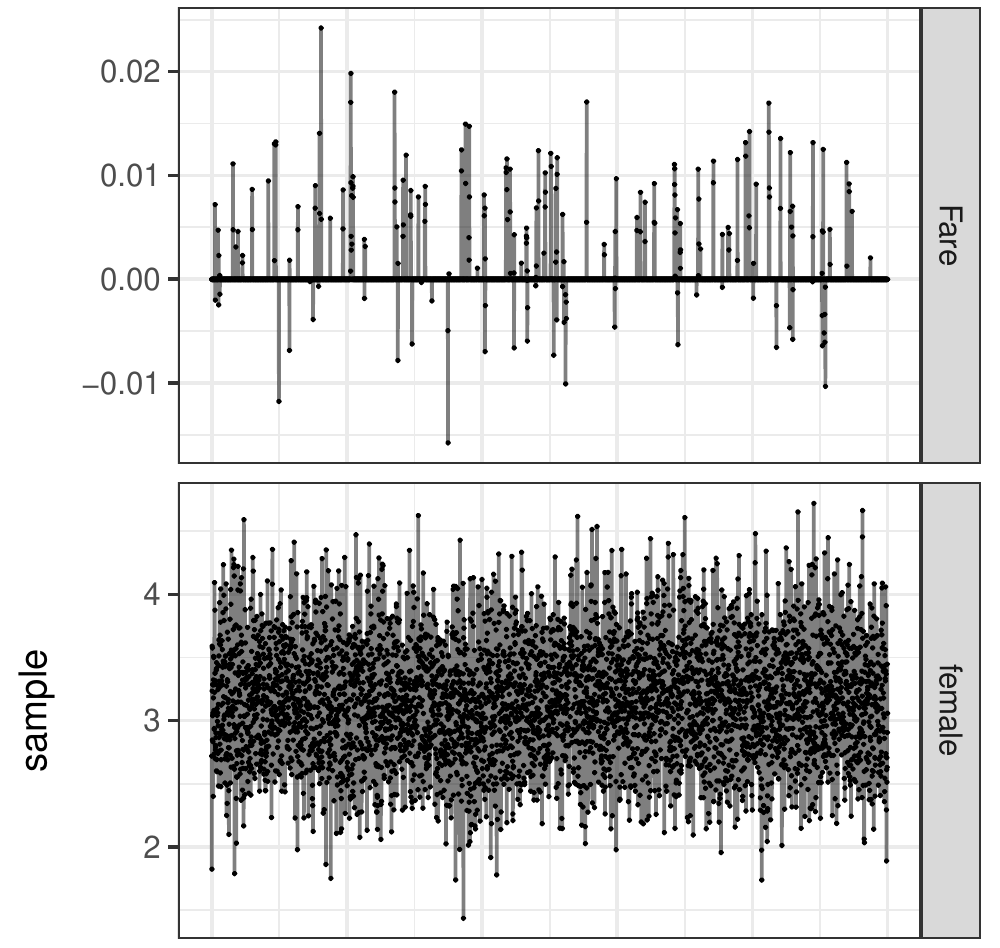}
  \end{minipage}
  \begin{minipage}[c]{0.45\textwidth}
    \includegraphics[width=\textwidth, clip=true, trim={0 2435 0 160}]{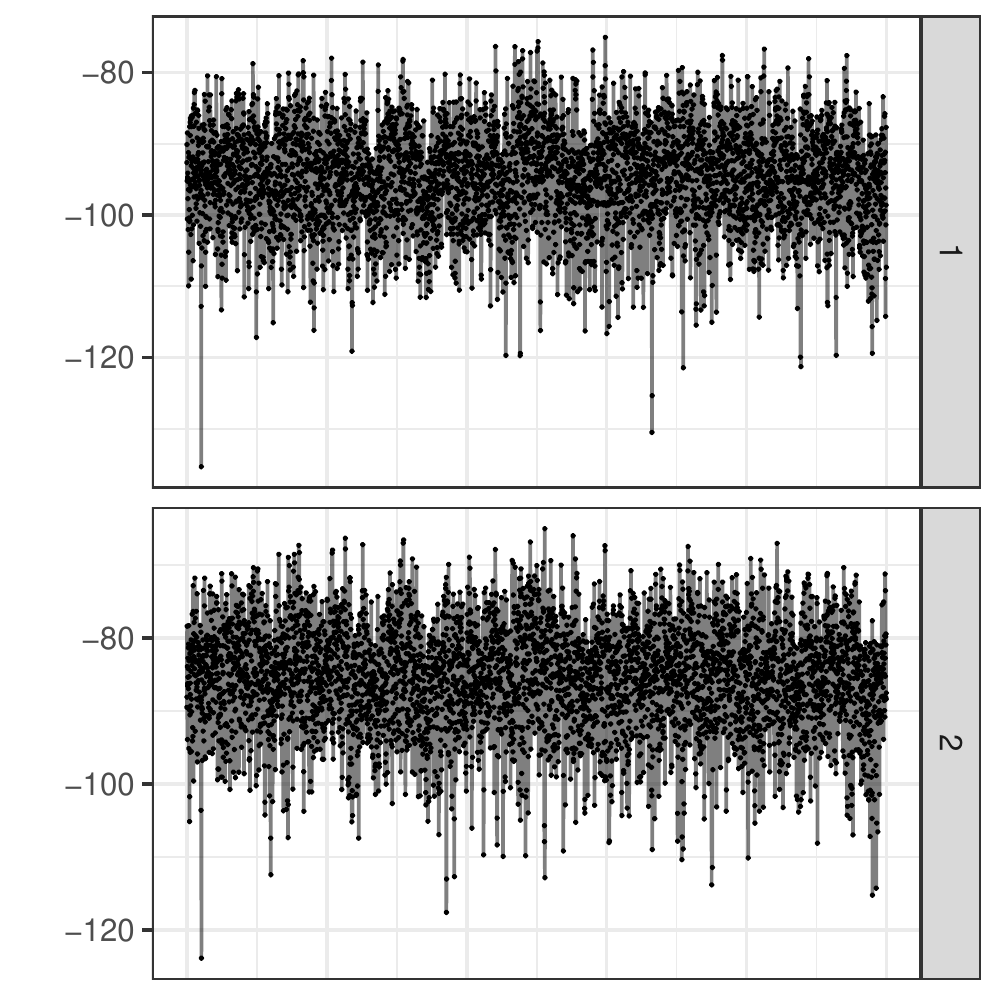}
  \end{minipage}
  \caption{Trace plots for a subset of various random variables in the spike and slab model. Left: the  coefficients visit zero with positive probability as expected. Right: log densities for two of the 20 tempered chains used in PT. }
  \label{fig:tracess}
\end{figure}

Figure \ref{fig:tracess} shows a subset of automatically post-processed trace plots.
Summary statistics for \code{SpikeSlabClassification} model's \code{parameters} are shown below (with truncated values):

\begin{center}
\begin{tabular}{llrrrrrrr}
  \hline
  index  & features & mean & sd & min & median & max & HDI.lower & HDI.upper \\
  \hline
  \hline
  1 & Age & $-0.02$ &$ 0.02 $&$ -0.11 $&$ -0.01 $&$ 0.01 $&$ -0.05 $& $0$ \\
  2 & child & $1.11$ &$ 1.04 $&$ -1.76 $&$ 0.97 $&$ 5.23 $&$ -0.11 $& $2.79$ \\
  3 & Fare & $0.00$ &$ 0.00 $&$ -0.01 $&$ 0 $&$ 0.02 $&$ 0 $& $0$ \\
  4 & female & $3.15$ &$ 0.44 $&$ 1.43 $&$ 3.14 $&$ 4.89 $&$ 2.43 $& $3.88$ \\
  5 & Par..Aboard & $0.01$ &$ 0.13 $&$ -0.79 $&$ 0 $&$ 0.89 $&$ -0.21 $&$ 0.25$ \\
  6 & Pclass & $-0.50$ &$ 0.29 $&$ -1.62 $&$ -0.51 $&$ 0.19 $&$ -0.88 $&$ 0$ \\
  7 & Sib..Aboard & $-0.71$ &$ 0.26 $&$ -1.90 $&$ -0.70 $&$ 0.07 $&$ -1.14 $&$ -0.27$ \\
  \hline
\end{tabular}
\end{center}

\newpage
\section{Internal architecture}

This section documents the high-level implementation decisions and trade-offs involved in the language construction. They may be skipped at first reading. 

\subsection{Language infrastructure}

\proglang{Blang} is developed using \proglang{Xtext}, a mature framework for programming language design supported by the Eclipse Foundation and TypeFox.
Thanks to the \proglang{Xtext} infrastructure, \proglang{Blang} incorporates a feature set comparable to many modern full-fledged multi-paradigm language: functional, generic and object programming, static typing.
\proglang{Blang} also automatically inherits state-of-the-art language development tools including a graphical integrated development environment (IDE) which leverages static types to provide insight into large \proglang{Blang} projects.
The IDE also has a full-feature debugger, and plug-ins have been tested to perform profiling and code coverage analysis. 

\subsection{Choice of compilation target}

Under the hood, \proglang{Blang} is compiled into \proglang{Java}, which in turn is compiled into Java Virtual Machine (JVM) bytecode.
This \emph{transpilation} step does not have marked effect on amortized compilation time since we use compilers supporting incremental compilation. 

This is the default model in \proglang{Xtext}, which, in addition to greatly simplifying \proglang{Xtext} development by using most of the provided default behaviour, has for the user's perspective two advantages related to performance and production deployment.
First, code running on modern JVM is fast.
For example, on the leading crowd-sourced language performance benchmark,\footnote{\url{https://benchmarksgame-team.pages.debian.net/benchmarksgame/which-programs-are-fastest.html}} as of June 2019, the geometric mean performance of \proglang{Java} is lower than \proglang{C++}, but higher than \proglang{Julia}, which itself outperforms the more common statistical computing choices such as \proglang{R} and \proglang{Python} by an order of magnitude or more.
The performance gains of advanced compilers such as \proglang{Java} and \proglang{Julia} over \proglang{R} and \proglang{Python} are especially important when dealing with combinatorial spaces where vectorization is generally not possible.
Other performance advantages include the JVM's high-performance multi-threading capacity and  garbage collection algorithms, which greatly facilitated the development of advanced Monte Carlo algorithms, for example for the parallel computation and memory management of particle genealogies.
The second advantage is related to production deployment.
\proglang{Java} is currently the most used language according to the TIOBE index as of June 2019, and this state may ease deployment of \proglang{Blang} software into existing production environments.

An often cited downside of using \proglang{Java} is its verbosity.
In our context, one specific concern is that more boilerplate code is typically needed to access high-performance computing libraries such as linear algebra libraries or random number generators.
Fortunately, \proglang{Blang} and \proglang{Xtend} avoid the key issues that make \proglang{Java} code verbose: checked exception, bad default behaviour for constructor/accessors, and redundant type declaration.
This brings \proglang{Blang} and \proglang{Xtend} code to a length similar to even non-statically typed language while preserving the advantages of the static type system.
We use \proglang{Blang} and \proglang{Xtend} advanced language features combined with allowed operator overloading to wrap existing dense and sparse matrix libraries into \pkg{xlinear}, a new linear algebra library written in \proglang{Xtend}, which provide succinct linear algebra expressions to \proglang{Blang}.
Similarly, we wrap existing random generation libraries into convenient \proglang{Xtend} extension methods. 

\subsection{Choice of sampler state representation} \label{app:samplerStateRep}

The state of the sampler is modified in place.
A priori, this choice appears in conflict with a popular doctrine in software engineering which is to avoid mutability and instead use functional-style idioms on immutable data structures.
While we agree these functional patterns are often tremendously helpful, in the context of our samples' state representation, we found mutable data structures more useful for three reasons.
First, the way we precompute a factor graph for efficient inference, via scoping analysis, assumes that certain references in the object graph stay invariant.
These invariant objects allow us to gain information on the scope and hence dependencies.
With functional style programming, we would trade immutability of the values into more mutability of the references making this scoping analysis complex.
Second, since the state objects are assembled and used in a completely automated way (via \proglang{Java} reflection), the user simply does not face the traps of mutable data structures in this specific context.
Third, there are computational complexity advantages to using mutable data structure: for example accessing or modifying array cost $O(1)$ instead of $O(\log n)$ for their functional copy on write counterparts.

\newpage
\section{Library dependencies} \label{app:dep}
\proglang{Blang}'s standard library uses its own language, and as such the majority of the dependencies were developed for \proglang{Blang}, and are handled automatically through \pkg{Maven} and Gradle.\footnote{\url{https://gradle.org/}}
Aside from libraries developed for \proglang{Blang}, \pkg{briefj}, \pkg{inits}, \pkg{bayonet}, \pkg{rejfree}, \pkg{binc}, \pkg{xlinear}, and \pkg{pxviz},\footnote{All of which are hosted on \url{https://github.com/UBC-Stat-ML/}.} the language depends on three additional, external libraries:
\pkg{Cloning}\footnote{\url{https://mvnrepository.com/artifact/uk.com.robust-it/cloning/1.9.6}}, \pkg{JGraphT}\footnote{\url{https://mvnrepository.com/artifact/org.jgrapht/jgrapht-core/0.9.0}}, and \pkg{Xbase}\footnote{\url{https://wiki.eclipse.org/Xbase}}.
For users who require automatic post-processors, \proglang{R} with packages \pkg{dplyr} and \pkg{ggplot2} are required.
Figure \ref{app:depstable} summarizes each of the aforementioned packages, while the remainder of this section expands on a select few that have been referenced earlier in the paper.\footnote{An exhaustive list of dependencies used by the \pkg{blangSDK} package and their versions can be obtained by typing \code{./gradlew dependencies} from the root of the \pkg{blangSDK} directory.}

\begin{figure}[H]
\centering
\begin{tabular}{l  l}
  \textbf{Library} & \textbf{Description} \\ [0.5em]
  \hline \\
  \text{\pkg{briefj}} & Utilities for writing succinct \proglang{Java} code. \\
  \text{\pkg{inits}} & A framework to organize inputs and outputs of scientific simulations. \\
  \text{\pkg{bayonet}} & Various low-level utilities for probabilistic inference.\\
  \text{\pkg{binc}} & An interface for calling binary programs from \proglang{Java} applications. \\
  \text{\pkg{xlinear}} & Linear algebra package for \proglang{Xtend} and \proglang{Java}. \\
  \text{\pkg{pxviz}} & A visualization library. \\ [1em]
  \hline \hline \\
  \text{\pkg{Cloning}} & Deep cloning library for \proglang{Java}.\\
  \text{\pkg{JGraphT}} & Graph theory data structures and algorithms for optimizing samplers. \\
  \text{\pkg{MathCommons}} & The Apache Commons Mathematics Library. \\
  \text{\pkg{Xbase}} & Used as the base language for the DSL.\\ [1em]
   \hline
\end{tabular}
\caption{A summary of library dependencies (automatically downloaded during installation, along with the transitive closure of these dependencies).}
\label{app:depstable}
\end{figure}

\subsection[bayonet]{\pkg{bayonet}} \label{app:bayonet}
The \pkg{bayonet} \url{https://github.com/UBC-Stat-ML/bayonet} library contains utilities for performing probabilistic inference.
\proglang{Blang} uses \pkg{bayonet.distribution.Random} as a replacement for \pkg{java.util.Random} for random number generation. This alternative is compatible with both \proglang{Java} and \pkg{Math Commons} random types.
\pkg{bayonet.math.SpecialFunctions} provides several statistical utility functions that are used heavily in \proglang{Blang}.

\subsection[inits]{\pkg{inits}} \label{app:inits}
\pkg{inits} \url{https://github.com/UBC-Stat-ML/inits} is a framework for performing scientific simulations, and can be viewed as a dependency injection framework tailored to complex and hierarchical command-line arguments.
\proglang{Blang}'s CLI argument setup is automatically handled by \pkg{inits}.

\subsection[xlinear]{\pkg{xlinear}} \label{app:xlinear}
\proglang{Blang}'s linear algebra is based on \pkg{xlinear} \url{https://github.com/UBC-Stat-ML/xlinear}, which itself relies on \pkg{Apache Commons}, \pkg{parallel COLT}, and \pkg{JEigen}.
The simple API of \pkg{xlinear} and the operator overloading functionality is what is leveraged in \proglang{Blang} to augment the \code{DenseMatrix} and \code{SparseMatrix} types into \code{DenseSimplex} and \code{DenseTransitionMatrix}.



\newpage
\section{Output format}\label{app:outputFormat}
\subsubsection{Output organization}
Every \proglang{Blang} execution creates a unique directory. The path is outputted to standard out at the end of the program's execution/run. The latest run is also softlinked at \code{results/latest}.

The directory has the following structure:
\begin{itemize}
\item \code{arguments-details.txt}: a detailed list of all arguments and options.
\item \code{arguments.tsv}: arguments used in current run.
\item \code{executionInfo}:  information for reproducibility (JVM arguments, version of the code, standard out, etc).
\item \code{init}: information about the initialization process.
\item \code{monitoring}: diagnostics for  samplers.
\item \code{samples}: samples from the target distribution. By default each random variable in the running model is output for each iteration (to disable this for some variables, e.g., those that are fully observed, use \code{-{}-excludeFromOutput}).
\item \code{logNormalizationEstimate.csv}: estimate of the natural logarithm of the probability of the data (also known as the log of the normalization constant of the prior times the likelihood, integrating over the latent variables). 
\end{itemize}

Additional files and directories if \code{-{}-postProcessor DefaultPostProcessor} is specified:
\begin{itemize}
\item \code{ess}: information for ess and energy for each chain.
\item \code{monitoringPlots}: sampler diagnostic plots
\item \code{posteriorPlots}: posterior densities and probability mass functions.
\item \code{summaries}: summary statistics such as posterior means, HDIs, etc.
\item \code{tracePlots}: trace plots for the random variables, log-density, and energy for each chain with burn-in samples discarded.
\item \code{tracePlotsFull}: trace plots with all samples included.
\end{itemize}

\subsubsection{Format of the samples}\label{sec:tidily}
Posterior samples are stored in Tidy CSV files.
For e.g., two samples for a \code{java.util.List} of three \code{RealVar}'s would look like:
\begin{center}
\begin{tabular}{c c c c c}
\hline
  \textbf{index} & \textbf{sample} & \textbf{value} \\
\hline
\hline
  $0$ & $0$ & $0.453$ \\
  $1$ & $0$ & $0.386$ \\
  $2$ & $0$ & $0.886$ \\
  $0$ & $1$ & $0.520$ \\ 
  $1$ & $1$ & $0.345$ \\ 
  $2$ & $1$ & $0.940$ \\ 
\hline
\end{tabular}
\end{center}
By default, the method \code{toString} is used to create the last column (value).
How can this be modified to encompass arbitrary data types?
For example, how do we output an object from permutation space (as in Section~\ref{sec:customSamplers}) as a Tidy CSV like below:

\begin{center}
\begin{tabular}{c c c c}
\hline
  \textbf{index} & \textbf{permutation\_index} & \textbf{sample} & \textbf{value} \\
\hline
\hline
  $0$ & $0$ & $0$ & $2$ \\
  $0$ & $1$ & $0$ & $0$ \\
  $0$ & $2$ & $0$ & $1$ \\ 
  $1$ & $0$ & $0$ & $1$ \\ 
  \vdots &\vdots &\vdots & \vdots \\
\hline
\end{tabular}
\end{center}

This behaviour can be customized to adhere to the Tidy philosophy by implementing the interface \code{TidilySerializable} for a class of arbitrary data type.\footnote{\api{inits}{blang/inits/experiments/tabwriters/TidilySerializable}{blang.inits.experiments.tabwriters.TidilySerializable}}
The method \code{serialize} is invoked and passed an instance of \code{Context}.\footnote{See the Context design pattern.}
Using \code{context.recurse(Object child, Object key, Object value )}, we can instruct the sampler to parse and output the custom data type:

\longline

\code{Permutation.xtend} \vspace*{-10pt}

\longline
\begin{lstlisting}
override void serialize(Context context) {
  for (int i : 0 ..< componentSize)
    context.recurse(connections.get(i), "permutation_index", i)
} 
\end{lstlisting}
\longline

The \code{child} argument is the value to write.
The \code{key} is the name of the key, for example \code{permutation_index}.
The \code{value} is the value of the key, for example index \code{i} of value in object.

Additional examples can be found in \code{TestTidySerializer.xtend}.\footnote{\url{https://github.com/UBC-Stat-ML/inits/blob/master/src/test/java/blang/inits/TestTidySerializer.xtend}}

\subsubsection{Output options}
The following command-line arguments can be used to tune the output:
\begin{itemize}
\item \code{-{}-excludeFromOutput}: space-separated list of random variables to exclude from output.
\item \code{-{}-experimentConfigs.managedExecutionFolder}: set to false in order to output in the current folder instead of in the unique folder created in results/all.
\item \code{-{}-experimentConfigs.recordExecutionInfo}: set to false to skip recording the reproducibility information in executionInfo.
\item \code{-{}-experimentConfigs.recordGitInfo}: set to false to skip git repository lookup for the code.
\item \code{-{}-experimentConfigs.saveStandardStreams}: set to false to skip recording the standard out and err.
\item \code{-{}-experimentConfigs.tabularWriter}: by default set to \code{CSV}. Can set to \code{Spark} to organize Tidy output into a hierarchy of directories each having a CSV (with less columns, as many columns in this format can now be inferred from the names of the parent directories). In certain scenarios this could save disk space. Inter-operable with Spark.
\end{itemize}

\newpage

\section[List of probability distributions in Blang's library]{List of probability distributions in \proglang{Blang}'s library}\label{app:list-distributions}
\maketitle

\subsection{Discrete distributions}\label{app:discretedists}
\vspace{2em}

Random variables in this section are integer-valued, hence \code{IntVar}s.

\code{Bernoulli}: Any random variable taking values in
\{0, 1\}.

\begin{itemize}

\item
  \code{param\ RealVar\ probability}: Probability $p
   \in [0, 1]$ that the realization is
  one.
\end{itemize}

\code{BetaBinomial}: A sum of $n$ IID
Bernoulli variables, with a marginalized Beta prior on the success
probability. Values in $(0, 1, 2,
 \dots, n)$.

\begin{itemize}

\item
  \code{param\ IntVar\ numberOfTrials}: The number
  $n$ of Bernoulli variables being summed.
  $n \textgreater{} 0$
\item
  \code{param\ RealVar\ alpha}: Higher values brings mean closer to
  one.  $\alpha \textgreater 0$
  
\item
  \code{param\ RealVar\ beta}: Higher values brings mean closer to
  zero. $\beta \textgreater 0$
  
\end{itemize}

\code{Binomial}: A sum of $n$ iid
Bernoulli variables. Values in \{0, 1, 2,
 \dots, n\}.

\begin{itemize}

\item
  \code{param\ IntVar\ numberOfTrials}: The number
  $n$ of Bernoulli variables being summed.
  $n \textgreater 0$
\item
  \code{param\ RealVar\ probabilityOfSuccess}: The parameter
  $p  \in [0, 1]$ shared by
  all the Bernoulli variables (probability that they be equal to 1).
\end{itemize}

\code{Categorical}: Any random variable over a finite set
\{0, 1, 2,  \dots,
$n-1$\}.

\begin{itemize}

\item
  \code{param\ Simplex\ probabilities}: Vector of probabilities
  $(p_0, p_1,  \dots,
  p_{n-1})$ for each of the
  $n$ integers.
\end{itemize}

\code{DiscreteUniform}: Uniform random variable over the contiguous
set of integers $\{m, m+1,
 \dots, M-1\}$.

\begin{itemize}

\item
  \code{param\ IntVar\ minInclusive}: The left point of the set
  (inclusive). $m \in(-\infty, M)$
\item
  \code{param\ IntVar\ maxExclusive}: The right point of the set
  (exclusive). $M  \in (m,
   \infty)$
\end{itemize}
\code{Geometric}: The number of unsuccessful Bernoulli trials until a
success. Values in $\{0, 1, 2,
 \dots\}$

\begin{itemize}

\item
  \code{param\ RealVar\ p}: The probability of success for each
  Bernoulli trial.
\end{itemize}

\code{HyperGeometric}: Hyper-geometric distribution with population $N$
and population satisfying certain condition $K$ and drawing $n$ samples.

\begin{itemize}

\item
  \code{param\ IntVar\ numberOfDraws}: number of samples $n$
\item
  \code{param\ IntVar\ population}: number of population $N$
\item
  \code{param\ IntVar\ populationConditioned}: number of population
  satisfying condition $K$
\end{itemize}

\code{NegativeBinomial}: Number of successes in a sequence of iid
Bernoulli until (r) failures occur. Values
in $\{0, 1, 2,
 \dots\}$.

\begin{itemize}

\item
  \code{param\ RealVar\ r}: Number of failures until experiment is
  stopped (generalized to the reals). $r \textgreater
  0$
\item
  \code{param\ RealVar\ p}: Probability of success of each experiment.
  $p  \in (0, 1)$
\end{itemize}

\code{Poisson}: Poisson random variable. Values in $\{0,
1, 2,  \dots\}$.

\begin{itemize}

\item
  \code{param\ RealVar\ mean}: Mean parameter
  $\lambda$.
  $\lambda \textgreater 0$
\end{itemize}

\code{YuleSimon}: An exponential-geometric mixture.

\begin{itemize}

\item
  \code{param\ RealVar\ rho}: The rate of the mixing exponential
  distribution.
\end{itemize}

\vspace{2em}
\subsection{Continuous distributions} \label{app:continuousdists}
\vspace{2em}

Random variables in this section are real-valued, hence \code{RealVar}s.

\code{Beta}: Beta random variable on the open interval
(0, 1).

\begin{itemize}

\item
  \code{param\ \ RealVar\ alpha}: Higher values brings mean closer to
  one. $\alpha \textgreater 0$
  
\item
  \code{param\ \ RealVar\ beta}: Higher values brings mean closer to
  zero. $\beta \textgreater 0$
\end{itemize}

\code{ChiSquared}: Chi Squared random variable. Values in
$(0,  \infty)$.

\begin{itemize}

\item
  \code{param\ \ IntVar\ nu}: The degrees of freedom
  $\nu$. 
   $\nu \textgreater 0$
\end{itemize}

\code{ContinuousUniform}: Uniform random variable over a close
interval $[m, M]$.

\begin{itemize}

\item
  \code{param\ \ RealVar\ min}: The left end point
  $m$ of the interval. $m
  \in (\infty, M)$
\item
  \code{param\ \ RealVar\ max}: The right end point of the interval.
  $M  \in (m,
   \infty)$
\end{itemize}

\code{Exponential}: Exponential random variable. Values in
$(0,  \infty)$.

\begin{itemize}

\item
  \code{param\ \ RealVar\ rate}: The rate
  $\lambda$, inversely
  proportional to the mean.  $\lambda
  \textgreater 0$
\end{itemize}

\code{F}: The F-distribution. Also known as Fisher-Snedecor
distribution. Values in $(0, \infty)$.

\begin{itemize}

\item
  \code{param\ RealVar\ d1,\ d2}: The degrees of freedom
  $d_1$ and $d_2$
  . $d_1, d_2 \textgreater 0$
\end{itemize}

\code{Gamma}: Gamma random variable. Values in $(0,
 \infty)$.

\begin{itemize}

\item
  \code{param\ \ RealVar\ shape}: The shape
  $\alpha$ is proportional to
  the mean and variance. $\alpha
  \textgreater 0$
\item
  \code{param\ \ RealVar\ rate}: The rate
  $\beta$ is inverse
  proportional to the mean and quadratically inverse proportional to the
  variance. $\beta \textgreater 0$
\end{itemize}

\code{Gompertz}: The Gompertz distribution. Values in
$[0,  \infty)$.

\begin{itemize}

\item
  \code{param\ RealVar\ shape}: The shape parameter
  $\nu$.
  $nu \textgreater 0$
\item
  \code{param\ RealVar\ scale}: The scale parameter
  $b$. $b \textgreater 0$
  )
\end{itemize}

\code{Gumbel}: The Gumbel Distribution. Values in $
 \mathbb{R}$.

\begin{itemize}

\item
  \code{param\ RealVar\ location}: The location parameter
  $\mu$. 
   $\mu  \in  \mathbb{R}$
\item
  \code{param\ RealVar\ scale}: The scale parameter
  $\beta$. 
   $\beta \textgreater 0$
\end{itemize}

\code{HalfStudentT}: HalfStudentT random variable. Values in
$(0,  \infty)$.

\begin{itemize}

\item
  \code{param\ RealVar\ nu}: A degree of freedom parameter
  $\nu$. 
   $\nu \textgreater 0$
\item
  \code{param\ RealVar\ sigma}: A scale parameter
  $\sigma$. 
   $\sigma \textgreater{} 0$
\end{itemize}

\code{Laplace}: The Laplace Distribution over
$\mathbb{R}$.

\begin{itemize}

\item
  \code{param\ RealVar\ location}: The mean parameter.
\item
  \code{param\ RealVar\ scale}: The scale parameter $b$, equal to the square root of half of the variance.
  $b \textgreater 0$
\end{itemize}

\code{Logistic}: A random variable with a logistic probability
distribution function. Values in 
 $\mathbb{R}$.

\begin{itemize}

\item
  \code{param\ RealVar\ location}: The centre of the PDF. Also the
  mean, mode and median. $\mu
   \in  \mathbb{R}$
\item
  \code{param\ RealVar\ scale}: The scale parameter.
  $s \textgreater 0$
\end{itemize}

\code{LogLogistic}: A log-logistic distribution is the probability
distribution of a random variable.

\begin{itemize}

\item
  \code{param\ RealVar\ scale}: The scale parameter
  $\alpha$ and also the
  median. $\alpha \textgreater 0$
\item
  \code{param\ RealVar\ shape}: The shape parameter
  $\beta$.
  $\beta \textgreater 0$
\end{itemize}

\code{Normal}: Normal random variables. Values in
$\mathbb{R}$.

\begin{itemize}

\item
  \code{param\ RealVar\ mean}: Mean
  $\mu$.
  $\mu  \in
   \mathbb{R}$
\item
  \code{param\ RealVar\ variance}: Variance
  $\sigma^2$.
  $\sigma^2 \textgreater
  0$
\end{itemize}

\code{StudentT}: Student T random variable. Values in
$\mathbb{R}$.

\begin{itemize}

\item
  \code{param\ RealVar\ nu}: The degrees of freedom
  $\nu$. 
   $\nu \textgreater 0$
\item
  \code{param\ RealVar\ mu}: Location parameter
  $\mu$.
  $\mu  \in
   \mathbb{R}$
\item
  \code{param\ RealVar\ sigma}: Scale parameter
  $\sigma$.
  $\sigma \textgreater 0$
\end{itemize}

\code{Weibull}: The Weibull Distribution. Values  in
$(0,  \infty)$.

\begin{itemize}

\item
  \code{param\ RealVar\ scale}: The scale parameter
  $\lambda$. $\lambda  \in (0, \infty)$
\item
  \code{param\ RealVar\ shape}: The shape parameter
  $k$. $k  \in
  (0, \infty)$
\end{itemize}

\vspace{2em}
\subsection{Multivariate distributions} \label{app:multidists}
\vspace{2em}

\code{Dirichlet}: The Dirichlet distribution over vectors of
probabilities $(p_0, p_1,  \dots,
p_{n-1})$. $p_i  \in (0,
1)$,  $\sum_i p_i = 1$.
Random variables with this distribution are of type \code{Simplex}.

\begin{itemize}

\item
  \code{param\ \ Matrix\ concentrations}: Vector
  $(\alpha_0,  \alpha_1,
   \dots,  \alpha_{n-1})$
  such that increasing the $i$th component
  increases the mean of entry $p_i$.
\end{itemize}

\code{MultivariateNormal}: Arbitrary linear transformations of
$n$ iid standard normal random variables.
Random variables with this distribution are of type \code{Matrix}.

\begin{itemize}

\item
  \code{param\ Matrix\ mean}: An $n  \times
  1$ vector
  $\mu$.
  $\mu  \in
   \mathbb{R}^n$
\item
  \code{param\ CholeskyDecomposition\ precision}: Inverse covariance
  matrix  $\Lambda$, a positive
  definite $n  \times n$
  matrix.
\end{itemize}

\code{NormalField}: A mean-zero normal, sparse-precision Markov random
field.
Random variables with this distribution are of type \code{Plated<RealVar>}.

\begin{itemize}

\item
  \code{param\ Precision\ precision}: Precision matrix structure.
\end{itemize}

\code{SimplexUniform}: $n$ dimensional
Dirichlet with all concentrations equal to one.
Random variables with this distribution are of type \code{Simplex}.

\begin{itemize}

\item
  \code{param\ Integer\ dim}: The dimensionality
  $n$. $n \textgreater 0$
\end{itemize}

\code{SymmetricDirichlet}: $n$ dimensional
Dirichlet with all concentrations equal to
$\frac{\alpha}{n}$.
Random variables with this distribution are of type \code{Simplex}.

\begin{itemize}

\item
  \code{param\ Integer\ dim}: The dimensionality
  $n$. $n \textgreater 0$
\item
  \code{param\ RealVar\ concentration}: The shared concentration
  parameter $\alpha$ before
  normalization by the dimensionality.
  $\alpha \textgreater 0$
\end{itemize}

\vspace{3em}
\subsection*{Miscellaneous} \label{app:miscdist}

\code{LogPotential}: A utility to
handle undirected models (or random fields). 

\begin{itemize}

\item
  \code{param\ RealVar\ logPotential}: The log of the current value of
  this potential.
\end{itemize}

\newpage
\section{Frequently used functions}\label{app:functions}

Any \proglang{Java} function can be called in \proglang{Blang}. The functions in Figure \ref{app:javafuncs} and Figure \ref{app:bayonetfuncs} are automatically and statically imported for easy access.
The functions below are the most useful of those imported and in addition to the functions, \proglang{Blang} also imports two fields from \pkg{java.lang.Math}, which are \code{E} and \code{PI}. \\

\begin{figure}[H]
\centering
\begin{tabular}{l  l }
\hline
  Function & Description \\ [0.5ex]
  \hline\hline
  \code{abs(double value)} & absolute value \\
  \code{acos(double a)} & arccosine \\
  \code{asin(double a)} & arcsine \\
  \code{atan(double a)} & arctangent \\
  \code{cbrt(double a)} & cube root \\
  \code{ceil(double a)} & ceiling \\
  \code{cos(double a)} & cosine \\
  \code{cosh(double a)} & hyperbolic cosine \\
  \code{exp(double a)} & exponential base $e$ \\
  \code{floor(double a)} & floor \\
  \code{log(double a)} & logarithm base $e$ \\
  \code{log10(double a)} & logarithm base $10$ \\
  \code{max(double a, double b)} & maximum of \code{a} and \code{b} \\
  \code{min(double a, double b)} & minimum of \code{a} and \code{b}  \\
  \code{pow(double a, double b)} & \code{a} to the power \code{b} \\
  \code{signum(double a)} & signum function \\
  \code{sin(double a)} & sine \\
  \code{sinh(double a)} & hyperbolic sine \\
  \code{sqrt(double a)} & square root \\
  \code{tan(double a)} & tangent \\
  \code{tanh(double a)} & hyperbolic tangent \\ [2ex]
  \hline
  \hline
  Imported Fields & Description \\ [0.5ex]
  \hline\hline
  \code{E} & \proglang{Java}'s \code{double} value for $e$ \\
  \code{PI} & \proglang{Java}'s \code{double} value for $\pi$ \\ [1ex]
\hline
\end{tabular}
\caption{Functions imported from \pkg{java.lang.Math}. Note that all trigonometric operations use angles expressed in radians and that the return type of all functions listed above are \code{double}.}
\label{app:javafuncs}
\end{figure}

\begin{figure}[H]
\resizebox{\textwidth}{!}{
\begin{tabular}{l  l }
\hline
  Function & Description \\ [0.5ex]
  \hline\hline
  \code{erf(double a)} & error function \\
  \code{inverseErf(double a)} & inverse error function \\
  \code{logistic(double a)} & standard logistic function \\
  \code{logit(double a)} & standard logit function \\
  \code{logBinomial(int n, int k)} & logarithm of ${n \choose k}$  \\
  \code{lnGamma(double alpha)} & logarithm of the gamma function of \code{alpha} \\
  \code{logFactorial(int input)} & logarithm of the factorial of \code{input} \\
  \code{multivariateLogGamma(int dim, double a)} & logarithm of the multivariate gamma function \\ [1ex]
\hline
\end{tabular}
}
\caption{Functions imported from \pkg{bayonet.math.SpecialFunctions}. All return types are \code{double}.}
\label{app:bayonetfuncs}
\end{figure}

\begin{figure}[H]
\centering
\resizebox{\textwidth}{!}{
\begin{tabular}{l  p{0.4\linewidth}  l}
  \hline
  Function & Description & Return Type \\ [0.5em]
  \hline\hline
  \code{latentInt()} & unobserved integer variable (initialized at zero) & \code{IntScalar} \\
  \code{latentReal()} & unobserved real variable (represented as a double, initialized at zero) & \code{RealScalar} \\
  \code{fixedInt(int value)} & fixed (constant or conditioned upon) integer scalar & \code{IntConstant} \\
  \code{fixedReal(double value)} & fixed real scalar & RealConstant \\
  \code{latentIntList(int size)} & size specifies the length of the list & \code{List<IntVar>} \\
  \code{latentRealList(int size)} & size specifies the length of the list & \code{List<RealVar>} \\
  \code{fixedIntList(int ... entries)} & list where the integer valued entries are fixed to the provided values & \code{List<IntVar>} \\
  \code{latentVector(int n)} & an n-by-1 latent dense vector (initialized at zero) & \code{DenseMatrix} \\
  \code{fixedVector(double ... entries)} & an n-by-1 fixed dense vector & \code{DenseMatrix} \\ [1em]
\hline
\end{tabular}
}
\caption{Functions used to initialize random variables.}
\label{app:rvarfuncs}
\end{figure}

\begin{figure}[H]
\centering
\resizebox{\textwidth}{!}{
\begin{tabular}{p{0.415\linewidth}  p{0.35\linewidth}  l}
\hline
  Function & Description & Return Type \\ [0.5em]
  \hline\hline
  \code{latentMatrix(int nRows, int nCols)} & an n-by-m latent dense matrix (initialized at zero) & \code{DenseMatrix} \\
  \code{fixedMatrix(double [][] entries)} & a constant dense matrix & \code{DenseMatrix} \\
  \code{latentSimplex(int n)} & latent n-by-1 matrix with entries summing to one (initialized at uniform) & \code{DenseSimplex} \\
  \code{fixedSimplex(double ... probs)} & creates a constant simplex, also checks the provided list of number sums to one & \code{DenseSimplex} \\
  \code{fixedSimplex(DenseMatrix probs)} & creates a constant simplex, also checks the provided vector sums to one & \code{DenseSimplex} \\
  \code{latentTransitionMatrix(int nStates)} & latent n-by-n matrix with rows summing to one & \code{DenseTransitionMatrix} \\
  \code{fixedTransitionMatrix(DenseMatrix probs)} & creates a constant transition matrix, also checks the provided rows all sum to one & \code{DenseTransitionMatrix} \\
  \code{fixedTransitionMatrix(double [][] probs)} & creates a constant transition matrix, also checks the provided rows all sum to one. & \code{DenseTransitionMatrix} \\ [1em]
\hline
\end{tabular}
}
\caption{Frequently used functions in \proglang{Blang} to complement the set of functions from the \pkg{xlinear} library.}
\label{app:linalgfuncs}
\end{figure}

\begin{figure}[H]
\centering
\resizebox{\textwidth}{!}{
\begin{tabular}{p{0.415\linewidth}  p{0.35\linewidth}  l}
\hline
  Function & Description & Return Type \\ [0.5em]
  \hline\hline
  \code{getRealVar(Matrix m, int row, int col)} & View a single entry of a \code{Matrix} as a \code{RealVar} & \code{Realvar} \\
  \code{getRealVar(Matrix m, int index)} & View a single entry of a 1-by-n or n-by-1 Matrix as a \code{RealVar}. & \code{RealVar} \\
  \code{asBool(int i)} & Returns \code{false} for \code{0} and \code{true} for \code{1} & \code{boolean} \\
  \code{asBool(IntVar i)} & Returns \code{false} for \code{0} and \code{true} for \code{1} & \code{boolean} \\
  \code{isBool(int i)} & Returns \code{true} if \code{i} is \code{0} or \code{1} and \code{false} otherwise. & \code{boolean} \\
  \code{isBool(IntVar i)} & Returns \code{true} if \code{i} is \code{0} or \code{1} and \code{false} otherwise. & \code{boolean} \\
  \code{asInt(boolean b)} & Returns \code{0} for \code{false} and \code{1} for \code{true} & \code{int} \\
  \code{asList(Plated<T> plated, Plate<Integer> plate)} & Returns a the plated variable as a list. & \code{List<T>} \\
  \code{asCollection(Plated<T> plated, Plate<Integer> plate)} & Returns a the plated variable as a collection. & \code{Collection<T>} \\
  \code{asMap(Plated<T> plated, Plate<Integer> plate)} & Returns a the plated variable as a map. & \code{Map<K, T>} \\
  \code{generator(java.util.Random random)} &  Upgrade a \code{java.util.Random} into to the type of \code{Random} \proglang{Blang} uses, \code{bayonet.distributions.Random}. & \code{Random} \\
  \code{setTo(Matrix one, Matrix another)} &  Copy the contents of a matrix into another one. & \code{void} \\
  \code{sum(Iterable<? extends Number> numbers)} & & \\
  \code{increment(Map<T, Double> map, T key, double value)} &  Increment an entry of a map to double, setting to the value if the key is missing. &  \code{void} \\
  \code{isClose(double n1, double n2)} &  Check if two numbers are within 1e-6 of each other. & \code{boolean} \\ [1em]
\hline
\end{tabular}
}
\caption{Extension methods (automatically) imported from \pkg{blang.types.ExtensionUtils}.}
\label{app:extenfuncs}
\end{figure}

\begin{figure}[H]
\centering
\resizebox{\textwidth}{!}{
\begin{tabular}{p{0.35\linewidth}  p{0.415\linewidth}  l}
\hline
  Function & Description & Return Type \\ [0.5em]
  \hline\hline
  \code{getName()} &  Human-readable name for the plate, typically automatically extracted from a \code{DataSource} column name. & \code{ColumnName} \\
  \code{indices(Query parentIndices)} &  Get the indices available given the indices of the parent (enclosing) plates. The parents can be provided in any order. & \code{Collection<Index<K>>} \\
  \code{index(K key)} &  Get the index given a \code{key}. & \code{Index<K>} \\
  \code{ofIntegers(ColumnName columnName, int size)} &  a plate with indices \code{0, 1, 2, ... size-1} & \code{Plate<Integer>} \\
  \code{ofStrings(ColumnName columnName, int size)} &  a plate with indices \code{category\_0, category\_1, ...} & \code{Plate<String>} \\
  \code{ofStrings(String columnName, int size)} &  a plate with indices \code{category\_0, category\_1, ...} & \code{Plate<String>} \\ [1em]
  \hline
  \code{get(Index<?> ... indices)} & get the random variable or parameter indexed by the provided indices. The indices can be given in any order. & \code{T} \\ [0.5em]
  \code{entries()} & list all variables obtained through \code{get(...)} so far. Each returned entry contains the variable as well as the associated indices (\code{Query}). & \code{Collection<Entry<Query, T>>} \\
  \code{slice(Index<?> ... indices)} & a view into a subset of the plated variable. & \code{Plated<T>} \\
  \code{latent(ColumnName name, Supplier<T> supplier)} & use the provided lambda expression to initialize several latent variables. & \code{<T> Plated<T>} \\ [1em]
\hline
\end{tabular}
}
\caption{Functions and methods related to \code{Plates} (above) and \code{Plated} (below) types.}
\label{app:platefuncs}
\end{figure}

\end{appendix}

\end{document}